\documentclass[12pt]{article}
\usepackage{amsmath}
\usepackage{slashed}
\usepackage{amssymb}
\usepackage{epsfig}
\usepackage{graphicx}
\usepackage{multirow}
\usepackage{color}
\usepackage[normal]{subfigure}
\usepackage{rotating}
\usepackage{hyperref}
\usepackage[margin=0.9in]{geometry}
\usepackage[table]{xcolor}
\usepackage{enumitem}
\usepackage[utf8x]{inputenc}
\usepackage[compress,numbers,sort]{natbib}
\usepackage{authblk}
\usepackage{colortbl}
\usepackage{pdflscape}
\usepackage{xspace}
\definecolor{LightCyan}{rgb}{0.88,1,1}
\definecolor{piggypink}{rgb}{0.99, 0.87, 0.9}

\usepackage{ucs}
\usepackage{amsmath}
\usepackage{amsfonts}
\usepackage{amssymb}
\usepackage{eucal}               
\usepackage{mathrsfs}          
\usepackage[sans]{dsfont}     
\usepackage[Symbol]{upgreek}  

\usepackage{array}             
\usepackage{booktabs}       
\usepackage{colortbl}          
\usepackage{multirow}
\usepackage{longtable}

\usepackage{cleveref}

\usepackage{fancyhdr}

\usepackage[normalem]{ulem}             
\usepackage{cancel}           

\usepackage{slashed}
\usepackage{simplewick}     
\usepackage{feynmf}           

\usepackage{color}
\definecolor{grigio}{cmyk}{0,0,0,0.1}
\definecolor{rosa}{cmyk}{0,0.1,0.1,0.02}
\definecolor{rosino}{cmyk}{0,0.05,0.05,0.02}
\definecolor{rosas}{cmyk}{0,0.3,0.25,0.05}
\definecolor{celeste}{cmyk}{0.1,0,0,0.02}
\definecolor{giallino}{cmyk}{0,0,0.1,0.02}
\definecolor{rosso}{cmyk}{0,1,1,0.4}
\definecolor{rossos}{cmyk}{0,1,1,0.55}
\definecolor{rossoc}{cmyk}{0,1,1,0.2}
\definecolor{blu}{cmyk}{1,1,0,0.3}
\definecolor{blus}{cmyk}{1,1,0,0.5}
\definecolor{bluc}{cmyk}{1,1,0,0.1}
\definecolor{blucc}{cmyk}{0.7,0.5,0,0}
\definecolor{viola}{cmyk}{0,1,0,0.6}
\definecolor{viola2}{cmyk}{0,1,0.2,0.6}
\definecolor{verde}{cmyk}{0.92,0,0.59,0.25}
\definecolor{verdec}{cmyk}{0.92,0,0.59,0.15}
\definecolor{verdes}{cmyk}{0.92,0,0.59,0.4}
\definecolor{verdino}{cmyk}{0.12,0,0.09,0.02}
\definecolor{giallo}{cmyk}{0,0,1,0}
\definecolor{gialloverde}{cmyk}{0.44,0,0.74,0}
\definecolor{Titolo}{rgb}{0.752941176,0.576470588,0.992156863}
\definecolor{altro}{rgb}{0.094117647,0.650980392,0.643137255}
\definecolor{Peanuts}{rgb}{0.2, 0.4, 0.6}
\definecolor{Pean1}{rgb}{0.6, 0.8, 0.4}
\definecolor{BHO}{rgb}{0.2, 0.8, 1}
\definecolor{Daria}{rgb}{0, 0.9412, 0}
\definecolor{UniPi}{rgb}{0.2549, 0.4627, 0.6275}
\definecolor{UniPidue}{rgb}{0.3216, 0.5804, 0.7882}
\definecolor{rossoCP3}{cmyk}{0,.88,.77,.40}
\definecolor{verdeCP3}{rgb}{0.09765625, 0.57421875, 0.1015625}
\definecolor{bluCP3}{rgb}{0, 0.23, 0.67}

\newcommand{\bs}[1]{\boldsymbol #1}

\newcommand{\Eqs}[2]{Eqs.~\eqref{#1} and \eqref{#2}}

\newcommand{\eg}{e.g.~}

\newcommand{\Eq}[1]{Eq.~\eqref{#1}}
\newcommand{\Fig}[1]{Fig.~\ref{#1}}
\newcommand{\Figs}[2]{Figs.~\ref{#1} and \ref{#2}}
\newcommand{\Sec}[1]{Sec.~\ref{#1}}
\newcommand{\App}[1]{Appendix \ref{#1}}
\newcommand{\Ref}[1]{Ref.~\cite{#1}}           

\newcommand{\Mel}{\mathscr{M}}	

\newcommand{\be}{\begin{equation}}
\newcommand{\ee}{\end{equation}}
\newcommand{\ud}{\text{d}}

\font\bb=bbmss10 scaled 1200         

\newcommand{\ER}{E_\text{R}}

\newcommand{\vesc}{v_\text{esc}}
\newcommand{\vmin}{v_\text{min}}

\newcommand{\LHC}{{\sf LHC}}
\newcommand{\LUX}{{\sf LUX}}
\newcommand{\LZ}{{\sf LZ}}
\newcommand{\ZEPLIN}{{\sf ZEPLIN}}
\newcommand{\ATLAS}{{\sf ATLAS}\xspace}
\newcommand{\PICO}{{\sf PICO-2L}\xspace}

\newcommand{\bma}{\begin{pmatrix}}
\newcommand{\ema}{\end{pmatrix}}

\newcommand{\runDM}{\textsc{runDM}\xspace}

\font\bb=bbmss10 scaled 1200
\newcommand{\unop}{\mbox{\bb 1}}

\ifx\pdfoutput\undefined
\usepackage[dvips,bookmarks]{hyperref}    
\else
\usepackage{hyperref}    
\fi
\hypersetup{colorlinks, bookmarksopen, bookmarksnumbered,
citecolor=verdeCP3, linkcolor=bluCP3, pdfstartview=FitH, urlcolor=rossoCP3}

\numberwithin{equation}{section}

\begin{document}

\begin{flushright}
{\footnotesize
{\sc SCIPP 16/07}
}
\end{flushright}
\color{black}

\begin{center}

{\Huge\bf You can hide but you have to run: \\ [2mm] direct detection with vector mediators}

\medskip
\bigskip\color{black}\vspace{0.6cm}

{
{\large\bf Francesco D'Eramo}$\, ^{a, b}$,
{\large\bf Bradley J. Kavanagh}$\, ^{c, d}$,
{\large\bf Paolo Panci}$\, ^e$
}
\\[7mm]
{\it $^a$ \href{http://www.physics.ucsc.edu/}{Department of Physics, University of California Santa Cruz}, \\ 1156 High St., Santa Cruz, CA 95064, USA} \\ [2mm]
{\it $^b$ \href{http://scipp.ucsc.edu/}{Santa Cruz Institute for Particle Physics}, \\ 1156 High St., Santa Cruz, CA 95064, USA} \\ [2mm]
{\it $^c$ \href{http://ipht.cea.fr/}{Laboratoire de Physique Th\'{e}orique et Hautes Energies}, \\  
CNRS, UMR 7589, 4 Place Jussieu, F-75252, Paris, France} \\ [2mm]
{\it $^d$ \href{http://ipht.cea.fr/}{Institut de Physique Th\'{e}orique, Universit\'e Paris Saclay}, \\  
CNRS, CEA, F-91191 Gif-sur-Yvette, France} \\ [2mm]
{\it $^e$ \href{http://www.iap.fr}{Institut d'Astrophysique de Paris}, UMR 7095 CNRS, Universit\'e Pierre et Marie Curie, \\ 98 bis Boulevard Arago, Paris 75014, France}\\[2mm]
\end{center}

\bigskip

\centerline{\large\bf Abstract}
\begin{quote}

We study direct detection in simplified models of Dark Matter (DM) in which interactions with Standard Model (SM) fermions are mediated by a heavy vector boson. We consider fully general, gauge-invariant couplings between the SM, the mediator and both scalar and fermion DM. We account for the evolution of the couplings between the energy scale of the mediator mass and the nuclear energy scale. This running arises from virtual effects of SM particles and its inclusion is not optional. We compare bounds on the mediator mass from direct detection experiments with and without accounting for the running. In some cases the inclusion of these effects changes the bounds by several orders of magnitude, as a consequence of operator mixing which generates new interactions at low energy. We also highlight the importance of these effects when translating \LHC\ limits on the mediator mass into bounds on the direct detection cross section. For an axial-vector mediator, the running can alter the derived bounds on the spin-dependent DM-nucleon cross section by a factor of two or more. Finally, we provide tools to facilitate the inclusion of these effects in future studies: general approximate expressions for the low energy couplings and a public code \runDM to evolve the couplings between arbitrary energy scales.

\end{quote}

\newpage
\tableofcontents


\section{Introduction}

Astrophysical and cosmological observations over the last four decades have accumulated indisputable evidence for dark matter (DM)~\cite{Bertone:2004pz}. This important component of our universe, five times more abundant than baryonic matter, cannot be accounted for by any Standard Model (SM) degree of freedom. The question of the origin and composition of DM is hence among the most urgent in particle physics~\cite{Feng:2010gw}. A stable Weakly Interacting Massive Particle (WIMP) with weak-scale mass and cross section provides a compelling solution. WIMPs naturally appear in well-motivated beyond the SM frameworks~\cite{Goldberg:1983nd,Ellis:1983ew,Servant:2002aq,Cheng:2002ej,Birkedal:2006fz}. Interactions with SM fields, responsible in the early universe for relic DM production through thermal freeze-out~\cite{Lee:1977ua,Scherrer:1985zt,Srednicki:1988ce}, today allow different and complementary experimental search strategies. 

Crucially, each search strategy probes DM interactions at different energy scales. In order to properly explore complementarity, one must carefully make the connection between physics at high and low energy. Such a scale connection is achieved by employing techniques from Effective Field Theory (EFT), performing a Renormalization Group (RG) analysis of DM interactions. This RG Evolution (RGE) typically introduces mixing between different DM-SM interactions, affecting the size of couplings, or even inducing new couplings which do not appear in a na\"ive comparison \cite{Kopp:2009et,Hill:2011be,Frandsen:2012db,Haisch:2013uaa,Hill:2013hoa,Vecchi:2013iza,Kopp:2014tsa,Crivellin:2014qxa,Crivellin:2014gpa,Hill:2014yxa,D'Eramo:2014aba,Berlin:2015njh,D'Eramo:2016mgv}. These effects do not depend on the properties of the Dark Sector and are not optional; they arise solely from RGE of the couplings via SM loops.

Our focus here is on direct detection (DD) experiments \cite{Goodman:1984dc,Drukier:1986tm}, which search for $\mathcal O(\sim \, \rm keV)$ low energy DM-nucleon recoils, meaning that the separation of scales can be large compared with high energy probes such as the Large Hadron Collider (\LHC). DD experiments also involve matrix elements which are extremely sensitive to the Lorentz structure of the effective operator under consideration~\cite{Goodman:1984dc}, meaning that operator mixing can have a substantial impact.

This work analyzes theories of DM where interactions with SM fermions are mediated by a massive vector boson. At energies below the mediator mass, DM-SM couplings are described by higher-dimensional contact interactions between spin-1 currents, similarly to the low-energy theory for SM weak interactions. The scale connection in this class of models can be performed by solving the RG equations derived in Ref.~\cite{D'Eramo:2014aba}. Although the authors of \Ref{D'Eramo:2014aba} only considered fermionic DM $\chi$ interacting through the vector current $\overline{\chi} \gamma^\mu \chi$, the RGE is driven by the SM current part of the contact interactions, and thus their RG equations are valid for other DM spins and interaction structures. We expand and generalize that work in two key ways. First, we consider a wider range of DM spins and interaction structures, constructing the most general model for SM fermions interacting with scalar or fermion DM through the exchange of a vector mediator. Second, we do not limit the analysis to standard spin-dependent (SD) and spin-independent (SI) DM-nucleon interactions. We also consider higher-order non-relativistic DM-nucleon interactions \cite{Fitzpatrick:2012ix}, which in some cases give rise to limits where none are expected in the standard SI/SD framework. 

We present the results of our study as follows. In \Sec{sec:SimplifiedModels}, we introduce the simplified model framework for vector mediators, coupling the mediator to SM fermions without breaking SM electroweak gauge invariance. DD rates are computed by following the general procedure outlined in \Sec{sec:RGE}.
The key results of this paper are then summarized in \Cref{fig:universalQ_fermionDM,fig:universalQ_scalarDM,fig:universalL_fermionDM,fig:universalL_scalarDM,fig:universalThird_fermionDM,fig:universalThird_scalarDM} of \Sec{sec:benchmarks}, where we give DD limits on the mass of the mediator for three different sets of benchmark couplings. We quantify the size of the RGE effects by presenting our results both with and without accounting for them. These limits can be easily generalized to any choice of couplings by following the prescription given in \Sec{sec:rescaling}. Complementarity with \LHC\ bounds is explored in \Sec{sec:LHC}; Figure~\ref{fig:LHClimits} shows an example of how limits on simplified models from the \LHC\ can be translated into the $(m_\chi, \, \sigma_{\chi N})$ plane of direct detection and how the RGE affects this translation. 
Finally, our conclusions are given in \Sec{sec:conclusions}. 

We supplement our paper with four appendices. In \App{app:analytical} we give approximate analytical expressions for the low-energy couplings, which correspond to the output of a fixed-order calculation. These formulae are useful to get an order of magnitude estimate of the RGE effects. The Reader interested in a more refined analysis can obtain the results of the RGE by using the code \runDM, released together with this work. The code, which is available at \href{https://github.com/bradkav/runDM/}{this http URL}, evolves the effective couplings from the high to the low energy scale as described in \App{app:code}. More details about the nuclear scale matching between a theory of quarks and gluons and a theory of nucleons are provided in \App{app:NRoperators}. Finally, the projected exclusion limits from \LZ\ are obtained as described in \App{app:LZ}.


\section{Simplified Models for Vector Mediators}
\label{sec:SimplifiedModels}

We focus on theories where the DM field is a SM gauge singlet, and we consider both scalars and fermions. We assume the DM interactions with SM fields to be mediated by a massive spin-1 particle $V$. This type of mediator can have a variety of origins in UV physics, including a gauge boson of a new spontaneously broken gauge interaction or a composite resonance of a new confining interaction. We remain agnostic about its UV origin and keep our analysis general by working within a simplified model framework (for a recent review see \Ref{DeSimone:2016fbz}). This allows us to study DM phenomenology in terms of a handful of masses and couplings~\cite{Dudas:2009uq,Goodman:2011jq,An:2012va,Frandsen:2012rk,Dreiner:2013vla,Buchmueller:2013dya,Alves:2013tqa,Arcadi:2013qia,Lebedev:2014bba,Abdullah:2014lla,Bell:2014tta,Buchmueller:2014yoa,Harris:2014hga,Alves:2015pea,Jacques:2015zha,Chala:2015ama,Alves:2015mua,Alves:2015dya} (with the further advantage of having the mediator in the spectrum) and allows us to combine mono-jet searches with complementary searches for resonances or mono-$V$ events~\cite{Gupta:2015lfa,Autran:2015mfa,Bai:2015nfa}.

The general Lagrangian reads~\footnote{Within the spirit of a simplified model analysis, we study theories with neither kinetic nor mass mixing between $V$ and SM neutral gauge bosons. Barring unnatural cancellations between these additional contributions and the RG effects we quantify, the direct detection limits found in this work are the most conservative.}
\be
\mathcal{L} = \mathcal{L}_{\rm SM} + \mathcal{L}_{\rm DM} - \frac{1}{4} V^{\mu\nu} V_{\mu\nu} + \frac{1}{2} m_V^2 V^\mu V_\mu + 
J^\mu_{\rm DM} \, V_\mu + J^\mu_{\rm SM} \, V_\mu \ .
\label{eq:LagGeneral}
\ee
The first term is just the SM Lagrangian, whereas the DM kinetic and mass terms depend on the spin of the DM itself
\be
\mathcal{L}_{\rm DM} = \left\{ 
\begin{array}{lccccl}
\left| \partial_\mu \phi \right|^2 - m_\phi^2 \left| \phi \right|^2 & & & & & \text{complex scalar DM} \vspace{0.1cm} \\
\mathcal{K}_\chi \;  \overline{\chi} \left( i \slashed{\partial} - m_\chi \right) \chi & & & & & \text{fermion DM} \ .
\end{array}
\right.
\ee
We only consider complex scalar DM; as explained below the DD rate for real scalar DM coupled to a vector mediator is below the reach of current and future experiments. On the other hand, both Dirac and Majorana are viable options for fermion DM. We ensure that the fermion field is canonically normalized by appropriately choosing the coefficient 
\be
\mathcal{K}_\chi = \left\{ 
\begin{array}{lccccl}
1 & & & & & \text{Dirac DM}  \\
1 / 2 & & & & & \text{Majorana DM}
\end{array}
\right. \ .
\label{eq:Kchidef}
\ee

The mediator couples to spin-1 currents. The DM current depends on the DM spin
\be
J^\mu_{\rm DM} = \left\{ 
\begin{array}{lccccl}
c_{\phi} \; \phi^\dag \, \overleftrightarrow{\partial}_\mu \phi & & & & & \text{complex scalar DM}  \\ 
c_{\chi A} / 2  \; \overline{\chi} \gamma^\mu \gamma^5 \chi   & & & & & \text{Majorana DM} \\
c_{\chi V} \; \overline{\chi} \gamma^\mu \chi +  c_{\chi A} \; \overline{\chi} \gamma^\mu \gamma^5 \chi   & & & & & \text{Dirac DM} \ .
\end{array}
\right.
\label{eq:JmuDM}
\ee
Here, we define the anti-symmetric combination
\be
 \phi^\dag \, \overleftrightarrow{\partial}_\mu \phi \equiv  \phi^\dag \; \partial_\mu \phi - \phi \; \partial_\mu \phi^\dag \ .
\ee
The symmetric combination $\partial_\mu(\phi^\dagger\phi)$ gives a coupling to quarks which is suppressed by the quark mass \cite{DelNobile:2013sia}, leading to a DD scattering rate which is strongly suppressed. We have checked that the symmetric scalar interaction gives no limits on the mediator mass from either \LUX\ or \LZ\ in the mass range considered here ($m_V > 1 \,\mathrm{GeV}$) and we therefore neglect it. This is also the reason why we do not consider the case of a real scalar DM particle, since there would be no detectable DD signal. For Majorana DM we can only couple to the axial-vector current, since the vector current vanishes identically as a consequence of the self-conjugation property of the Majorana field. Finally, for Dirac DM the mediator can couple to both the vector and the axial-vector current. We normalize the Majorana current with an overall factor of $1/2$, in such a way that  the limits we derive for Dirac and Majorana DM are directly comparable.

The mediator interactions with SM fields involve more independent couplings. We reduce the possible options by focusing on simplified models where the mediator interacts only with SM fermions, which are listed as Weyl fields with definite chirality in Tab.~\ref{tab:SMfields} with their associated gauge quantum numbers. We make sure to preserve the SM electroweak gauge invariance, as we want the simplified model to be consistent at the energy scales probed by the \LHC, substantially larger than the scale of electroweak symmetry breaking (EWSB). This is achieved by coupling the mediator $V$ to a spin-1 current made of the SM fields in Tab.~\ref{tab:SMfields}. The most general gauge invariant current, without mixing fields from different generations, reads
\be
 J^\mu_{\rm SM} = \sum_{i = 1}^3 \left[ c_{q}^{(i)} \; \overline{q_L^i} \gamma^\mu q_L^i + c_{u}^{(i)} \;  \overline{u_R^i} \gamma^\mu u_R^i +  c_{d}^{(i)} \;  \overline{d_R^i} \gamma^\mu d_R^i + c_{l}^{(i)} \; \overline{l_L^i} \gamma^\mu l_L^i + c_{e}^{(i)} \; \overline{e_R^i} \gamma^\mu e_R^i \right] \ .
\label{eq:JmuSM}
\ee
We therefore have in principle $5 \times 3 = 15$ independent couplings. 

\begin{table} 
\begin{center}
\begin{tabular}{ c || c c c c c}
 &  $q_L^i$ &  $u_R^i$ & $d_R^i$ & $l_L^i$ &$e_R^i$ \\ 
\hline  \hline 
$SU(3)_c$ & $\bs 3$ & $ \bs 3$ & $ \bs 3$ & $\bs 1$ & $\bs 1$  \\
$SU(2)_L$ & $\bs 2$ & $\bs 1$ & $\bs 1$ & $\bs 2$ & $\bs 1$ \\
$U(1)_Y$  & $+1/6$ & $+2/3$ & $-1/3$ & $-1/2$ & $-1$  \\
\end{tabular}
\end{center}
\caption{SM fermions with their gauge charges. The index $i$ runs over the three generations.}
\label{tab:SMfields}
\end{table}

An alternative and perhaps more common way to define this current is in terms of SM Dirac fermions, which do not have well defined electroweak quantum numbers and are massive after EWSB. The most general expression is then a superposition of vector and axial-vector currents
\be
\begin{split}
J^{\prime \,\mu}_{\rm SM} = & \, \sum_{i = 1}^3 \left[ c_{Vu}^{(i)} \; \overline{u^i} \gamma^\mu u^i +
c_{Vd}^{(i)} \; \overline{d^i} \gamma^\mu d^i + c_{Ve}^{(i)} \; \overline{e^i} \gamma^\mu e^i  \right] + \\ & 
\sum_{i = 1}^3 \left[  c_{Au}^{(i)} \; \overline{u^i} \gamma^\mu \gamma^5 u^i +
c_{Ad}^{(i)} \; \overline{d^i} \gamma^\mu \gamma^5 d^i + c_{Ae}^{(i)} \; \overline{e^i} \gamma^\mu \gamma^5 e^i  \right] \ .
\end{split}
\label{eq:JprimemuSM}
\ee
However, the simplified model Lagrangian in \Eq{eq:LagGeneral} with SM currents as defined in \Eq{eq:JprimemuSM} should be used with caution for certain collider processes. As an example, Refs.~\cite{Bell:2015sza,Bell:2015rdw,Haisch:2016usn} pointed out a violation of unitarity in the mono-$W$ scattering cross section as a consequence of the lack of gauge  invariance. Nevertheless, scattering processes in DD experiments involve energy scales considerably smaller than the EWSB scale, and the potential lack of electroweak gauge invariance is not a concern. Both currents in \Eqs{eq:JmuSM}{eq:JprimemuSM} are valid choices. The latter has $18$ independent couplings, a number larger than the $15$ for the former. This mismatch can be understood under the assumption that the UV physics of the mediator respects electroweak gauge invariance. If this is the case, the couplings in \Eq{eq:JprimemuSM} are not independent and are given by
\be
\begin{split}
c_{V,A \, u}^{(i)} = & \, \frac{ \pm c_{q}^{(i)} + c_{u}^{(i)}}{2}  \ ,  \\ 
c_{V,A \, d}^{(i)} =  & \, \frac{ \pm c_{q}^{(i)} + c_{d}^{(i)}}{2}  \ , \\
c_{V,A \, e}^{(i)} = & \, \frac{ \pm c_{l}^{(i)} + c_{e}^{(i)}}{2}   \ .
\end{split}
\label{eq:VAfromGaugeInv}
\ee


\section{Low-energy couplings and direct detection rates}
\label{sec:RGE}

The simplified model Lagrangian we have presented in Sec.~\ref{sec:SimplifiedModels} allows for a practical exploration of \LHC\ phenomenology without relying upon a specific model. In order to compare exclusion bounds from collider and DD experiments, we need to carefully compute DD rates from the same Lagrangian. We outline this procedure here.

Evaluating the rate for DM elastic scattering off target nuclei requires the knowledge of nuclear matrix elements. These hadronic quantities are evaluated at the nuclear scale $\mu_N \sim 1 \, {\rm GeV}$, where the theory looks quite different: the mediator effects can be approximated by contact interactions, and the only SM degrees of freedom accessible are light quarks ($u$, $d$ and $s$), gluons and photons. The connection between the simplified model valid at the collider energy and the low-energy EFT valid at the nuclear scale is achieved through two main steps:
\begin{itemize}
\item {\bf Integrate-out the mediator:} we always consider mediator masses significantly larger than the momentum exchanged in DD processes. Thus we can safely take the limit of a contact interaction between DM and SM fermions described by the dimension 6 operators
\be
\mathcal{L}^{(m_V)}_{\rm EFT} = - \frac{J_{{\rm DM}\, \mu} \; J^\mu_{\rm SM} }{m_V^2} \ .
\label{eq:EFT2}
\ee
These interactions, arising upon integrating-out the mediator, describe an EFT with dimensionless couplings (Wilson coefficients) defined at the renormalization scale $\mu = m_V$.
\item {\bf RG flow down to the nuclear scale:} the Wilson coefficients in \Eq{eq:EFT2}, defined at the mediator mass scale, are evolved to the nuclear scale by solving the system of RG equations derived in \Ref{D'Eramo:2014aba}. Here, we summarize the main steps of this procedure. The RGE is divided into two different regimes, above and below the EWSB scale, which we define as the mass of the $Z$ boson. Above the EWSB scale, for values of the renormalization scale $\mu$ in the range $m_Z < \mu < m_V$, the RGE is performed in an EFT with unbroken EW symmetry and containing the full SM degrees of freedom in the spectrum. The heavy SM degrees of freedom ($W$ and $Z$ gauge bosons, Higgs boson and top quark) are then integrated-out at the renormalization scale $m_Z$, and the EFT valid in the unbroken phase is matched onto a different EFT with only light SM degrees of freedom and broken EW symmetry. Finally, in the renormalization scale range $\mu_N < \mu < m_Z$, the effective couplings are evolved to the nuclear scale. For a light mediator $m_V < m_Z$, we only perform the RGE in the regime below the EWSB scale. We work in the mass independent renormalization scheme $\overline{\rm MS}$, and thus the renormalization scale $\mu$ appears only implicitly in the RG equations through the scale-dependent SM couplings. 

The SM couplings entering the anomalous dimension matrices derived in \Ref{D'Eramo:2014aba} are electroweak gauge couplings and fermion Yukawa couplings. The effects are therefore perturbative and there is no need to perform a resummation of logarithms. As we will see in Sec.~\ref{sec:benchmarks}, the inclusion of our effects may alter direct detection rates by orders of magnitude. This is not because we have large contributions from higher orders in perturbation theory, but rather due to the generation of new interactions through operator mixing. If the new interactions that are generated have a significantly larger matrix element for DM elastic scattering, the one-loop induced contribution can easily win over the tree-level one. For this reason, the leading effect is captured by the solution of a fixed-order calculation, which is given in \App{app:analytical}.  A more rigorous analysis can be performed by accounting for the evolution of the SM couplings entering the RG equations. We distribute the code \runDM together with this work to achieve this goal. In the unbroken phase we evolve the SM couplings following the results of \Ref{Buttazzo:2013uya}, whereas for the SM in the broken phase we RG evolve the quark masses and the electromagnetic coupling following Refs.~\cite{Xing:2011aa} and \cite{Pivovarov:2000cr}, respectively. The output of this RG procedure is an effective Lagrangian describing contact interactions between the DM particle and the SM fields, with Wilson coefficients defined at the nuclear scale $\mu_N$. As it turns out for vector mediators, contact interactions to light quarks are the only relevant ones for DD rates. The final result takes the form 
\be
\begin{split}
\mathcal{L}^{(\mu_N)}_{\rm EFT}  = & \, \mathcal{L}^{(\mu_N)}_V  +  \mathcal{L}^{(\mu_N)}_A \ , \\
\mathcal{L}^{(\mu_N)}_V = & \, - \frac{J^\mu_{{\rm DM}}}{m_V^2} \, 
\left[ \mathcal{C}_{V}^{(u)} \; \overline{u} \gamma^\mu u + \mathcal{C}_{V}^{(d)} \; \overline{d} \gamma^\mu d \right] \ , \\
 \mathcal{L}^{(\mu_N)}_A = & \, - \frac{J^\mu_{{\rm DM}}}{m_V^2} \, 
\left[ \mathcal{C}_{A}^{(u)} \; \overline{u} \gamma^\mu \gamma^5 u +
\mathcal{C}_{A}^{(d)} \; \overline{d} \gamma^\mu \gamma^5 d +  \mathcal{C}_{A}^{(s)}  \; \overline{s} \gamma^\mu \gamma^5 s \right] \ ,
\end{split}
\label{eq:LagEFTmuN}
\ee
where we separate quark vector and axial-vector currents and we only include contributions giving a DD rate. The analysis of \Ref{D'Eramo:2014aba} connects the couplings to light quarks in  \Eq{eq:LagEFTmuN} with the dimensionless coefficients at high energy appearing in \Eq{eq:JmuSM}. 

\end{itemize}

The effective Lagrangian in \Eq{eq:LagEFTmuN} describing DM interactions with quarks at the nuclear scale is our intermediate result. The differential cross section for DM-nucleus elastic scattering can be constructed via three additional steps:
\begin{itemize}
\item[$\diamond$] {\bf Dress the quark-currents to the nucleon level:} The quark-currents of \Eq{eq:LagEFTmuN} induce an effective Lagrangian for the nucleon field of the form
\be
\begin{split}
\mathcal{L}^{(\mu_N)}_{N} &= - \frac{J^\mu_{{\rm DM}}}{m_V^2}  \left[ \mathcal{C}_{V}^{(N)} \; \overline{N} \gamma^\mu N + \mathcal{C}_{A}^{(N)}  \; \overline{N} \gamma^\mu \gamma^5 N \right]\, ,
\end{split}
\label{eq:LagEFTnucleon}
\ee
where $N$ stands for nucleons: protons ($p$) and neutrons ($n$). The couplings are determined in the standard way by embedding the quarks within the nucleon, as reviewed in Ref.~\cite{DelNobile:2013sia}:
\be
\begin{split}
\mathcal{C}_{V}^{(p)} &= 2 \mathcal{C}_{V}^{(u)} + \mathcal{C}_{V}^{(d)}\\
\mathcal{C}_{V}^{(n)} &=  \mathcal{C}_{V}^{(u)} + 2\mathcal{C}_{V}^{(d)}\\
\mathcal{C}_{A}^{(N)} &=  \sum_{q = u,d,s} \mathcal{C}_{A}^{(q)} \Delta_q^{(N)}\, .
\end{split}
\label{eq:nucleoncouplings}
\ee
The axial charges $\Delta_q^{(N)}$ specify the contribution of the light quarks $q$ to the spin of the nucleon $N$. We use the values in the lower panel of Tab.~II in Ref.~\cite{Cheng:2012qr}.

The precise form of the nucleon-level Lagrangian in \Eq{eq:LagEFTnucleon} depends on the nature of the DM and the form of the DM current. Explicit expressions for the dimension-6 effective operators appearing in \Eq{eq:LagEFTnucleon} are given in Eqs.~(\ref{eq:LaglevelOp&coeffScalar}$-$\ref{eq:LaglevelOp&coeffDirac}) for the various possibilities we consider in this work. We notice that for complex scalar DM only interactions involving a DM vector current arise. Conversely, for Majorana DM, the DM vector current vanishes identically, therefore we can only have the axial-vector current of DM coupling to either the vector or axial-vector SM currents. For Dirac DM all combinations are instead possible. It is worth stressing here that interaction structures involving  only vector currents or only axial-vector currents preserve parity, while those involving both vector and axial-vector currents break it.  As will become clear later on, the latter types of interactions produce direct detection rates which are suppressed with respect to those induced by parity conserving operators. 


\item[$\diamond$] {\bf Calculate the non-relativistic (NR) DM-nucleon Matrix Element (ME):} We have now to evaluate the DM-nucleon ME  and reduce it to the NR limit. This can be done, as explicitly shown in Appendix~\ref{app:NRoperators}, by contracting \Eq{eq:LagEFTnucleon} with the initial and final states of the scattering process and then expanding the solution of the Dirac equation in the NR limit. The result can be expressed as a linear combination of NR operators, for which we give the explicit expressions and corresponding coefficients in Eqs.~(\ref{eq:MEOp&coeffScalar}$-$\ref{eq:MEOp&coeffDirac}). Full details of this matching between the Lagrangian in \Eq{eq:LagEFTnucleon} and NR DM-nucleon ME in Eq.~\eqref{eq:MEL} can be found in Refs.~\cite{Fitzpatrick:2012ix,Fitzpatrick:2012ib,DelNobile:2013sia}. 

We simply comment here on the NR structure of the nucleon-level Lagrangian. Regardless of the DM current type, we first note that the DM-nucleon ME always reduces to the sum of a leading order NR operator and at least one which is suppressed either by powers of the DM-nucleus relative velocity $v$ or by the recoil momentum $q_{\rm R}$. Focusing first on parity conserving structures in \Eq{eq:LagEFTnucleon}, we note that the purely vector current coupling between both DM and SM sectors leads to the standard spin-independent (SI) interaction ($\mathcal O_1^{\rm NR}=\unop$).  Meanwhile, the coupling between the axial-vector currents of DM and SM fermions gives rise to the standard spin-dependent (SD) interaction ($\mathcal O_4^{\rm NR}=\vec s_N \cdot \vec s_\chi$). On the other hand, parity violating structures (e.g.~those involving both vector and axial-vector currents) generate NR operators which are always suppressed. It is worth stressing here that as a consequence, a theory which is parity-violating in the UV could appear parity-conserving  at the nuclear energy scale; even if the running only induces a small coefficient for parity-conserving interactions, these interactions are unsuppressed and may therefore dominate.


\item[$\diamond$] {\bf Correct the DM-nucleon ME with the nuclear response functions:} Once we express the DM-nucleon ME  in terms of the relevant  degrees of freedom of the NR elastic collision, the spin-averaged amplitude squared for scattering off a target nucleus can be constructed in terms of a finite set of nuclear response functions as shown explicitly in Appendix~\ref{app:NRoperators}.  These response functions critically depend on the type of scattered nucleus. 
As briefly pointed out in the previous item, the leading order NR interactions are those coming from standard SI and SD contact interactions. In the zero momentum transfer limit, the nuclear response functions associated with such interactions are independent of $v$ and $q_{\rm R}$.  The other possible NR interactions triggered by \Eq{eq:LagEFTnucleon} still induce SI or SD  nuclear responses. They are however suppressed by powers of $v^2$ or $q_{\rm R}^2$ as one can see in Appendix A.2 of Ref.~\cite{Fitzpatrick:2012ix}.

Since we construct the most general model for SM fermions interacting with scalar or fermion DM through the exchange of a vector mediator, it is useful to outline here the main NR interactions and in turn nuclear response functions which arise in such a model (which are listed in full in Eqs.~(\ref{eq:MEOp&coeffScalar}$-$\ref{eq:MEOp&coeffDirac})). For scalar DM, in addition to the leading SI response, one can induce a velocity suppressed SD interaction. For Majorana DM, the leading response function is SD. However, a velocity suppressed SI and a momentum suppressed SD interaction can be triggered at the same time. The Dirac DM case, in addition to the NR nuclear responses of both scalar and Majorana DM,  also triggers a momentum suppressed SD interaction.

\end{itemize}

Once we follow the steps explained in this section,  we are able to connect a simplified model valid at high energy to its NR manifestation at the nuclear level. Hence, we can write the differential scattering cross section as shown in \Eq{eq:diffXS}. This is the most general form of the scattering cross section and allows us to evaluate the nuclear recoil rate for all  possible NR interactions induced by the Lagrangian in \Eq{eq:LagEFTnucleon}. 

As an example, we can immediately write down the total DM-nucleon cross sections for the piece of the Lagrangian in \Eq{eq:LagEFTnucleon} which induces standard SI and SD interactions. By integrating \Eq{eq:diffXS} up to the maximal recoil energy $\ER^{\rm max}=2 m_{\rm DM}^2 m_{\mathcal N}/(m_{\rm DM} + m_{\mathcal N})^2 v^2$ and replacing $\mathcal N \rightarrow N$, we see that in the zero momentum transfer limit\footnote{The SI nuclear response is enhanced by the coherent factor $A^2$ in the zero momentum transfer limit. Hence, for single nucleon scattering this is trivially equal to 1. The SD nuclear response is instead enhanced by the total nuclear spin. Its normalization for  single nucleon scattering  can be inferred from Eqs.~(58$-$60) of Ref.~\cite{Fitzpatrick:2012ix}. In the zero momentum transfer limit this is equal to 3/16.  } the DM-nucleon cross section reduces to the usual expression 
\be
\begin{split}
&\sigma^N_{\mathrm{SI}}  = \frac{\mu_N^2}{\pi} \frac{\left(c_{\phi(\chi V)} \mathcal{C}_{V}^{(N)} \right)^2}{ m_V^4} \ ,   \\
&\sigma^N_{\mathrm{SD}}   = \frac{3 \mu_N^2}{\pi} \frac{\left(c_{\chi A} \mathcal{C}_{A}^{(N)} \right)^2}{ m_V^4} \ ,
\end{split} 
\label{eq:usualSI&SD}
\ee
where $\mu_N=m_{\rm DM}m_N/(m_{\rm DM}+m_N)$ is the DM-nucleon reduced mass. We explicitly show the cross sections in \Eq{eq:usualSI&SD}, since these are the quantities for which limits are presented by direct detection collaborations. We will also use \Eq{eq:usualSI&SD} in Sec.~\ref{sec:LHC} in order to correctly explore the complementarity between low-energy DD searches and the \LHC\ bounds. However, we emphasise that, as explained above, these are not the only contributions to the DD cross section, but merely the ones which are typically focused on.


\section{Direct detection bounds on benchmark models}
\label{sec:benchmarks}

We are now ready to compare the predictions obtained as prescribed in Sec.~\ref{sec:RGE} with the experimental bounds. Before doing that, it is helpful to perform a counting of the parameters in our simplified model. We have two mass scales, the DM and the mediator mass, and in this section we will always present our results in the two-dimensional $(m_\mathrm{DM}, m_V)$ plane. We must therefore fix the remaining dimensionless couplings to the mediator. For Dirac (Majorana, complex scalar) DM we have the two (one) dimensionless couplings in \Eq{eq:JmuDM} parameterizing the interaction strength with the mediator. We consider models where only one of these DM couplings is switched on and it is equal to one. We thus have three cases for the DM couplings:
\be
J^\mu_{\rm DM} = \left\{ 
\begin{array}{lccccl}
\phi^\dag \, \overleftrightarrow{\partial_\mu} \phi & & & & & \text{complex scalar DM}  \\
 \overline{\chi} \gamma^\mu \chi & & & & & \text{Dirac DM vector coupling} \\ 
\mathcal{K}_\chi  \; \overline{\chi} \gamma^\mu \gamma^5 \chi  & & & & & \text{Dirac or Majorana DM axial-vector coupling} \,.
\end{array}
\right.
\label{eq:DMcurrentChoices}
\ee
The factor $\mathcal{K}_\chi$ is defined in \Eq{eq:Kchidef} and ensures that our limits are valid for both Dirac and Majorana DM. We have not yet specified the $15$ couplings to SM fermions appearing in \Eq{eq:JmuSM}, and we focus here on three benchmark models: flavor universal couplings to all quarks (Sec.~\ref{sec:BenchmarkI}), flavor universal couplings to all leptons (Sec.~\ref{sec:BenchmarkII}), and couplings only to third generation fermions (Sec.~\ref{sec:BenchmarkIII}). In each case we determine the region of parameter space excluded by current experimental bounds as well as the projected exclusion reach of future experiments. These three benchmarks will allow us to highlight the key effects which appear when the running of the couplings is considered.

The bounds derived in this Section are obtained by performing the rigorous RG analysis outlined in Sec.~\ref{sec:RGE}, by accounting for the SM couplings RGE and numerically solving the system of RG equations derived in \Ref{D'Eramo:2014aba}. 
This allows us to obtain the coupling to light quarks at the nuclear energy scale and from these the coefficients of the relevant NR DM-nucleon interactions. We then determine the limits from DD experiments using \textsc{NRopsDD }, available at \href{http://www.marcocirelli.net/NROpsDD.html}{this http URL} and described in Ref.~\cite{DelNobile:2013sia}.\footnote{We note that while the statistical tests defined in Ref.~\cite{DelNobile:2013sia} are only expected to hold approximately, we have checked explicitly that the limits obtained from \LUX\ match those reported in Ref.~\cite{Akerib:2013tjd} to within 5\%.} Calculation of the expected rate in the \LUX\ experiment~\cite{Akerib:2013tjd} is detailed in Addendum~1 of Ref.~\cite{DelNobile:2013sia}, while calculation of the rate and projected limits from the planned \LZ\ experiment \cite{Akerib:2015cja} is described in Appendix~\ref{app:LZ} of this work.

\subsection{Benchmark I: flavor universal couplings to quarks}
\label{sec:BenchmarkI}

The first benchmark model we consider is a vector mediator coupled to quarks only
\be
c_{l}^{(i)} = c_{e}^{(i)} = 0 \,, \qquad \qquad \qquad (i = 1,2,3)\,.
\ee
We assume the coupling to quarks to be flavor universal
\be
c_{q}^{(i)} = c_q \, , \qquad \qquad c_{u}^{(i)} = c_u \, , \qquad \qquad c_{d}^{(i)} = c_d \, , \qquad \qquad \qquad   (i = 1,2,3) \ ,
\ee
consistent with minimal flavor violation (MFV)~\cite{D'Ambrosio:2002ex}. These restrictions leave us with three independent couplings. It is helpful to rewrite the SM current for this benchmark model with the language of \Eq{eq:JprimemuSM} in terms of vector and axial-vector contributions
\be
\begin{split}
J^{\prime \,\mu}_{\rm SM} = & \,  \frac{c_q + c_u}{2}  \; \sum_{i = 1}^3  \overline{u^i} \gamma^\mu u^i +
\frac{c_q + c_d}{2} \;  \sum_{i = 1}^3  \overline{d^i} \gamma^\mu d^i + \\
& \, \frac{- c_q + c_u}{2} \;  \sum_{i = 1}^3  \overline{u^i} \gamma^\mu \gamma^5 u^i +
\frac{- c_q + c_d}{2} \; \sum_{i = 1}^3  \overline{d^i} \gamma^\mu \gamma^5 d^i  \ .
\end{split}
\ee
The coefficients ensure that the above expression is SM gauge invariant. We choose models where the couplings are either entirely vector or axial-vector. The lack of axial-vector couplings imposes the condition $c_q = c_u = c_d$. Likewise, getting rid of the vector currents requires $c_u = c_d = - c_q$. Hence we perform the analysis for the two choices of couplings
\begin{align}
\label{eq:bench1a} \text{Vector to quarks:} & \qquad   \qquad  c_u = c_d = c_q = 1 \, , \\
\label{eq:bench1b} \text{Axial-vector to quarks:} & \qquad   \qquad c_u = c_d = - c_q = 1  \,.
\end{align}
Equivalently, we consider the contact interactions obtained after integrating-out the mediator
\begin{align}
\text{Vector to quarks:} & \qquad   \qquad 
\mathcal{L}_{\rm EFT} = - \frac{1}{m_V^2} \, J_{{\rm DM}\, \mu} \; \sum_{i = 1}^3 \left[ \overline{u^i} \gamma^\mu u^i +
\overline{d^i} \gamma^\mu d^i  \right]  \, , \\
\label{eq:ben1axial} \text{Axial-vector to quarks:} & \qquad   \qquad 
\mathcal{L}_{\rm EFT} = - \frac{1}{m_V^2} \, J_{{\rm DM}\, \mu} \; \sum_{i = 1}^3 \left[ \overline{u^i} \gamma^\mu \gamma^5 u^i + \overline{d^i} \gamma^\mu \gamma^5 d^i  \right]  \,.
\end{align}
Other choices may violate SM gauge invariance. As an example, choosing axial-vector couplings only as in \Eq{eq:ben1axial} but with a relative minus sign between currents of up- and down-type quarks is not consistent with SM electroweak gauge invariance. 

\begin{figure}
\centering
\includegraphics[width=0.495\textwidth]{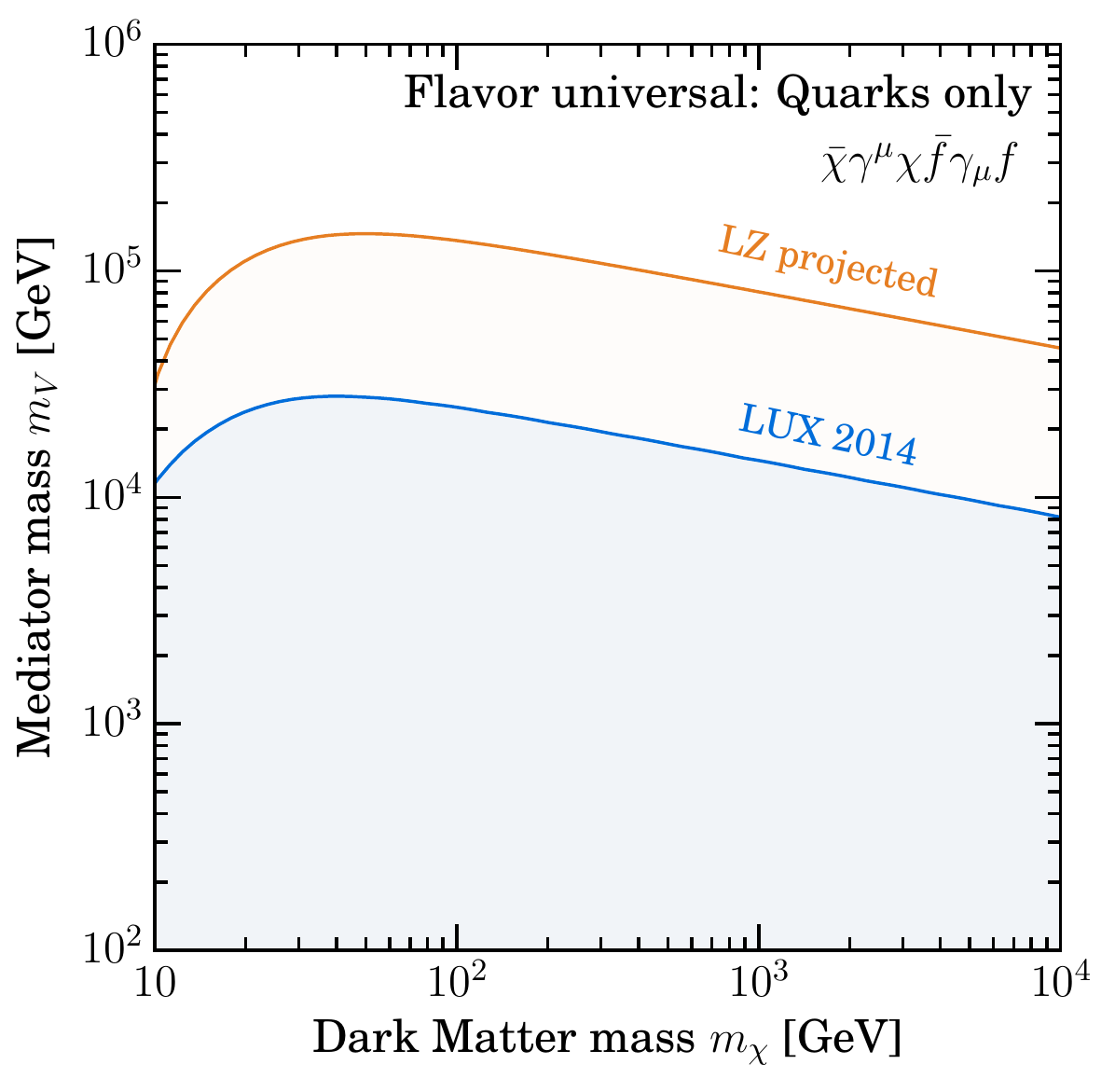} 
\includegraphics[width=0.495\textwidth]{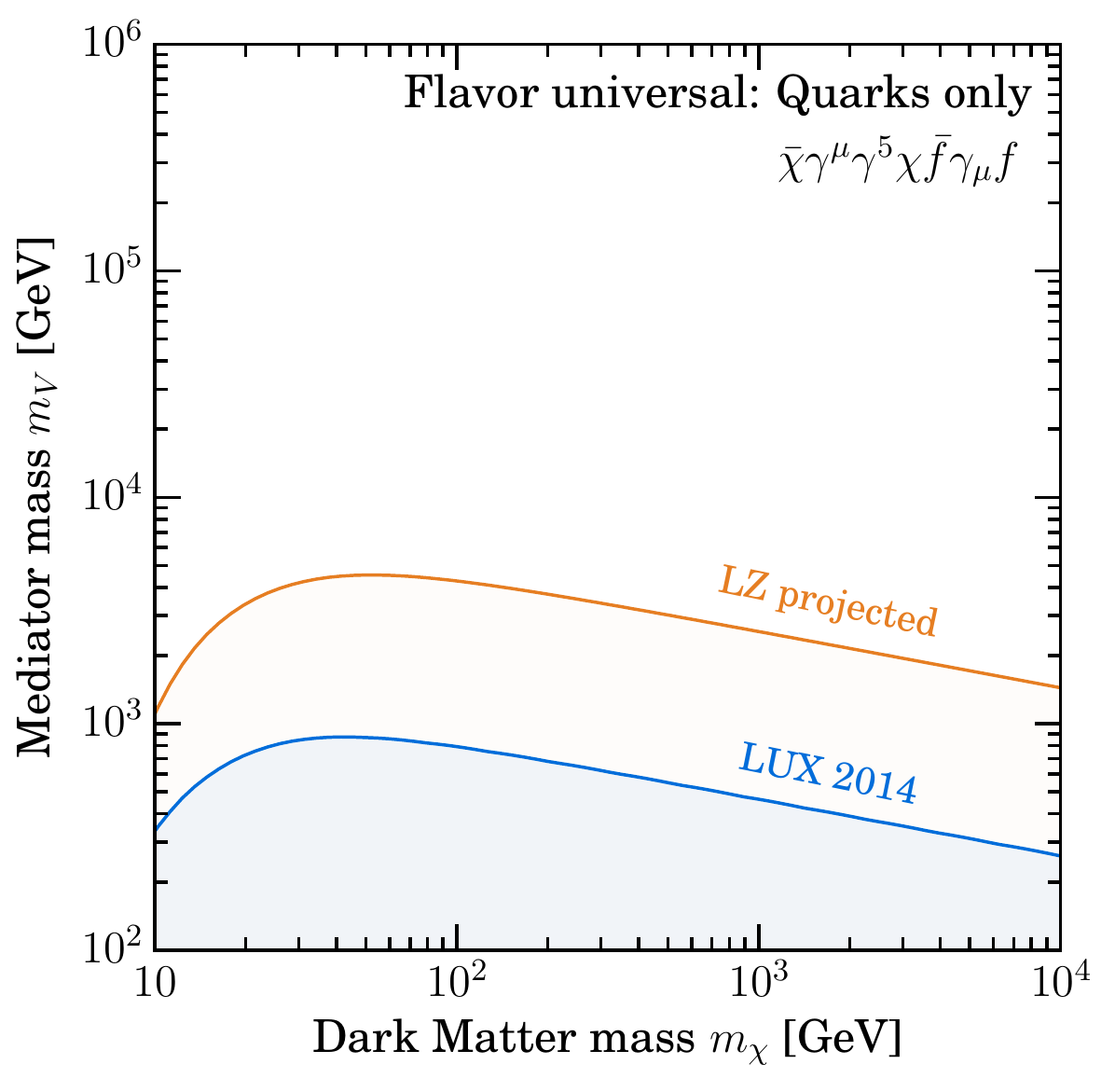}  \\
\includegraphics[width=0.495\textwidth]{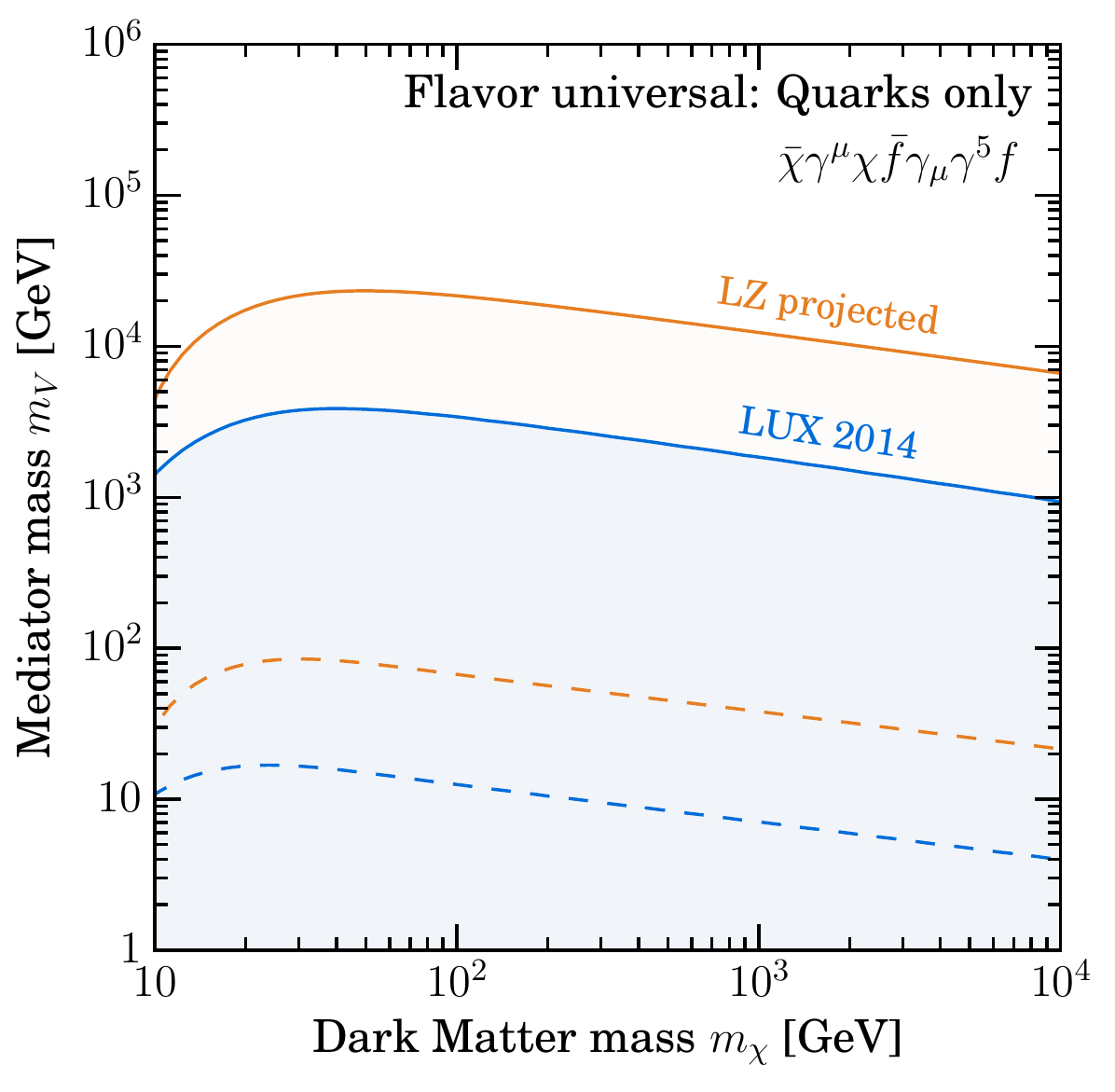}
\includegraphics[width=0.495\textwidth]{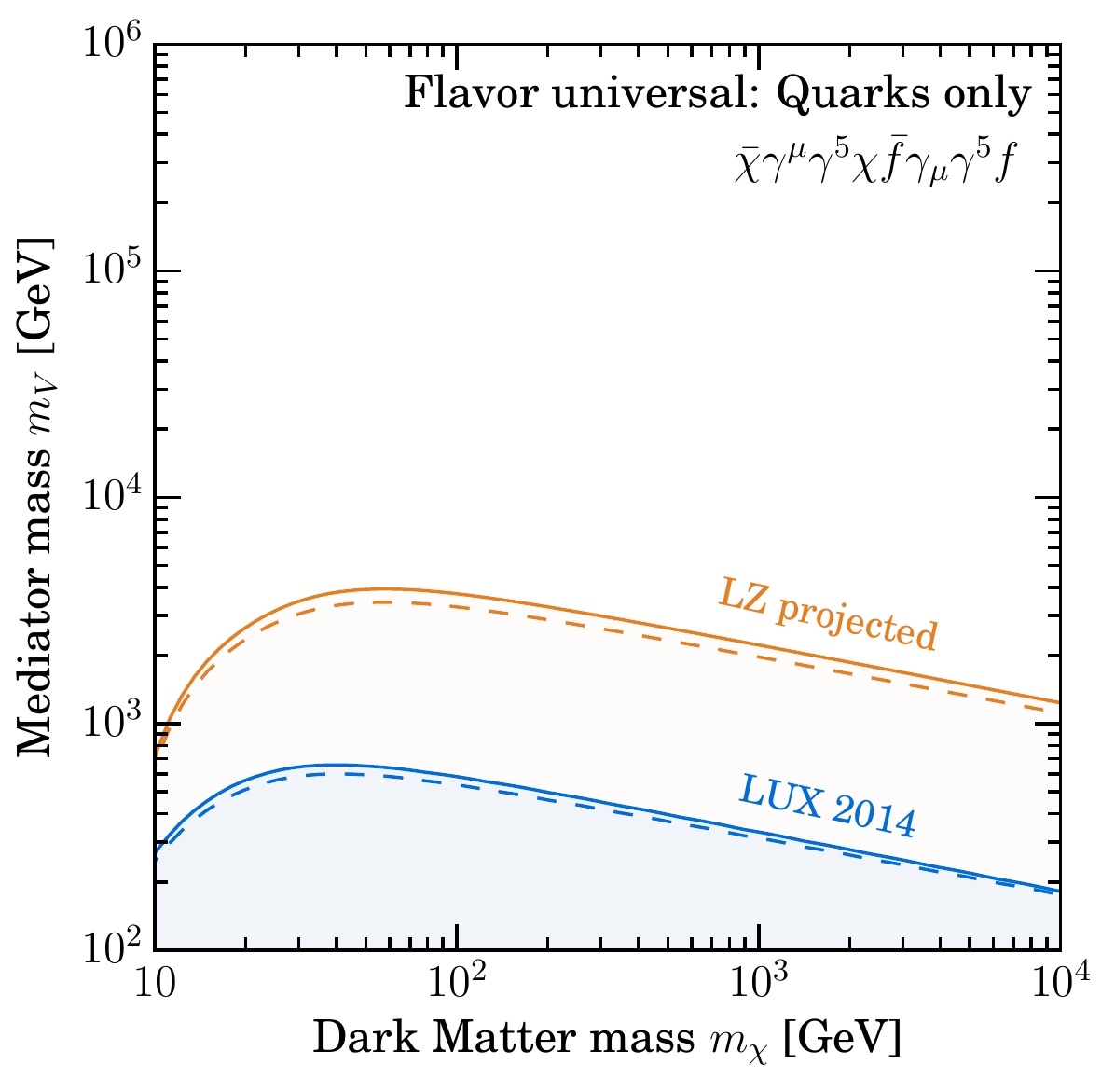} 
\caption{Experimentally excluded regions in the $(m_{\rm DM}, m_V)$ plane for the case of fermion DM and mediator with flavor universal coupling to quarks (Benchmark I). We shade the \LUX\  90\% excluded region (blue) and also show the projected \LZ\ exclusion (orange). In the two upper panels we consider a mediator coupling to the quark vector current, and on the left (right) we take the coupling to the DM (axial-)vector current. We do the same in the lower panels, where we consider a mediator coupled to the quark axial-vector current. Dashed lines indicate the exclusion limits when the running of the couplings is not considered. Where only solid lines are shown this means that the limits with and without running are indistinguishable.}
\label{fig:universalQ_fermionDM}
\end{figure}

The DM current can be chosen in three different ways as in \Eq{eq:DMcurrentChoices}, giving a total of 6 cases. Exclusion limits for these 6 benchmark models are shown in \Figs{fig:universalQ_fermionDM}{fig:universalQ_scalarDM} for fermion and scalar DM, respectively. We emphasize once more than these results are derived by numerically solving the RG system and accounting for the RGE of SM parameters. Despite the involved numerical analysis, it is possible to understand the features of the excluded regions by considering the approximate analytical solution given in \App{app:analytical}. The error made by employing these approximate expressions is quantified in \App{app:code}.

Before discussing the figures, then, we provide the approximate expressions for the Wilson coefficients of the effective Lagrangian in \Eq{eq:LagEFTmuN} describing DM interactions with light quarks at the nuclear scale. For universal vector coupling to quarks we have
\be
\begin{split}
\text{Vector to quarks:}  \qquad  & \mathcal{C}_{V}^{(u)} \simeq 
1 - 8 \frac{\alpha_{\rm em}}{9 \pi} \left[ \ln(m_V/m_Z) + \frac{1}{2} \ln(m_V/\mu_N)  \right] \ , \\
& \mathcal{C}_{V}^{(d)} \simeq 1 + \frac{4 \alpha_{\rm em}}{3 \pi} 
\left[ \ln(m_V/m_Z) + \frac{1}{2} \ln(m_V/\mu_N)  \right]  \,, \\
& \mathcal{C}_{A}^{(u)} \simeq \mathcal{C}_{A}^{(d)} \simeq \mathcal{C}_{A}^{(s)} \simeq 0 \,.
\end{split}
\label{eq:VectorQuarksAnalytical}
\ee
We see that there are no RG induced axial-vector currents with light quarks. However, the vector currents, with tree-level coefficients equal to one, receive radiative corrections from one-loop diagrams involving the electromagnetic coupling. These corrections are $\mathcal{O}(1 \%)$ for $m_V \sim 1 \, {\rm TeV}$ and scale logarithmically with the mediator mass. Although they should be included for a correct evaluation of the rate, they turn out to give only a small effect. 

The situation is dramatically different for the case of axial-vector couplings to quarks
\be
\begin{split}
\text{Axial-vector to quarks:}  \qquad  & \mathcal{C}_{V}^{(u)} \simeq 
\left(3 - 8 s_w^2 \right) \left[ \frac{\alpha_t}{2 \pi}  \ln(m_V / m_Z) - 
\frac{\alpha_b}{2 \pi}  \ln(m_V / \mu_N) \right]   \,, \\
& \mathcal{C}_{V}^{(d)} \simeq 
\left(3 - 4 s_w^2 \right)  \left[ - \frac{\alpha_t}{2 \pi}  \ln(m_V / m_Z) +
\frac{\alpha_b}{2 \pi} \ln(m_V / \mu_N)  \right]  \,, \\
& \mathcal{C}_{A}^{(u)} \simeq 1 - \frac{3 \alpha_t}{2 \pi}  \ln(m_V / m_Z) +  \frac{3 \alpha_b}{2 \pi}  \ln(m_V / \mu_N) \,, \\
& \mathcal{C}_{A}^{(d)} \simeq  \mathcal{C}_{A}^{(s)} \simeq 1 + \frac{3 \alpha_t}{2 \pi}  \ln(m_V / m_Z) - \frac{3 \alpha_b}{2 \pi}  \ln(m_V / \mu_N) \,.
\end{split}
\label{eq:AxialQuarksAnalytical}
\ee
The mixing into the light quark vector currents is non-vanishing in this case. The dominant contribution comes from the top Yukawa, and for $m_V \sim 1 \text{ TeV}$ this gives an $\mathcal{O}(3-5\%)$ coupling to the light quark vector currents. Moreover, the size of the axial-vector coupling to light quarks is also affected. Again, due to the size and the RGE of the top Yukawa, this effect is larger than the approximate expressions in \Eq{eq:AxialQuarksAnalytical}, giving corrections of order $\mathcal{O}(10\%)$ for $m_V \sim 1 \, {\rm TeV}$.

\begin{figure}
\centering
\includegraphics[width=0.495\textwidth]{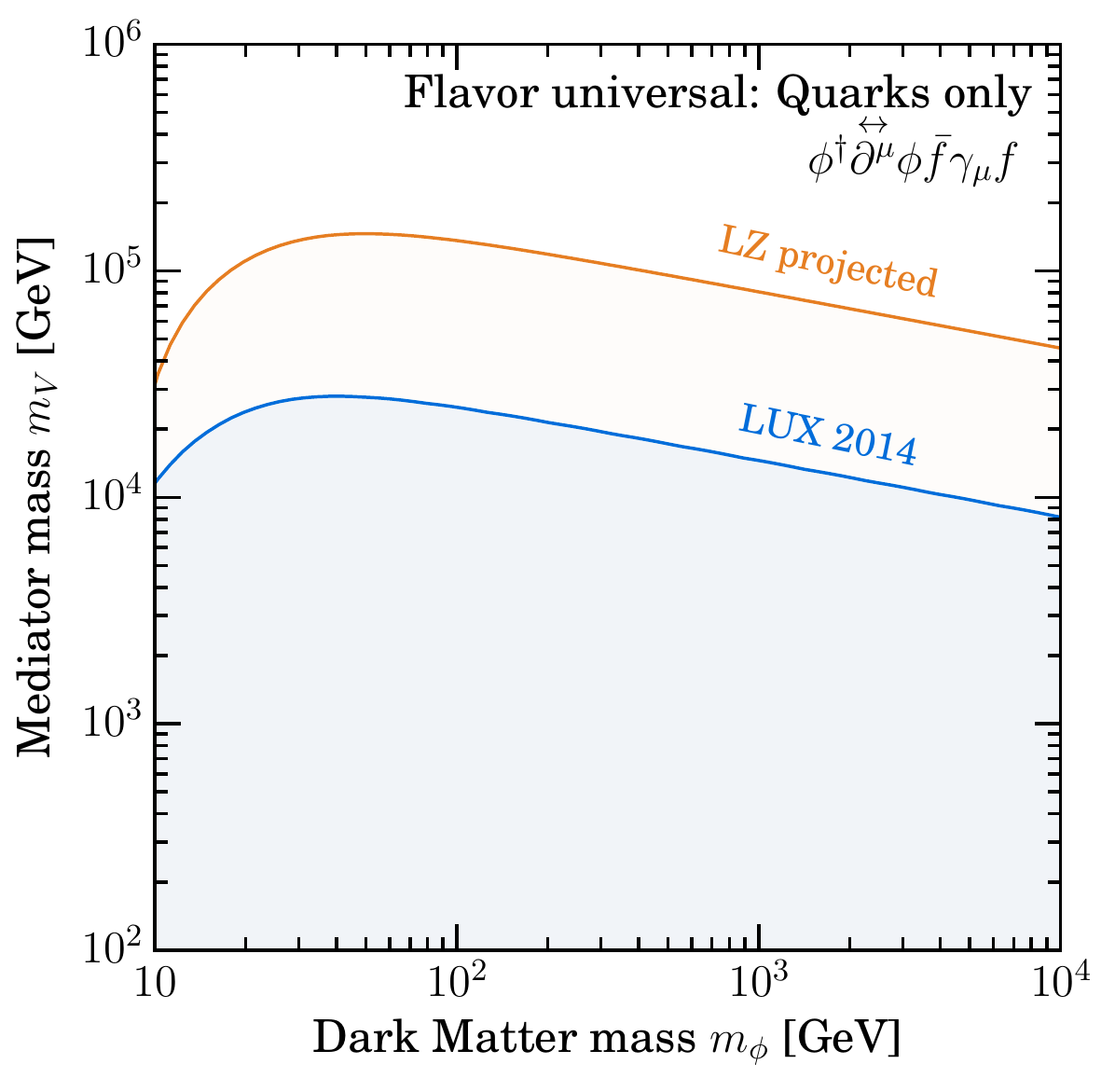} 
\includegraphics[width=0.495\textwidth]{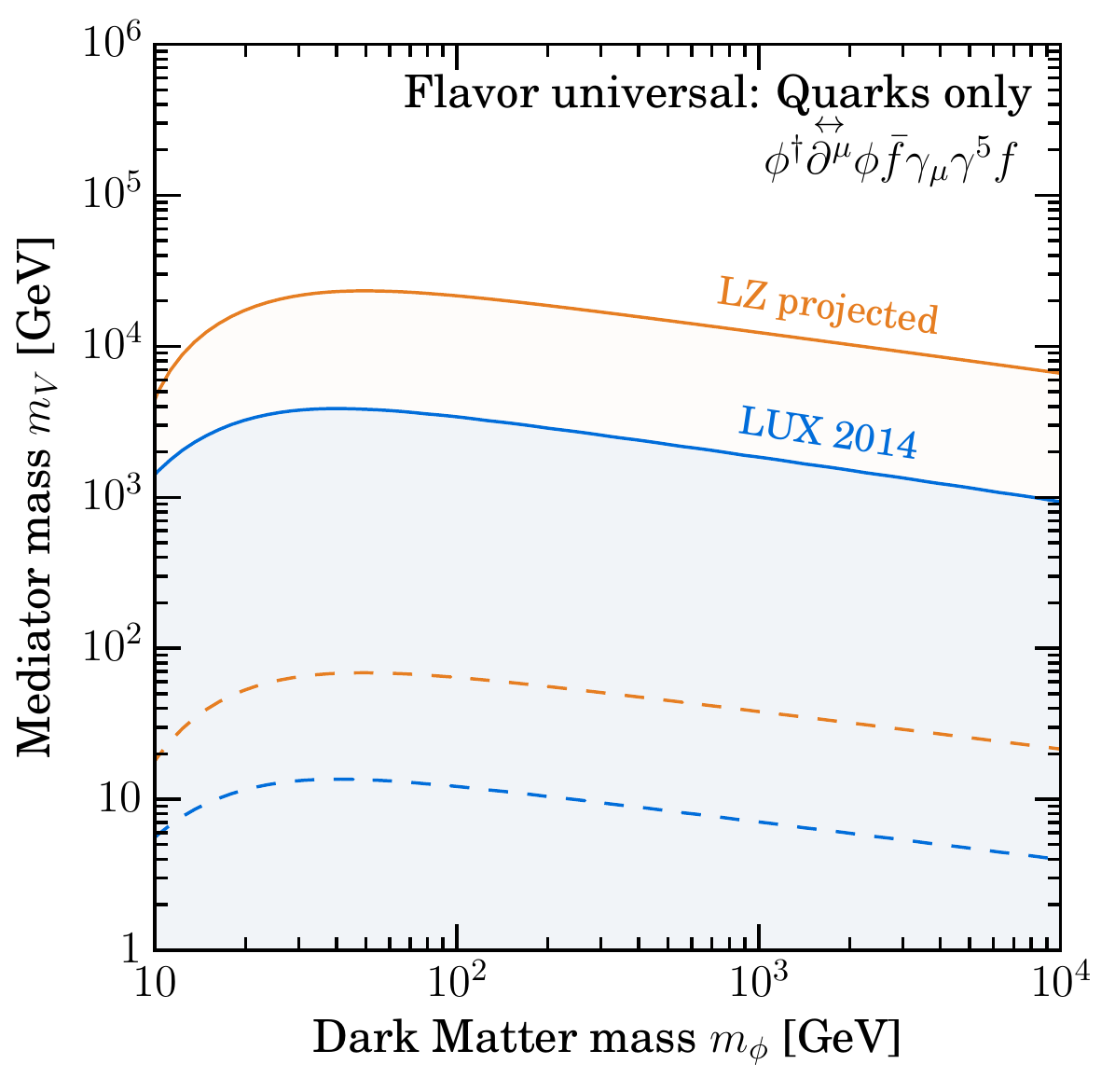} 
\caption{Same as \Fig{fig:universalQ_fermionDM} but for scalar DM.}
\label{fig:universalQ_scalarDM}
\end{figure}

We now turn to Fig.~\ref{fig:universalQ_fermionDM}, which shows the 90\% exclusion limits on the mediator mass for fermionic DM, obtained from the null results of the \LUX\ experiments (blue) as well as the projected limits from the \LZ\ experiment (orange). For the case of vector interactions in both the DM and quark sectors (top left), we obtain strong limits on the mediator mass, up to around $m_V \sim 20 \text{ TeV}$ from \LUX. In this benchmark, DM couples to the light quarks at tree level with a standard SI interaction, which is coherently enhanced with the square of the number of nucleons in the nucleus. For this benchmark, the limits with and without running are indistinguishable. As shown in \Eq{eq:VectorQuarksAnalytical}, corrections to the quark vector couplings are driven by electromagnetic interactions and are therefore small.

For the same reason, there are no strong running effects when we consider the axial-vector current of DM coupling to the quark vector current (top right). The quark vector current does not mix into the axial-vector current, meaning that the standard SD interaction is not induced. Instead, the DD rate is due to NR operators $\mathcal{O}_8^{\rm NR}$ and $\mathcal{O}_9^{\rm NR}$ (see \App{app:NRoperators}), which are suppressed by the DM velocity and the nuclear recoil momentum respectively. Thus, we obtain weaker limits from \LUX\ and \LZ\ when compared with the vector-vector case.

For the vector current of DM coupling to the axial-vector current of quarks (bottom left), we obtain very weak limits in the na\"ive analysis where we neglect operator mixing (dashed lines). Again, this is because the resulting interaction (mediated by operators $\mathcal{O}_7^{\rm NR}$ and $\mathcal{O}_9^{\rm NR}$ defined in \App{app:NRoperators}) is velocity and momentum suppressed. However, when we include the effects of mixing (solid lines), the limits are strengthened by around 2 orders of magnitude. As shown in \Eq{eq:AxialQuarksAnalytical}, a coupling to the quark vector current is induced. The origin of this mixing is the mediator coupling to the top quark, and the effect is proportional to the top Yukawa. The resulting vector-vector interaction gives rise to SI scattering, meaning that despite this rather small induced coupling, the DD rate is large due to the coherent enhancement. This effect was pointed out previously in Refs.~\cite{Crivellin:2014qxa,D'Eramo:2014aba} and demonstrates that models which appear to have suppressed DM-nucleon interactions can in fact lead to unsuppressed scattering and therefore much stronger limits.

For axial-vector couplings to both DM and quarks (bottom right), DM-nucleon scattering is mediated by the standard SD coupling at tree level. Because there is no coherent enhancement in this case, the limits are weaker than in the vector-vector case (top left). When the effects of running are included, the limits become stronger. Though the mixing of the quark axial-vector current with the vector current does lead to a new velocity-suppressed contribution, the dominant effect is still coming from the axial-vector couplings themselves. From the last two lines of \Eq{eq:AxialQuarksAnalytical}, we see that the coupling to the up quark is reduced, while the coupling to the down and strange quarks is enhanced. The origin of these one-loop corrections are again operator mixing proportional to the top Yukawa. This has the effect of increasing the SD coupling to neutrons, which increases the DD rate in Xenon targets, whose nuclear spin is dominated by the presence of an unpaired neutron. Note, however, that the running also has the effect of \textit{decreasing} the SD coupling to protons, meaning that for some targets (such as Fluorine) the DD rate will be reduced. Remarkably, this isospin-violation is entirely driven by loops of SM particles and it persists even if the mediator $V$ couples in an isospin-conserving way, as first pointed out in \Ref{Crivellin:2014qxa}. We will discuss this effect in more detail in Sec.~\ref{sec:LHC}, where we explore complementarity between DD and collider searches. 

Finally, we consider the limits on this benchmark for scalar DM, shown in Fig.~\ref{fig:universalQ_scalarDM}. In the left panel, for scalar DM coupling to the quark vector current, we obtain limits identical to those of fermionic DM vector-vector interactions. This is because both interactions reduce to the same non-relativistic operator, namely $\mathcal{O}_1^{\rm NR}$, the standard SI interaction. For scalar DM coupling to the quark axial-vector current, the results also appear similar to the DM vector-current case. In the scalar case, however, the limits without running (dashed lines) are slightly weaker, due to the fact that only the suppressed operator $\mathcal{O}_7^{\rm NR}$ contributes (see \Eq{eq:MEOp&coeffScalar} in Appendix~\ref{app:NRoperators}). However, once running is taken into account, the standard SI interaction is induced and the limits match those obtained for the fermionic DM vector current.

In this section, we have discussed a benchmark in which DM couples to all quarks in a flavor universal way through a vector mediator. In particular, this means that the interactions which mediate DM-nucleon scattering (the couplings to light quarks) appear at tree level in the high energy Lagrangian. However, there are two key effects which arise when the RGE is taken into account. First, the size of these tree level couplings may be altered, as seen in the lower right panel of Fig.~\ref{fig:universalQ_fermionDM} for standard SD interactions. Second, new couplings may be induced which, if they lead to unsuppressed DM-nucleon scattering, can lead to much stronger limits. Both of these effects are driven by mediator interactions with the top quark mixing onto operators with light quark. We note in particular that for each of the DM interaction structures we have considered (scalar, vector and axial-vector), the limits coming from couplings to quark vector currents and couplings to quark axial-vector currents are comparable once the running has been accounted for. Though in the standard picture (without running) only one of these interactions is assumed to dominate, here we have demonstrated that both are important and both must be included to give accurate limits on simplified models with vector mediators.

\subsection{Benchmark II: flavor universal couplings to leptons}
\label{sec:BenchmarkII}

We now consider the opposite case in which all couplings to quarks are switched off,
\be
c_{q}^{(i)} = c_{u}^{(i)} = c_{d}^{(i)} = 0 \,, \qquad \qquad \qquad   (i = 1,2,3) \,,
\ee
and interactions between DM and SM involve only lepton fields. We assume flavor universal couplings also in this case, and we are thus left with only two dimensionless couplings:
\be
c_{l}^{(i)} = c_l \,, \qquad \qquad  c_{e}^{(i)} = c_e \,, \qquad \qquad \qquad (i = 1,2,3) \,.
\ee
We present results for the two following choices 
\begin{align}
\text{Vector to leptons:} & \qquad   \qquad  c_e = c_l = 1 \,, \\
\text{Axial-vector to leptons:} & \qquad   \qquad c_e = - c_l = 1  \,,
\end{align}
namely for the effective interactions at the mediator mass scale of the form
\begin{align}
\text{Vector to leptons:} & \qquad   \qquad 
\mathcal{L}_{\rm EFT} = - \frac{1}{m_V^2} \, J_{{\rm DM}\, \mu} \; \sum_{i = 1}^3 \overline{e^i} \gamma^\mu e^i \,, \\
\text{Axial-vector to leptons:} & \qquad   \qquad 
\mathcal{L}_{\rm EFT} = - \frac{1}{m_V^2} \, J_{{\rm DM}\, \mu} \; \sum_{i = 1}^3 \overline{e^i} \gamma^\mu \gamma^5 e^i  \,.
\end{align}

\begin{figure}
\centering
\includegraphics[width=0.495\textwidth]{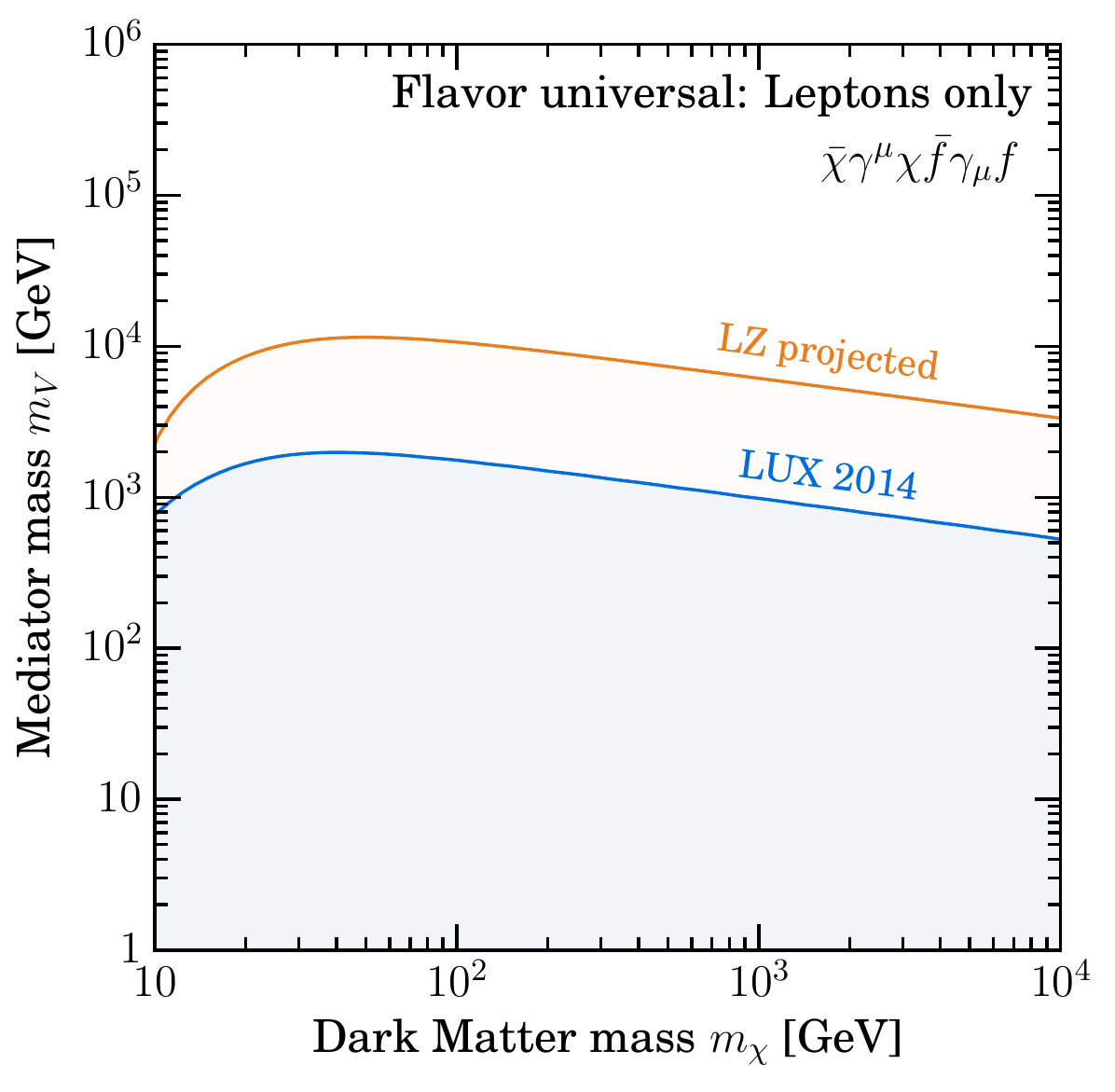} 
\includegraphics[width=0.495\textwidth]{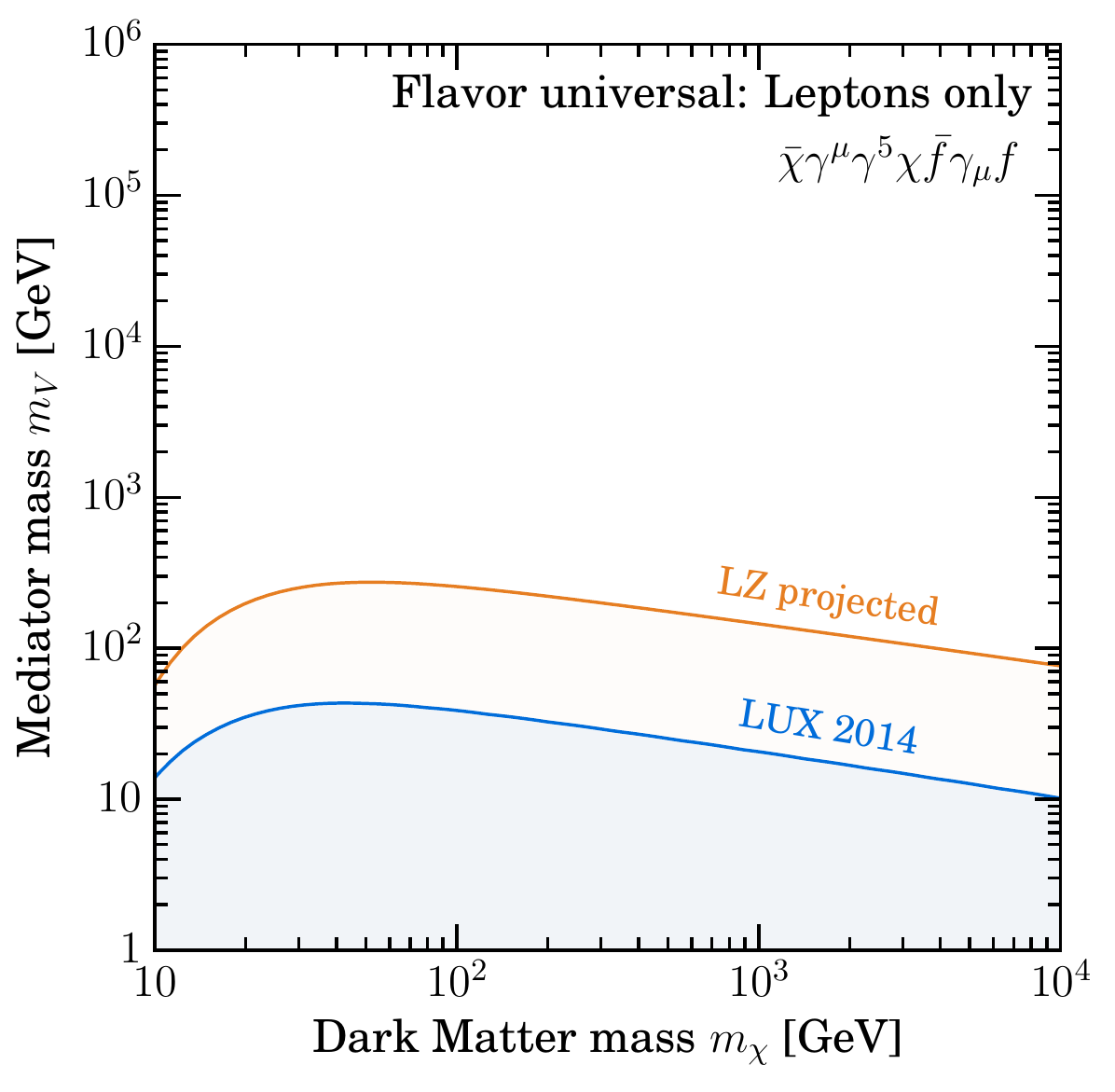}  \\
\includegraphics[width=0.495\textwidth]{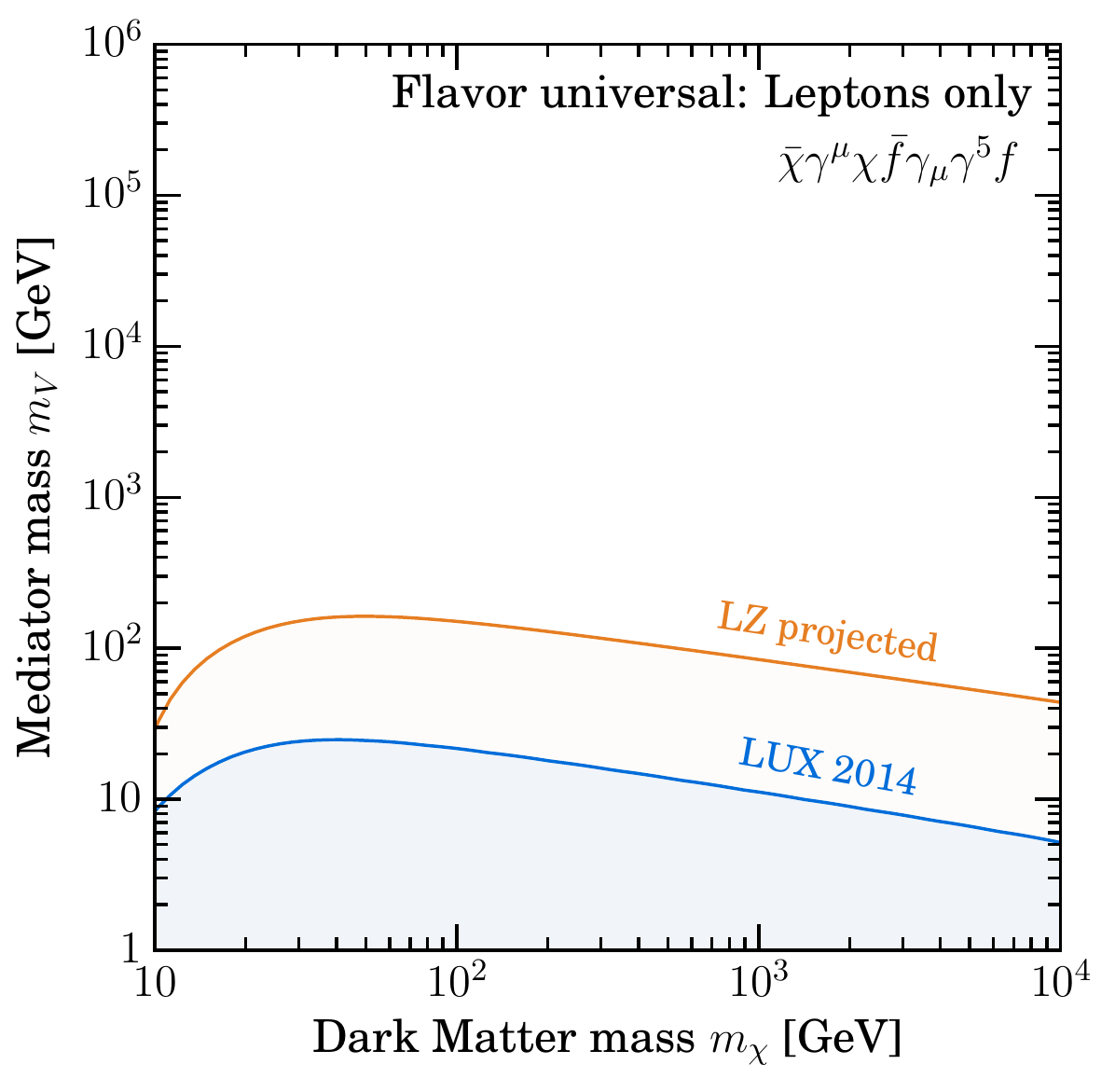}
\includegraphics[width=0.495\textwidth]{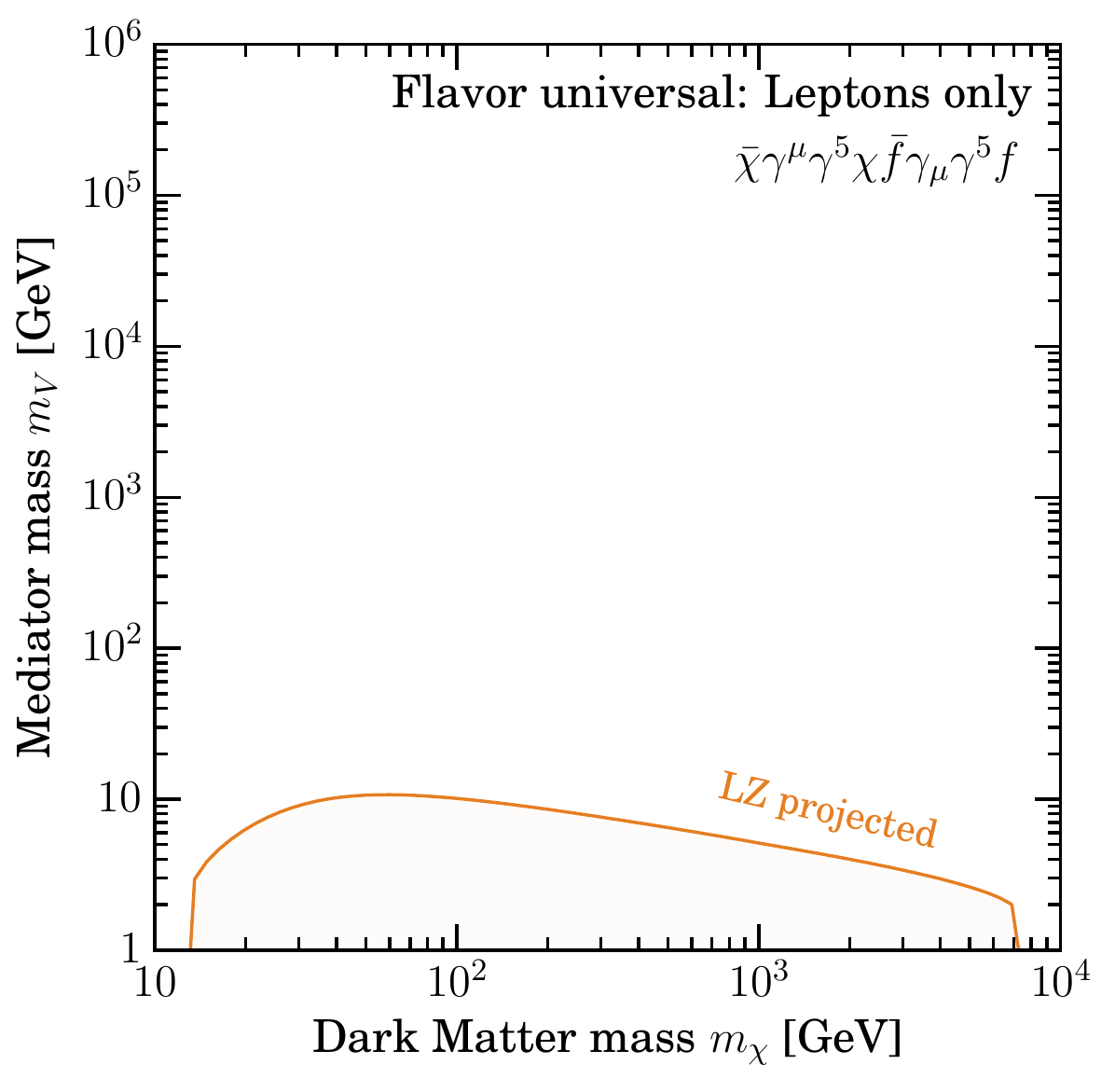} 
\caption{Same as \Fig{fig:universalQ_fermionDM} but for flavor universal couplings to leptons (Benchmark II). In all cases, no limits are obtained when running of the couplings is not taken into account.}
\label{fig:universalL_fermionDM}
\end{figure}

A peculiar feature of this benchmark is that DD rates are entirely due to RG effects. The contact interactions obtained by integrating-out the mediator particle do not involve light quarks. Thus, the DM-nucleon scattering cross section is only due to radiatively induced interactions with light quarks. The regions excluded in the $(m_\chi, m_V)$ plane for the 6 possible combinations are shown in \Figs{fig:universalL_fermionDM}{fig:universalL_scalarDM}. In order to understand these plots, it is helpful again to look at the approximate analytical solutions. For vector couplings to leptons we have
\be
\begin{split}
\text{Vector to leptons:}  \qquad  & \mathcal{C}_{V}^{(u)} \simeq \frac{4 \alpha_{\rm em}}{3 \pi} \ln(m_V / \mu_N)   \,, \\
& \mathcal{C}_{V}^{(d)} \simeq  - \frac{2 \alpha_{\rm em}}{3 \pi} \ln(m_V / \mu_N)   \,, \\
& \mathcal{C}_{A}^{(u)} \simeq \mathcal{C}_{A}^{(d)} \simeq \mathcal{C}_{A}^{(s)} \simeq 0 \,.
\end{split}
\label{eq:VectorLeptonsAnalytical}
\ee
As we have already found in \Eq{eq:VectorQuarksAnalytical}, there is no induced axial-vector couplings to light quarks if the mediator couples only to vector currents. However, unlike the case of \Eq{eq:VectorQuarksAnalytical}, there are only RG induced vector couplings to light quarks. These are the main source for DD rates. Likewise, for axial-vector couplings to leptons we have
\be
\begin{split}
\text{Axial-vector to leptons:}  \qquad  
& \mathcal{C}_{V}^{(u)} \simeq - \frac{\alpha_\tau}{6 \pi} \left(3 - 8 s_w^2 \right) \ln(m_V / \mu_N)   \,, \\
& \mathcal{C}_{V}^{(d)} \simeq  \frac{\alpha_\tau}{6 \pi} \left(3 - 4 s_w^2 \right) \ln(m_V / \mu_N)   \,, \\
& \mathcal{C}_{A}^{(u)} \simeq - \mathcal{C}_{A}^{(d)} \simeq  - \mathcal{C}_{A}^{(s)} \simeq \frac{\alpha_\tau}{2 \pi}  \ln(m_V / \mu_N) \,.
\end{split}
\label{eq:AxialLeptonsAnalytical}
\ee
The Yukawa coupling of the $\tau$ leptons generates one-loop suppressed couplings to both vector and axial-vector couplings of the low energy current. 

\begin{figure}
\centering
\includegraphics[width=0.495\textwidth]{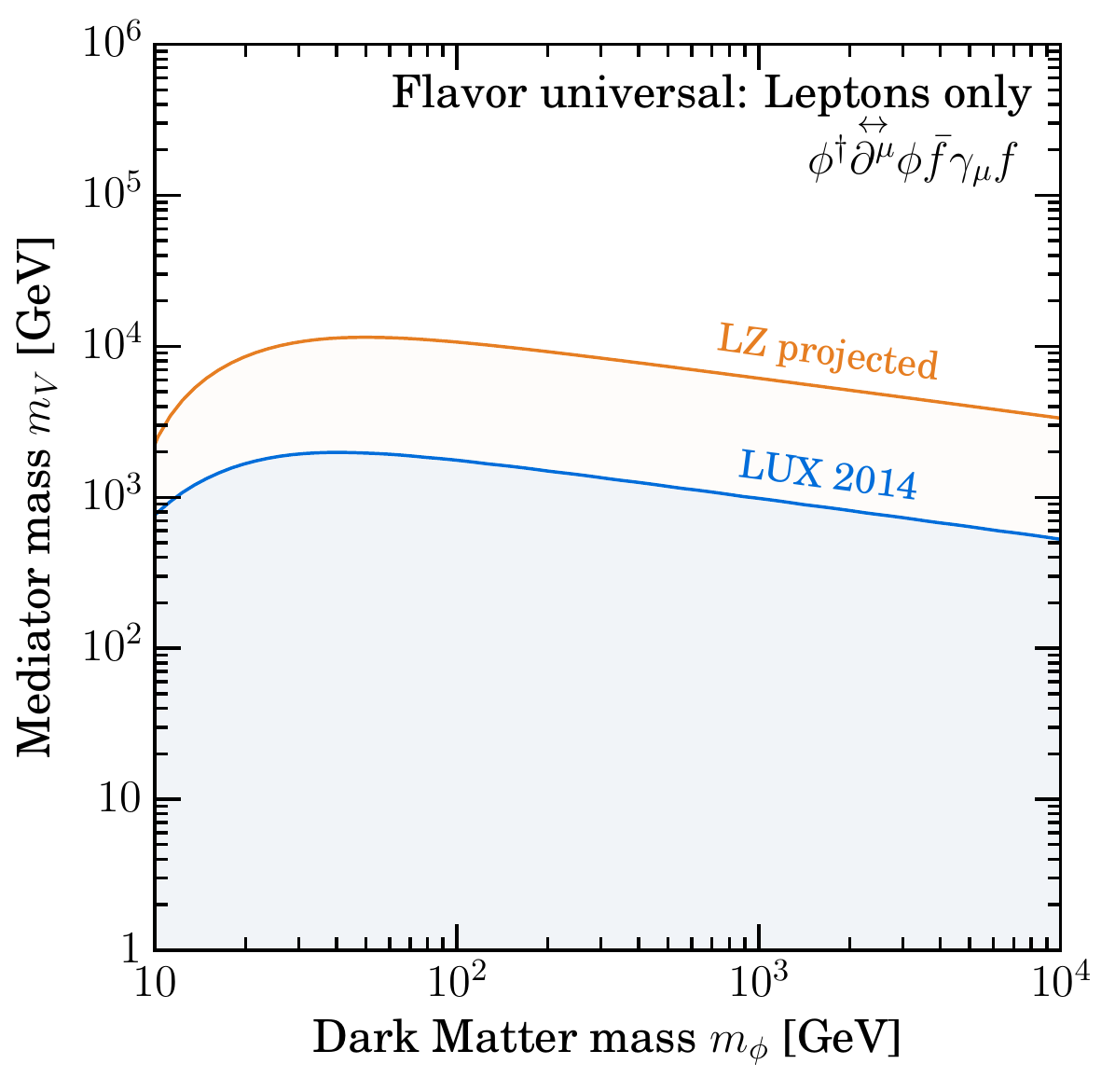} 
\includegraphics[width=0.495\textwidth]{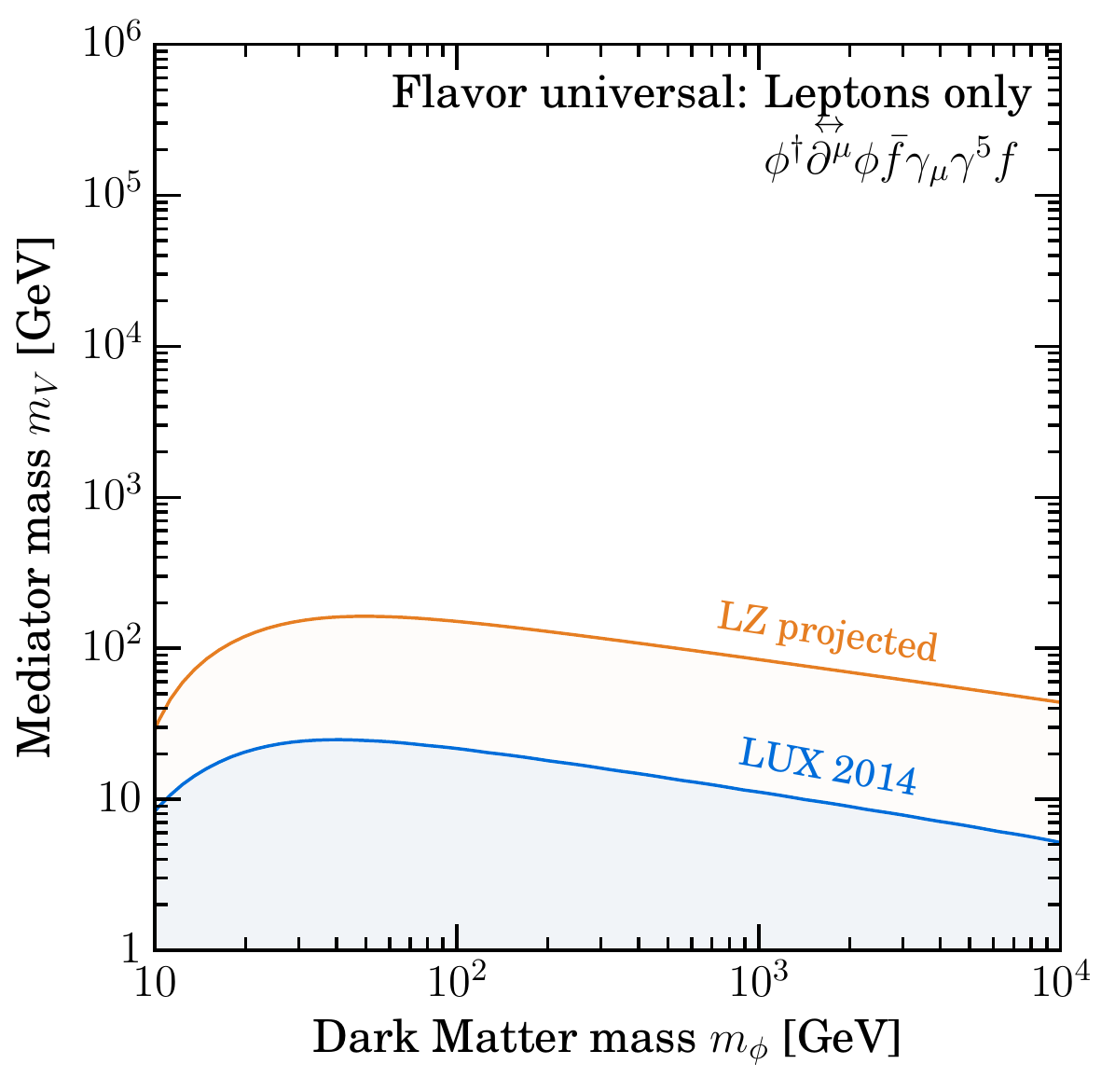} 
\caption{Same as \Fig{fig:universalL_fermionDM} but for scalar DM.}
\label{fig:universalL_scalarDM}
\end{figure}

Turning now to the limits obtained for this benchmark, we see from the top row of Fig.~\ref{fig:universalL_fermionDM} that for fermionic DM, constraints on the mediator mass are roughly an order of magnitude weaker for vector couplings to leptons compared to universal vector couplings to quarks. In the latter case, the running of the couplings had little impact, due to the smallness of the electromagnetic coupling compared to the tree level coupling to the light quarks. In the present case, as can been seen in \Eq{eq:VectorLeptonsAnalytical}, the radiatively induced coupling to light quarks is the only contribution. This RG-induced coupling is at the percent level, meaning that $m_V$ must be reduced by a factor of roughly 10 to achieve the same DD cross section as in the universal quark coupling case (as the cross section scales as $\sim\mathcal{C}^2/m_V^{4}$).

For couplings of fermionic DM to the axial-vector current of leptons (bottom row of Fig.~\ref{fig:universalL_fermionDM}), the limits are now roughly two orders of magnitude weaker than in the case of couplings to quarks. Once again, the only contribution to the DD rate is due to RG effects. However, the effect of the running is smaller for the lepton benchmark than for the quark benchmark. In the latter, the running is driven by the top Yukawa, while in the former, the running is driven predominantly by the smaller $\tau$ Yukawa (\Eq{eq:AxialLeptonsAnalytical}). 

For axial-vector couplings to both DM and leptons (lower right panel of Fig.~\ref{fig:universalL_fermionDM}), we note that we obtain no limits from \LUX\ above a mediator mass of 1 GeV and only very weak limits projected for \LZ. We also note that at low and high mass, the projected \LZ\ limits appear to drop off rapidly. For mediator masses close to 1 GeV, the running is too small to produce an observable DD signal and therefore no limits are obtained. However, we stress that the formalism used for the RGE (which involves resumming loop contributions proportional to  $\mathrm{Log} [m_V / \mu_N]$) is likely to break down as we approach $m_V = \mu_N$. The limits obtained with this formalism therefore cannot necessarily be trusted for $m_V \sim$ 1 GeV. 


Finally, for scalar DM (Fig.~\ref{fig:universalL_scalarDM}), the limits are very similar to those of fermionic DM with vector-current interactions, as discussed for Benchmark I. We highlight once again that these scalar DM interactions with leptons would lead to no DD constraints in a na\"ive treatment, but can give substantial constraints once RG effects are accounted for (especially in the case of coupling to lepton vector currents). 

\subsection{Benchmark III: couplings to third generation}
\label{sec:BenchmarkIII}

\begin{figure}
\centering
\includegraphics[width=0.495\textwidth]{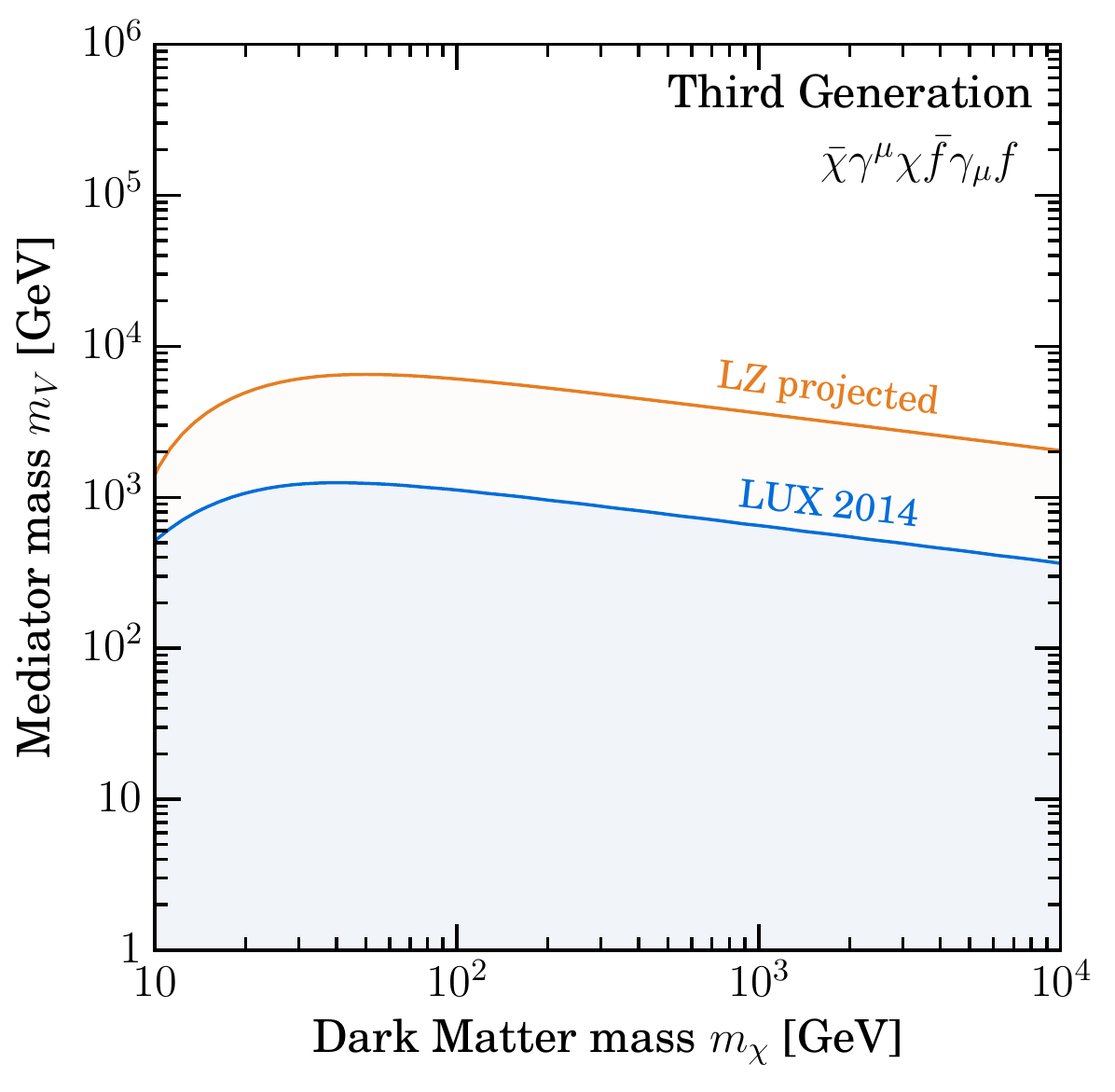} 
\includegraphics[width=0.495\textwidth]{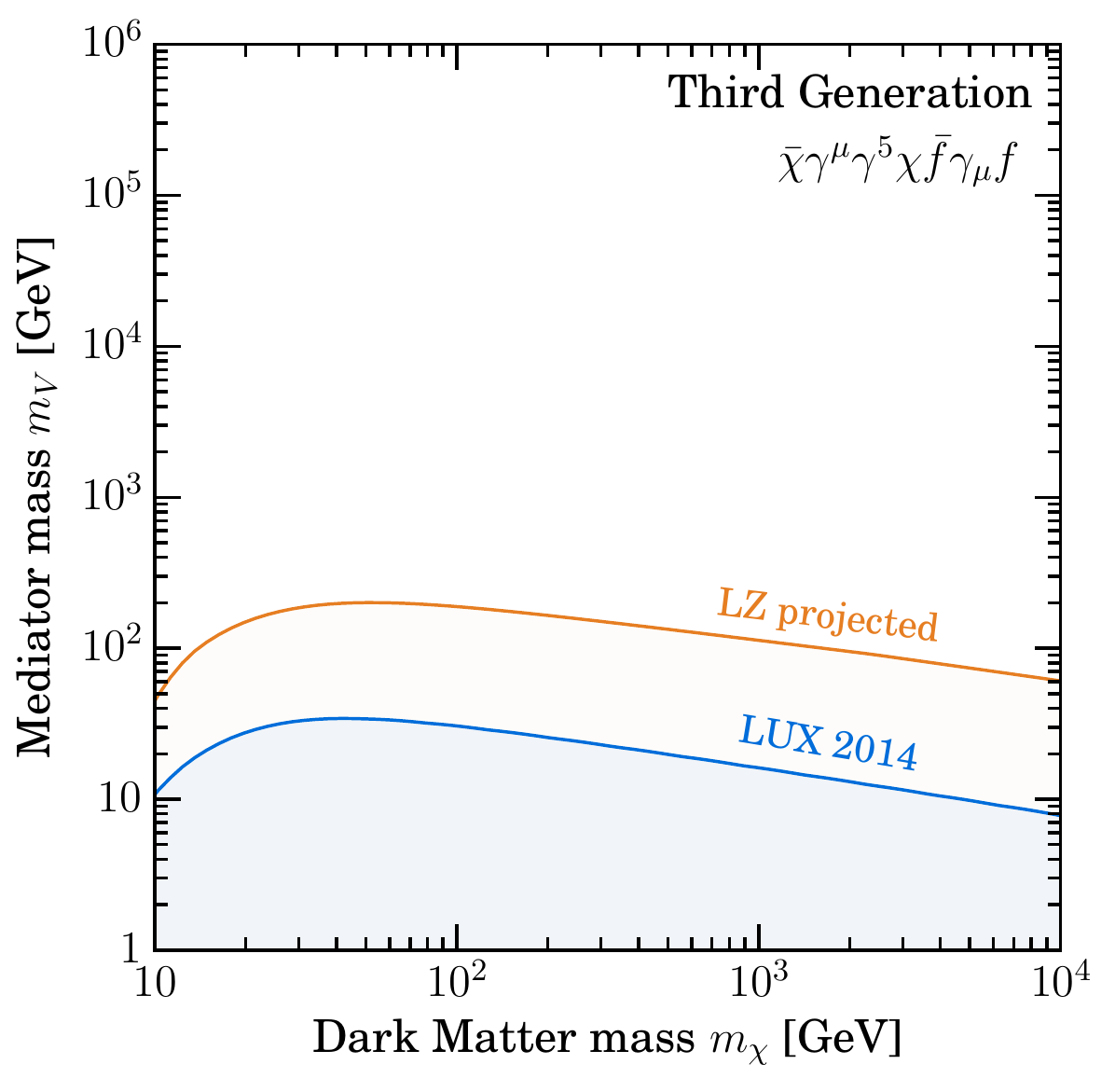}  \\
\includegraphics[width=0.495\textwidth]{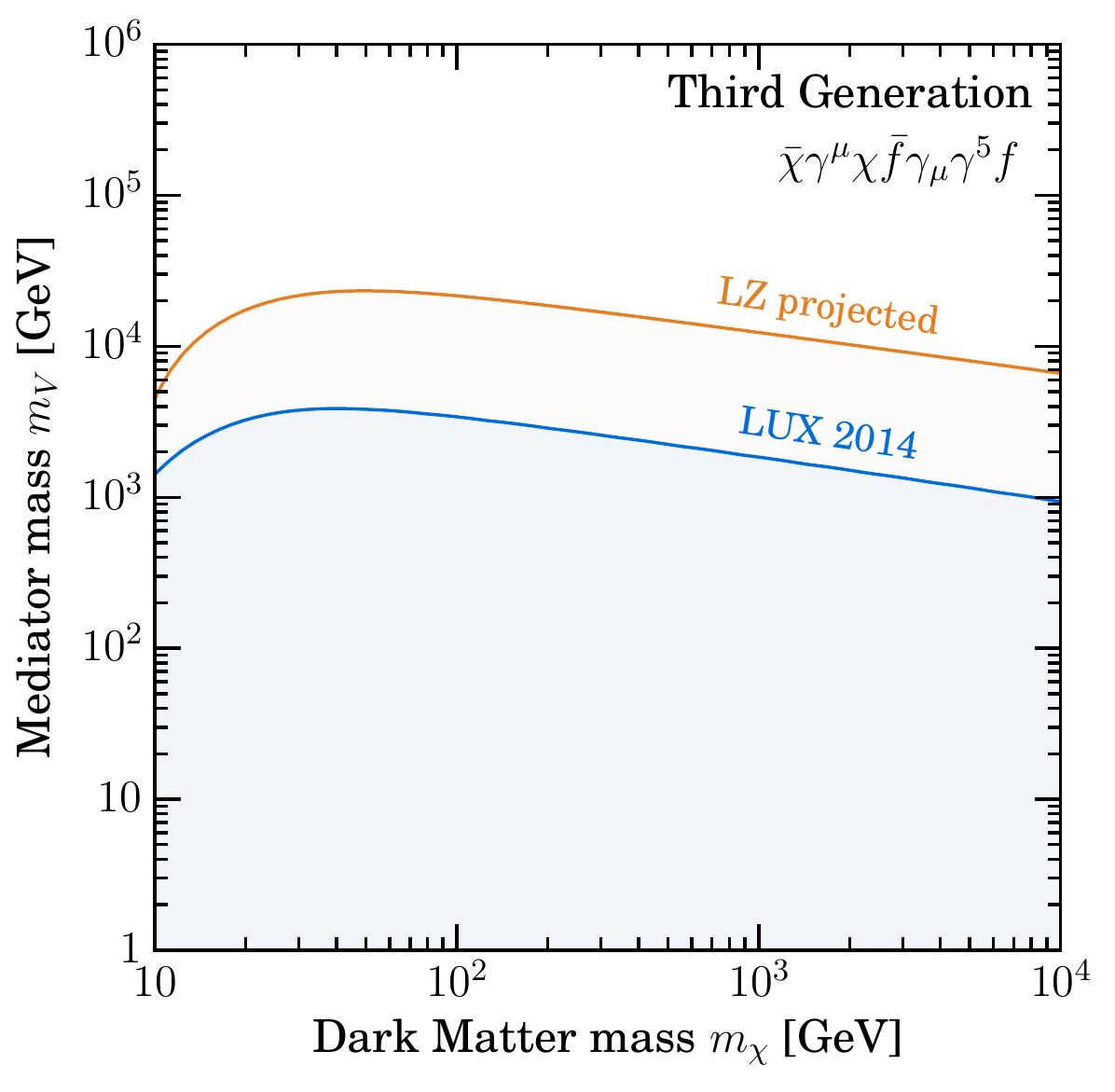}
\includegraphics[width=0.495\textwidth]{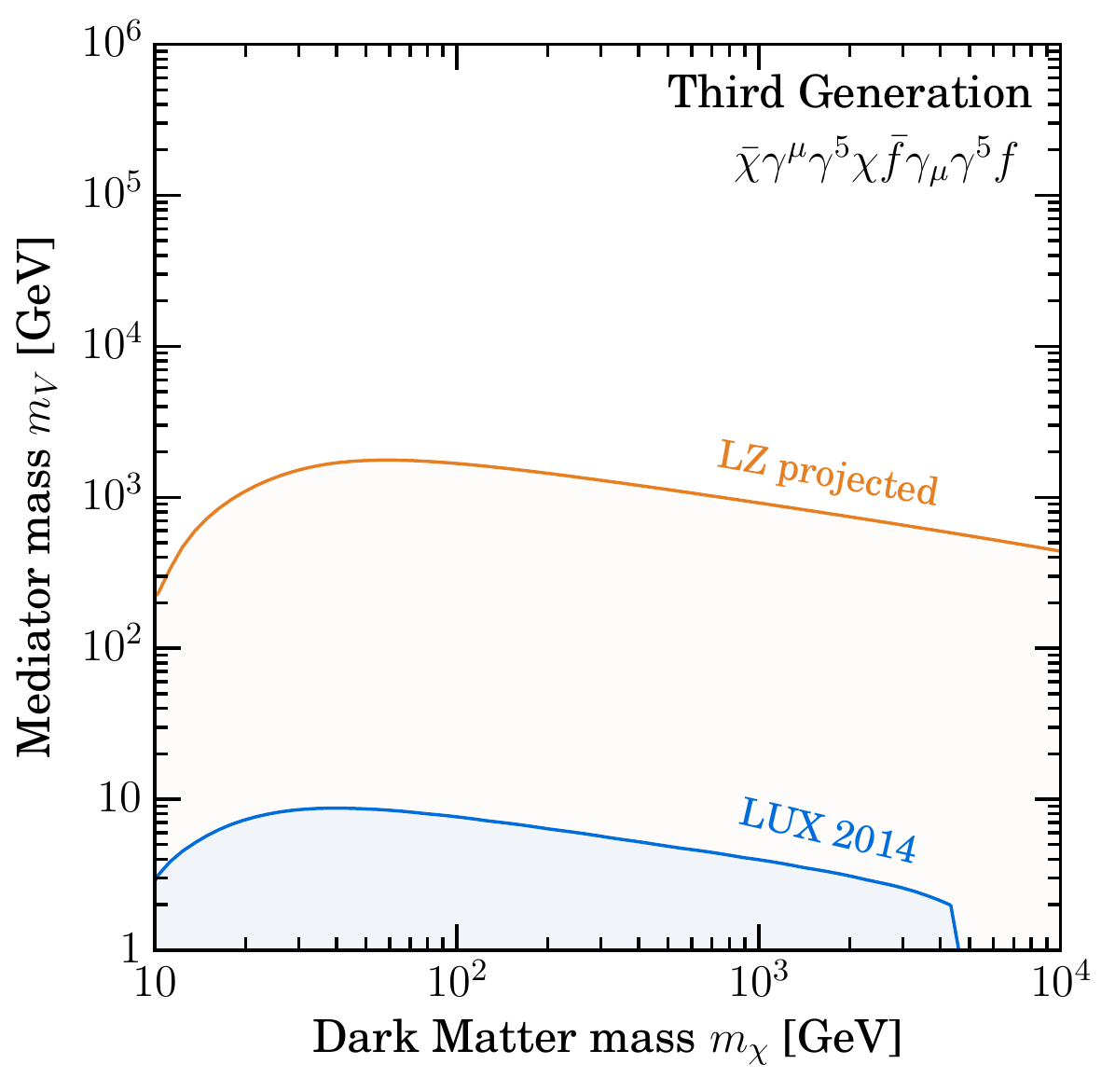} 
\caption{Same as \Fig{fig:universalQ_fermionDM} but for couplings to third generation fermions (Benchmark III). In all cases, no limits are obtained when running of the couplings is not taken into account.}
\label{fig:universalThird_fermionDM}
\end{figure}

Our last case involves coupling to only the third generation of SM fermions
\be
\begin{split}
c_{q}^{(i)} = & \, \delta^{i3} c_{q}^{(3)}  \,, \qquad\qquad c_{u}^{(i)} =  \delta^{i3} c_{u}^{(3)}  \,, \qquad\qquad c_{d}^{(i)} =  \delta^{i3} c_{d}^{(3)}  \,, \\
c_{l}^{(i)} = & \, \delta^{i3} c_l^{(3)} \,, \qquad\qquad  c_{e}^{(i)} =  \delta^{i3} c_e^{(3)}
 \qquad \qquad \qquad   (i = 1,2,3) \,.
\end{split}
\ee
We choose again two benchmark models with purely vector or axial-vector couplings
\begin{align}
\text{Vector to 3rd:} & \qquad   \qquad  c_{u}^{(3)} = c_{d}^{(3)} = c_{q}^{(3)} = c_{l}^{(3)} = c_{e}^{(3)} = 1 \,, \\
\text{Axial-vector to 3rd:} & \qquad   \qquad c_{u}^{(3)} = c_{d}^{(3)} = - c_{q}^{(3)} = - c_{l}^{(3)} = c_{e}^{(3)} = 1   \,.
\end{align}
The associated effective Lagrangian features only DM interactions with the top and bottom quarks and the $\tau$ lepton
\begin{align}
\text{Vector to 3rd:} & \qquad   
\mathcal{L}_{\rm EFT} = - \frac{1}{m_V^2} \, J_{{\rm DM}\, \mu} \; 
\left[ \overline{t} \gamma^\mu t + \overline{b} \gamma^\mu b + \overline{\tau} \gamma^\mu \tau  \right]  \,, \\
\text{Axial-vector to 3rd:} & \qquad   
\mathcal{L}_{\rm EFT} = - \frac{1}{m_V^2} \, J_{{\rm DM}\, \mu} \; 
\left[ \overline{t} \gamma^\mu \gamma^5 t + \overline{b} \gamma^\mu \gamma^5 b + \overline{\tau} \gamma^\mu \gamma^5 \tau \right]  \,.
\end{align}
The exclusion limits for these benchmarks are shown in \Figs{fig:universalThird_fermionDM}{fig:universalThird_scalarDM}, but we once again examine the approximate analytical expressions for the couplings before looking at the bounds in detail.

As was the case for the lepton benchmarks, this case has DD rates due entirely to radiatively induced interactions. An interesting effect happens for the case of vector couplings: above the weak scale the contribution from quarks is exactly canceled by the one from leptons. This can be understood by noticing that the effect is driven by electromagnetic loops of heavy fermions, with total contribution proportional to
\be
\sum_{f = t, b, \tau} N_c^{(f)} Q_f = 3 \times \left( \frac{2}{3} - \frac{1}{3}\right) - 1 = 0 \ .
\ee
Here, $Q_f$ is the electromagnetic charge of $f$ and the number of colors is $N_c^{(f)} = 3 \, (1)$ for quarks (leptons). We are only left with the RGE below the weak scale, non-vanishing now since the top quark is integrated-out. The net result is an induced coupling to vector currents of light quarks:
\be
\begin{split}
\text{Vector to 3rd:}  \qquad  & \mathcal{C}_{V}^{(u)} \simeq \frac{8 \alpha_{\rm em}}{9 \pi} \ln(m_Z / \mu_N)   \,, \\
& \mathcal{C}_{V}^{(d)} \simeq  - \frac{4 \alpha_{\rm em}}{9 \pi} \ln(m_Z / \mu_N)   \,, \\
& \mathcal{C}_{A}^{(u)} \simeq \mathcal{C}_{A}^{(d)} \simeq \mathcal{C}_{A}^{(s)} \simeq 0 \,.
\end{split}
\label{eq:VectorThirdAnalytical}
\ee
Although smaller than the couplings in the lepton case in \Eq{eq:VectorLeptonsAnalytical} due to less RG evolution, these contributions are still sizable and able to be probed by current and future experiments. Finally, we consider the axial-vector case. Since the effect is driven by SM Yukawa interactions, the low-energy couplings are just the sum of the RG induced ones in \Eqs{eq:AxialQuarksAnalytical}{eq:AxialLeptonsAnalytical}
\be
\begin{split}
\text{Axial-vector to 3rd:}  \qquad  & 
\mathcal{C}_{V}^{(u)} \simeq  \frac{\alpha_t}{2 \pi} \left(3 - 8 s_w^2 \right) \ln(m_V / m_Z) + \\ & \qquad \qquad 
- \left(3 - 8 s_w^2 \right) \left[ \frac{\alpha_b}{2 \pi}  +   \frac{\alpha_\tau}{6 \pi} \right]  \ln(m_V / \mu_N) \,, \\
& \mathcal{C}_{V}^{(d)} \simeq - \frac{\alpha_t}{2 \pi} \left(3 - 4 s_w^2 \right) \ln(m_V / m_Z) + \\ & \qquad \qquad
 \left(3 - 4 s_w^2 \right) \left[ \frac{\alpha_b}{2 \pi} + \frac{\alpha_\tau}{6 \pi} \right]  \ln(m_V / \mu_N)  \,, \\
& \mathcal{C}_{A}^{(u)} \simeq - \mathcal{C}_{A}^{(d)} \simeq - \mathcal{C}_{A}^{(s)} \simeq - \frac{3 \alpha_t}{2 \pi}  \ln(m_V / m_Z) +  \\ & \qquad \qquad \qquad  \frac{3 \alpha_b}{2 \pi}  \ln(m_V / \mu_N) + \frac{\alpha_\tau}{2 \pi}  \ln(m_V / \mu_N)  \, .
\end{split}
\ee

\begin{figure}
\centering
\includegraphics[width=0.495\textwidth]{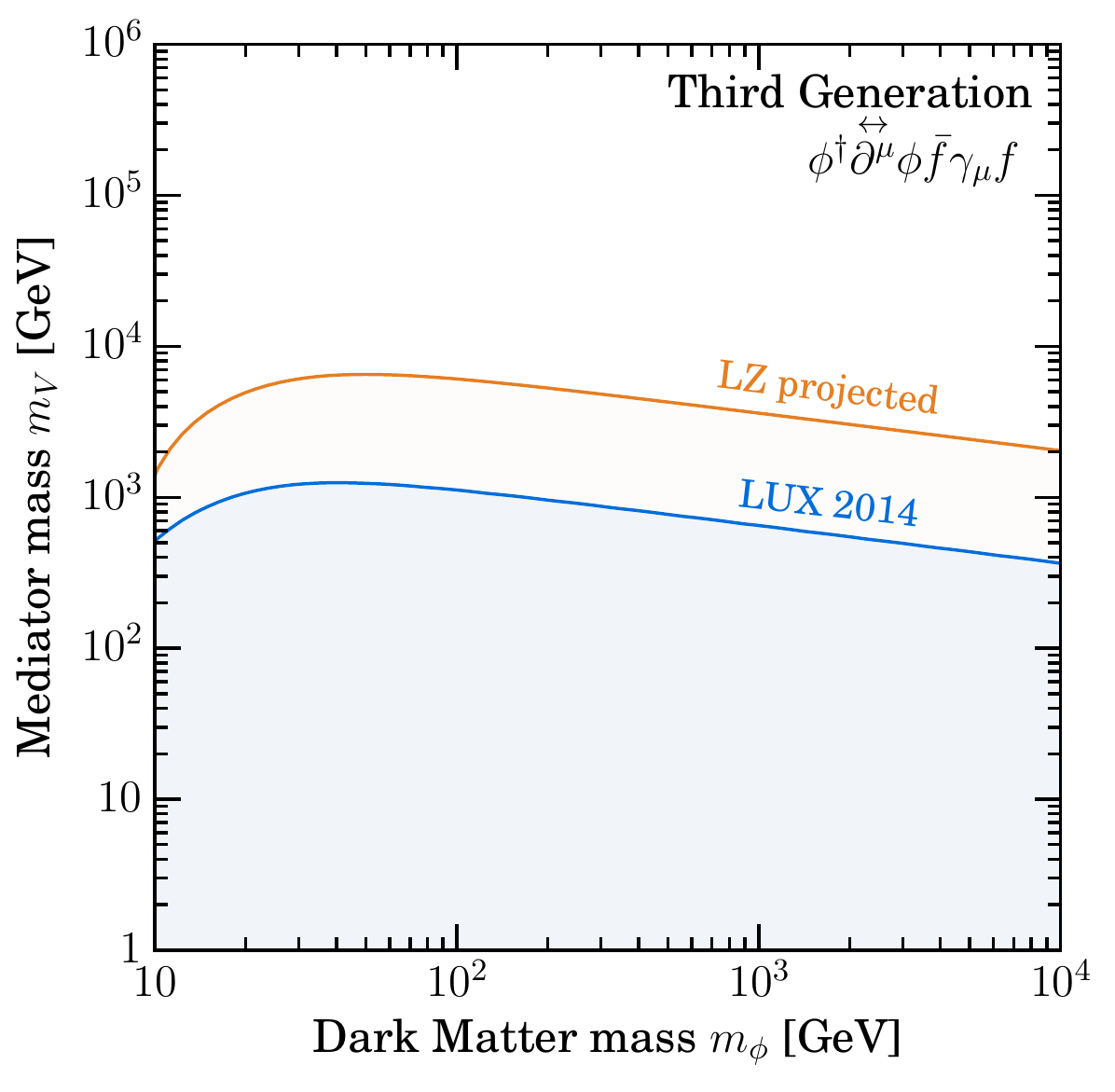} 
\includegraphics[width=0.495\textwidth]{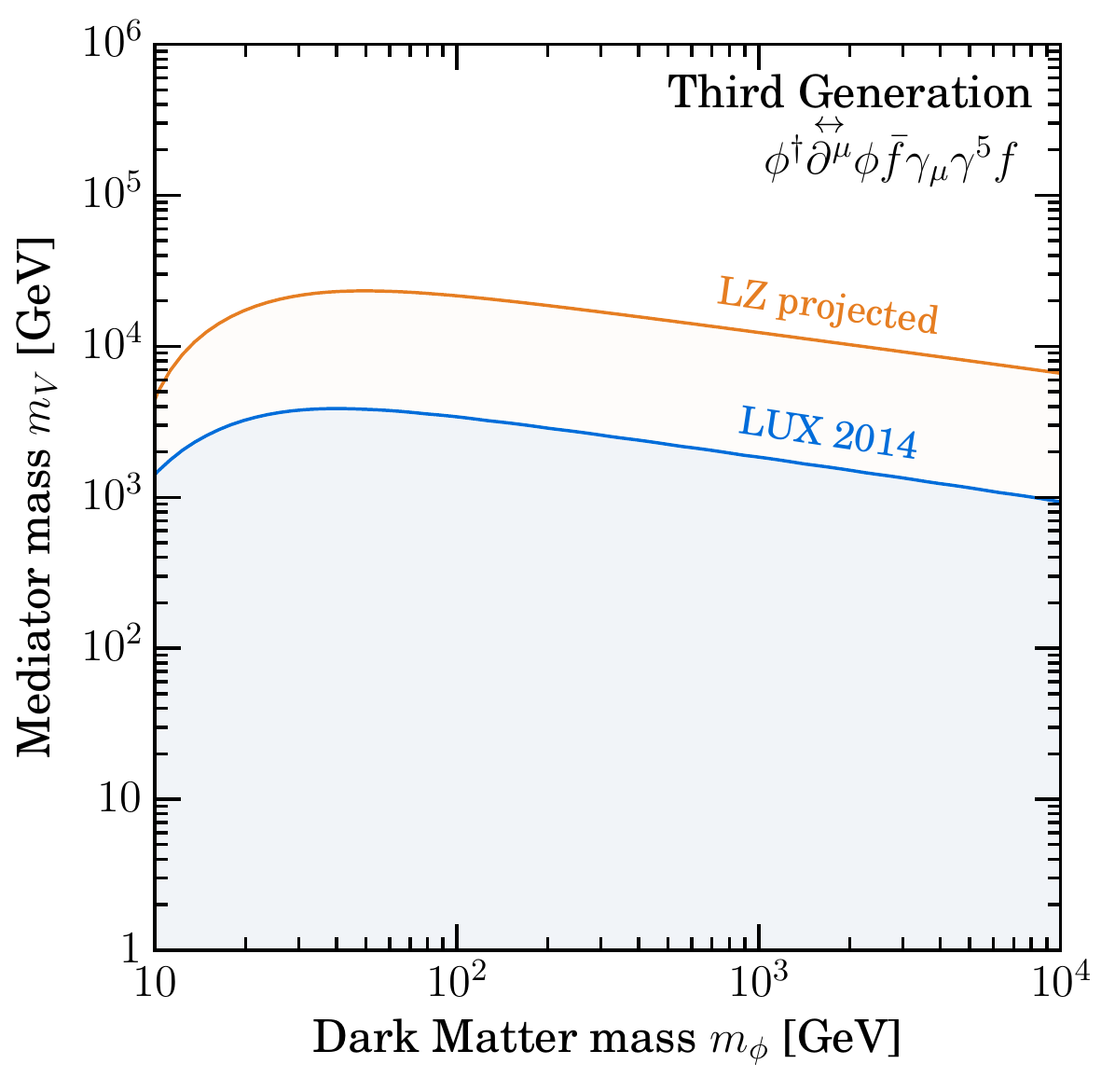} 
\caption{Same as \Fig{fig:universalThird_fermionDM} but for scalar DM.}
\label{fig:universalThird_scalarDM}
\end{figure}

The limits on the mediator mass in this benchmark are shown in Fig.~\ref{fig:universalThird_fermionDM} in the case of fermionic DM. For vector couplings to the third generation fermions (top row), the limits are slightly weaker than those obtained in the lepton-only benchmark (Fig.~\ref{fig:universalL_fermionDM}). As already discussed, there is no running above the weak scale in the former case, meaning that the radiatively induced couplings are smaller. 

In contrast, for axial-vector couplings to third generation fermions (bottom row of Fig.~\ref{fig:universalThird_fermionDM}), the constraints are substantially stronger when compared to the leptons-only benchmark. This is because the DM couples to all 3 heavy fermions ($t$, $b$, $\tau$) and not just the $\tau$ as in the leptons-only case. As the mixing is driven by the Yukawa couplings, this enhances the RG effects and strengthens the limits. This is perhaps most obvious in the case of scalar DM (Fig.~\ref{fig:universalThird_scalarDM}), where coupling to the axial-vector current now gives stronger constraints than coupling to the vector current. The DD rate in both cases is dominated by the same low-energy operator (the SI operator $\mathcal{O}_1^{\rm NR}$), indicating that the mixing into the light quark vector current must be larger in the case of axial-vector interactions at high-energy.

We finally comment on the fact that the \LZ\ projected limits are substantially stronger than the current \LUX\ limits in the case of axial-vector axial-vector couplings to the third generation fermions (lower right panel of Fig.~\ref{fig:universalThird_fermionDM}). A na\"ive comparison of the \LUX\ and \LZ\ exposures would suggest that \LZ\ should only be able to constrain mediator masses of $m_V \gtrsim 100 \,\, \mathrm{GeV}$. However, as the mediator mass increases above the weak scale, the effects of the running become more significant. The top quark has been integrated out below the $Z$ mass, while it is still in the spectrum (and contributes to the running) above the $Z$ mass. This increases the RG-induced couplings to the light quark axial-vector currents and thus leads to stronger limits on $m_V$.


\section{Rescaling and general limits}
\label{sec:rescaling}

We have so far focused only on a small number of benchmark scenarios. In this section we summarize the procedure for obtaining limits on general vector mediator models with arbitrary coefficients. The public code \runDM, released with this paper and available at  \href{https://github.com/bradkav/runDM/}{this http URL}, allows the user to evolve the Wilson coefficients of the simplified model between two arbitrary energy scales. The coefficients at the nuclear scale can then be fed into \textsc{NRopsDD}, available at \href{http://www.marcocirelli.net/NROpsDD.html}{this http URL}, which allows the user to derive bounds from current direct detection experiments.\footnote{As part of the \runDM release, we also provide an example mathematica notebook illustrating how to interface the two codes.} The procedure is as follows:


\begin{itemize}
\item[1a.] Define your favorite model according to the Lagrangian in Eq.~\eqref{eq:LagEFTaboveAPP}. This amounts to specifying a 16-dimensional array of Wilson coefficients defined in Eq.~\eqref{ed:cdef}, which can be used as input for the \runDM code.

\item[1b.]  Use the function  \texttt{DDCouplingsQuarks} to compute the low-energy couplings to light quarks ($u$, $d$, $s$) relevant for direct detection. These should then be dressed to the nucleon level by means of Eq.~\eqref{eq:nucleoncouplings}, and matched onto the NR operator coefficients (making sure to include all relevant operators, which may not be limited to the standard SI and SD interactions defined in Eq.~\eqref{eq:usualSI&SD}). In order to facilitate this, we also provide the function \texttt{DDCouplingsNR} which returns the NR coefficients $\mathfrak{c}_{\phi(\chi) \, i}^{(N)}$, defined in Appendix~\ref{app:NRoperators}, for each type of DM current considered in this work. These coefficients are functions of the coupling of the DM current defined in Eq.~\eqref{eq:DMcurrentChoices}, 16 Wilson coefficients of  Eq.~\eqref{eq:LagEFTaboveAPP}, the DM mass $m_\chi$ and the mediator mass $m_V$. For compactness, we define $\lambda \equiv (g_{\phi(\chi)}, \text{16 Wilson coefficients}, \, m_V)$. The function \texttt{DDCouplingsNR} automatically performs the embedding of the quark-level couplings to the nucleon level and computes the NR DM-nucleon matrix element according to the numbering of the NR operators defined in Appendix~\ref{app:NRoperators}.




\item[2a.] Calculate the benchmark DD coupling $\lambda_{\rm B}$ as a function of the parameters $\lambda$ and $m_\chi$:
\be\label{eq:QF}
\begin{split}
\sum_{i, j = 1, 7} \sum_{N, N' = p, n} \mathfrak{c}_{\phi \, i}^{(N)}(\lambda, m_\chi) \, \mathfrak{c}_{\phi \, j}^{(N')}(\lambda, m_\chi) \, \mathcal{Y}_{i, j}^{(N,N')}(m_\chi) &= \lambda_\text{B}(m_\chi)^2 \ , \quad \text{for scalar DM}\,, \\
\sum_{i, j = 4, 8, 9} \sum_{N, N' = p, n} \mathfrak{c}_{\chi \, i}^{(N)}(\lambda, m_\chi) \, \mathfrak{c}_{\chi \, j}^{(N')}(\lambda, m_\chi) \, \mathcal{Y}_{i, j}^{(N,N')}(m_\chi) &= \lambda_\text{B}(m_\chi)^2 \ , \quad \text{for Majorana DM}\,, \\
\sum_{i, j = 1, 4, 7, 8, 9} \sum_{N, N' = p, n} \mathfrak{c}_{\chi \, i}^{(N)}(\lambda, m_\chi) \, \mathfrak{c}_{\chi \, j}^{(N')}(\lambda, m_\chi) \, \mathcal{Y}_{i, j}^{(N,N')}(m_\chi) &= \lambda_\text{B}(m_\chi)^2 \ , \quad \text{for Dirac DM}\,.
\end{split}
\ee
Here $\mathcal{Y}_{i, j}^{(N,N')}(m_\chi)$ are the rescaling functions computed in \textsc{NRopsDD}, which relate the DD rates for different NR operators. The sum in $i, j$ is over the specific NR operators relevant in each case.

\item[2b.] Insert the above expression for $\lambda_{\rm B}$ into the test statistic function ${\rm TS}(\lambda_\text{B}, m_\chi)$,  provided in \textsc{NRopsDD}, for a given DD experiment. At this point the user possesses the test statistic ${\rm TS}(\lambda, m_\chi)$ and can derive a bound on $\lambda$ (or equivalently $m_V$) at the desired confidence level, \eg by drawing a contour plot of ${\rm TS} = 2.71$ for a 90\% CL. 

\end{itemize}


\section{Connecting direct detection and collider searches}
\label{sec:LHC}

We now turn to an example where the techniques we have developed are important for a correct exploration of the complementarity between different DM searches. Using simplified models, it is possible to map \LHC\ constraints on the mass of the mediator onto the $(m_\chi, \,\sigma_{\chi N})$ plane which is usually presented by DD experiments \cite{Boveia:2016mrp}. However, the large separation of scales and resulting RG effects are not typically included in such a mapping. As we will now show, these effects can have a non-trivial effect on the derived cross section limits.

The simplified models used in DM searches at the \LHC\ typically assume a universal coupling of the mediator to quarks, corresponding exactly to our Benchmark I, discussed in Sec.~\ref{sec:BenchmarkI}. When the mediator is assumed to couple to the vector current of both DM and quarks, we have seen that the DD rate will not be substantially affected by RG effects. This is because the rate is dominated by the tree level coupling to the light quark vector currents, for which the running is small. In this case, then, the comparison between \LHC\ and DD limits is relatively straightforward.

\begin{figure}
\centering
\includegraphics[width=0.495\textwidth]{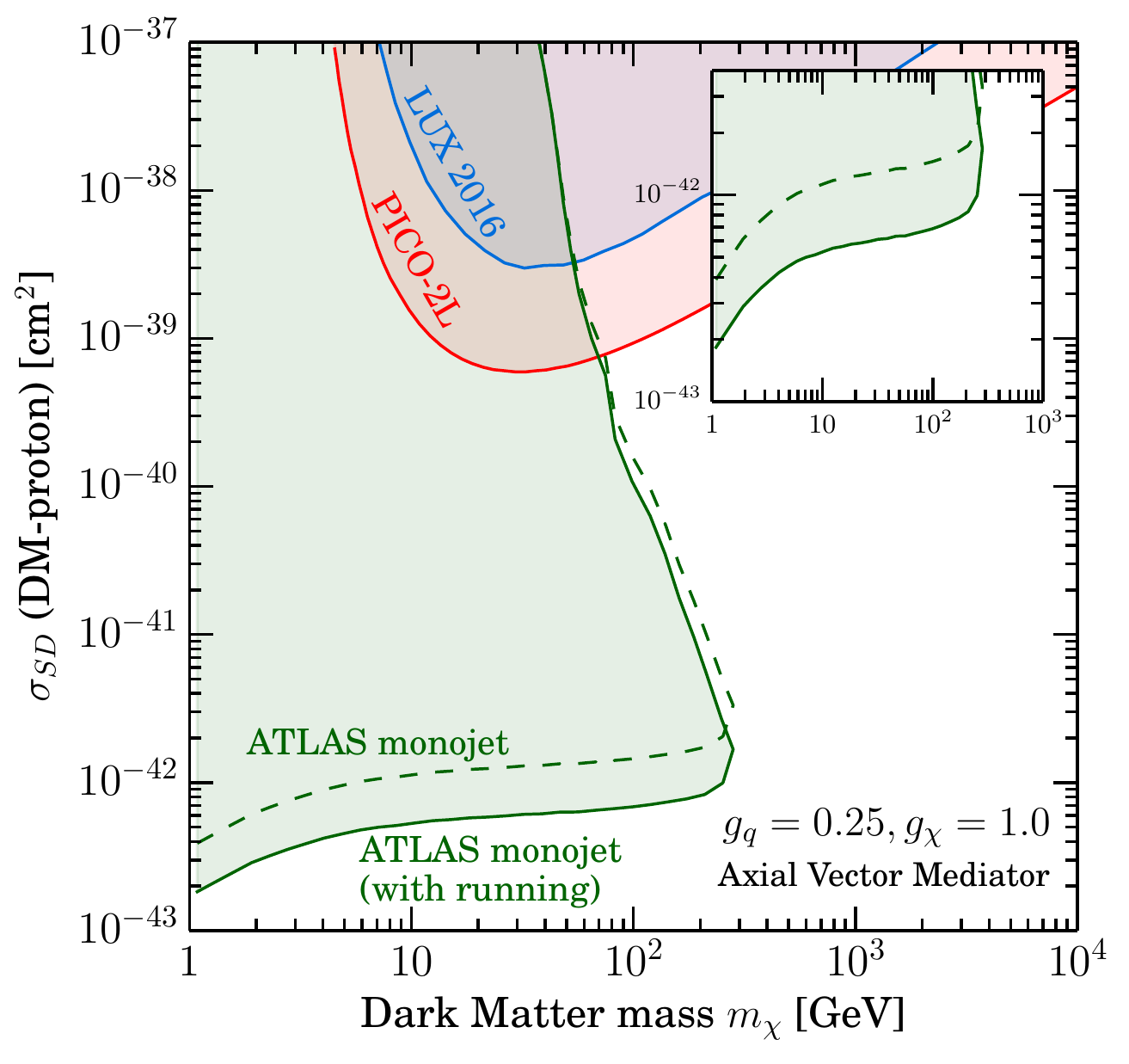} 
\includegraphics[width=0.495\textwidth]{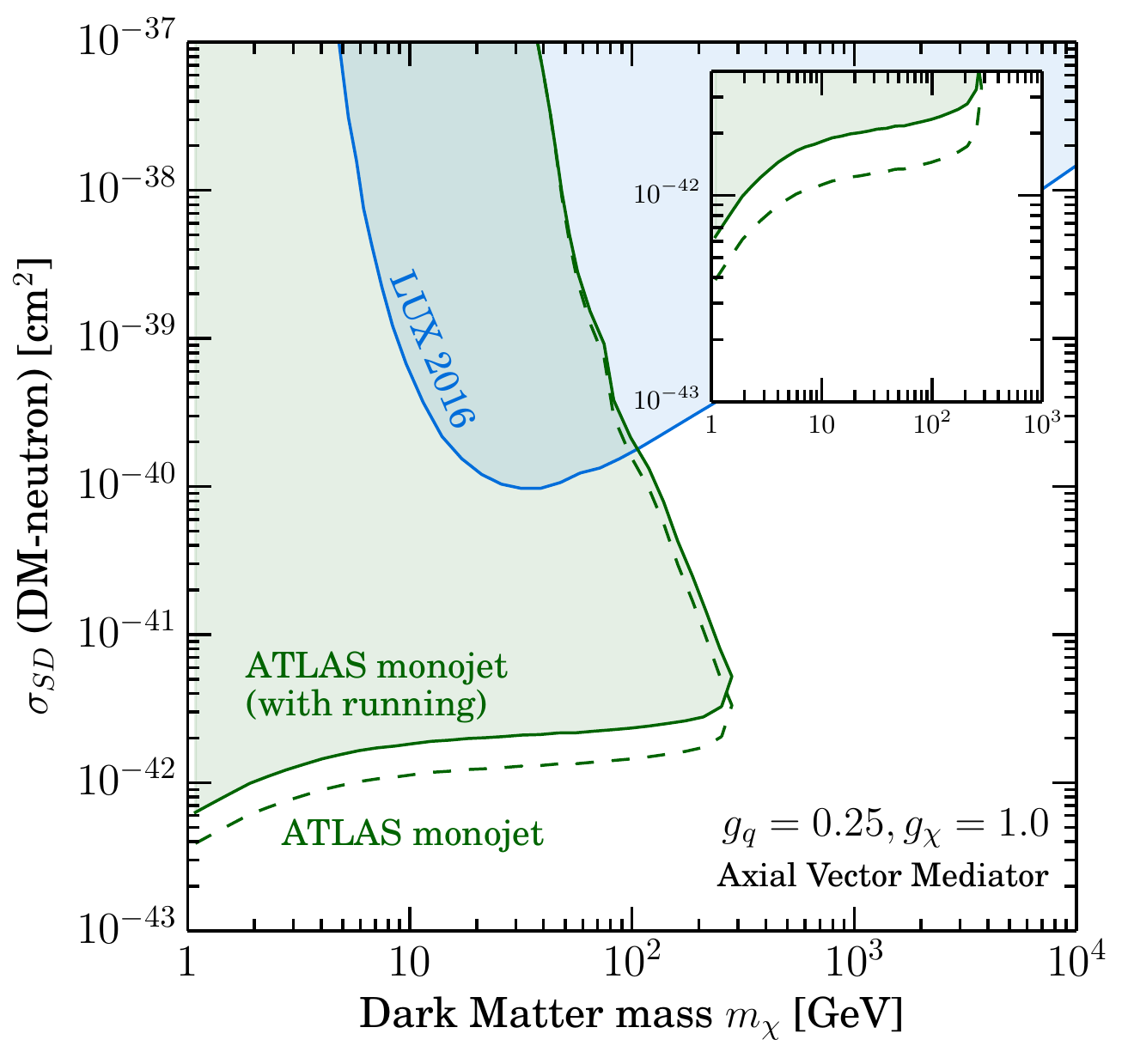} 
\caption{90\% confidence limits on the DM-proton (left) and DM-neutron (right) spin-dependent scattering cross section for Dirac DM. The dashed green lines show the limits from the \ATLAS monojet search at 13 TeV, as reported in Ref.~\cite{Aaboud:2016tnv}, assuming a universal coupling to all quarks through the axial-vector operator $\overline{\chi}\gamma^\mu \gamma^5 \chi \overline{q} \gamma_\mu\gamma^5 q$. Solid green lines show the correct limits when running of the couplings is taken into account. The inset shows a zoom-in of the \ATLAS limits at small cross section. Direct detection limits from \LUX\ \cite{Akerib:2013tjd} (blue) and \PICO \cite{Amole:2015lsj} (red) are also shown for comparison. The coupling $g_\chi$ used by the \LHC\ collaborations corresponds in this context to our $c_{\chi A}$ of \Eq{eq:JmuDM}.}
\label{fig:LHClimits}
\end{figure}

\begin{figure}
\centering
\includegraphics[width=0.495\textwidth]{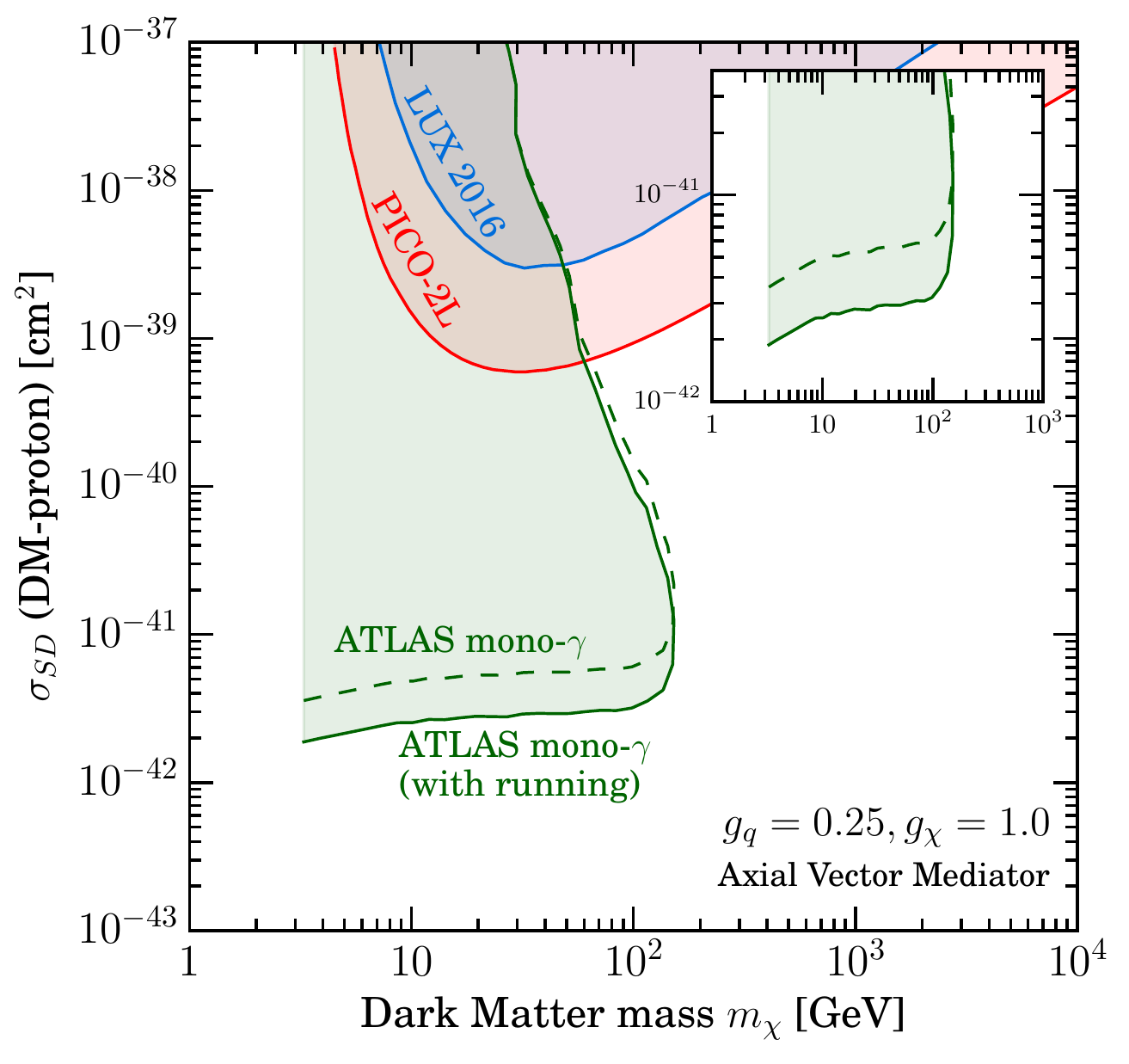} 
\includegraphics[width=0.495\textwidth]{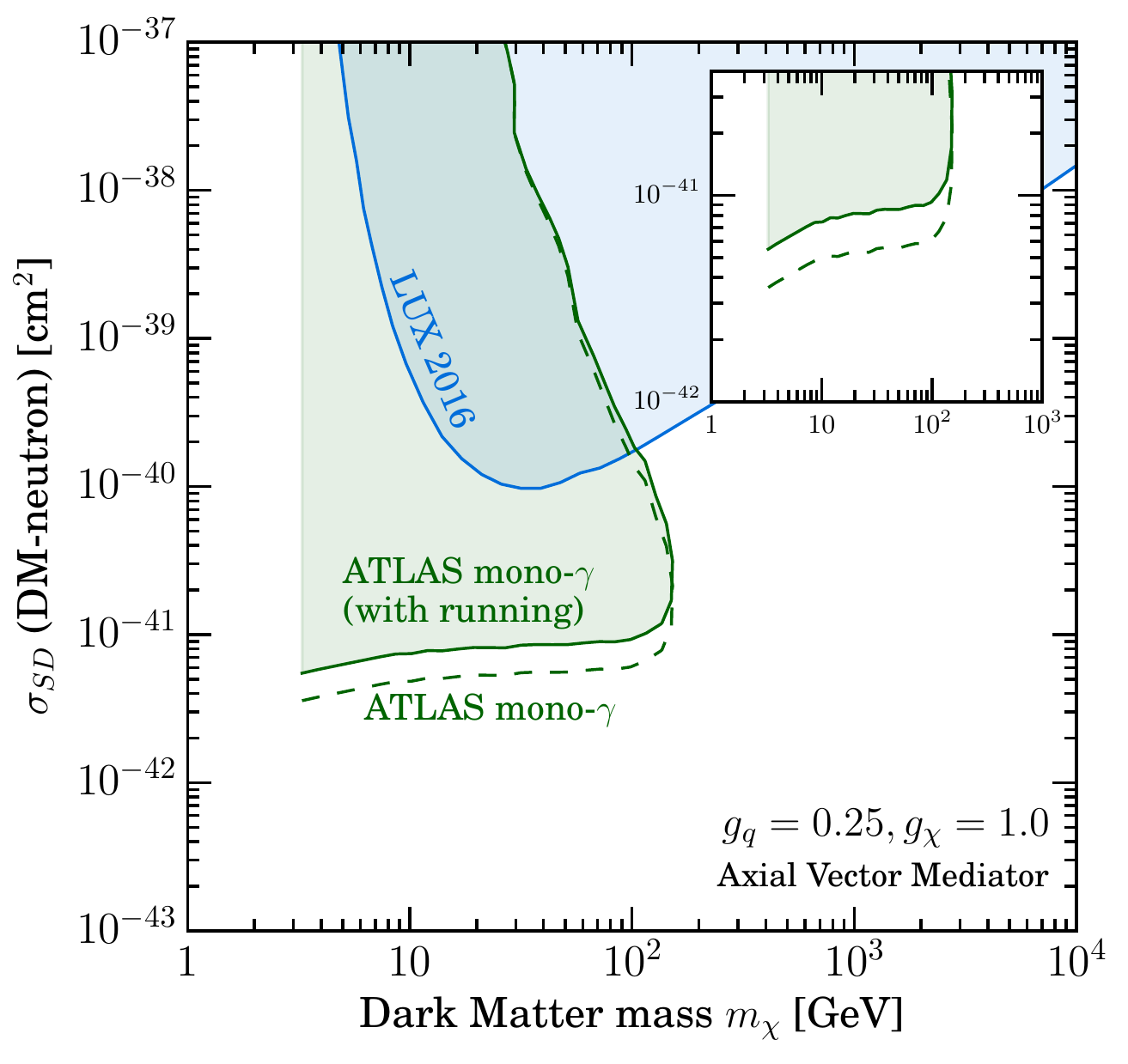} 
\caption{As Fig.~\ref{fig:LHClimits}, but showing limits in green (with and without running) from the \ATLAS monophoton search at 13 TeV, as reported in Ref.~\cite{Aaboud:2016uro}.}
\label{fig:LHClimits_monophoton}
\end{figure}

However, as shown in Sec.~\ref{sec:BenchmarkI}, for axial-vector current interactions there is a large mixing effect driven by the top Yukawa which leads to a larger effect on DD rates. As discussed in Sec.~\ref{sec:RGE}, interactions of the form $\overline{\chi}\gamma^\mu \gamma^5 \chi \overline{q} \gamma_\mu \gamma^5 q$ lead to a spin-dependent (SD) interaction which is not suppressed by powers of the momentum transfer or the DM speed. In Fig.~\ref{fig:LHClimits}, we show the 90\% limits on the DM-proton (left panel) and DM-neutron (right panel) SD cross sections reported by the \LUX\ \cite{Akerib:2013tjd} (blue) and \PICO \cite{Amole:2015lsj} (red) direct detection experiments. On the same plane, we show limits from a recent Run-2 monojet search at \ATLAS \cite{Aaboud:2016tnv}, interpreted in the context of an axial-vector mediator simplified model. The dashed green line shows the limit reported in Ref.~\cite{Aaboud:2016tnv}, without including the effects of running. The solid green line shows the limit when the RG effects discussed in this paper are included. The limits including running are obtained by first taking the reported limits (dashed line) and calculating the corresponding limit on the mediator mass.\footnote{The reported limits on the cross-section are obtained by digitizing Fig.~7 of Ref.~\cite{Aaboud:2016tnv}, using the publicly-available \textsc{WebPlotDigitizer} \cite{WebPlot}.}  We then run the couplings from this mediator mass down to the nuclear scale and calculate the SD cross section (see \Eq{eq:sigmaSD}). The difference is a factor of approximately 2 in the cross section, with the limits being strengthened in the DM-proton case and weakened in the DM-neutron case. We find similar results when we include the effects of running on \ATLAS limits from the monophoton search at 13 TeV \cite{Aaboud:2016uro}, as shown in Fig.~\ref{fig:LHClimits_monophoton}.

In order to understand this difference, we need the explicit expression for  the DM-nucleon SD cross section (in the zero momentum transfer limit) shown in Eq.~\eqref{eq:usualSI&SD}.  
%
Combining \Eqs{eq:AxialQuarksAnalytical}{eq:nucleoncouplings}, we see that at the nuclear energy scale, the axial couplings to nucleons can be written,

\begin{equation}
\mathcal{C}_{A}^{(N)} = g_q\left[ \sum_{q = u,d,s} \Delta_q^{(N)} \right] + \frac{3g_q}{2\pi}\left( \Delta_d^{(N)} + \Delta_s^{(N)} - \Delta_u^{(N)}\right) \left[ \alpha_t \ln(m_V / m_Z) - \alpha_b  \ln(m_V / \mu_N)   \right]\,,
\label{eq:sigmaSD}
\end{equation}
where the axial-charges $\Delta_q^{(N)}$ are defined in \Eq{eq:nucleoncouplings}. We have also explicitly included the coupling of the mediator to quarks $g_q$ to allow for the possibility that this may differ from unity. The first term (in square brackets) is the DM-nucleon coupling evaluated at the energy scale of the mediator and leads to equal couplings to protons and neutrons. The second term in Eq.~\ref{eq:sigmaSD} accounts for RG effects, which induce a correction to the up quark current with an opposite sign to the corrections for down and strange quarks. Noting that $\Delta_u^{(p)} = \Delta_d^{(n)}$ and $\Delta_d^{(p)} = \Delta_u^{(n)}$, we see that the corrections to the nucleon couplings will have roughly opposite signs for protons and neutrons.\footnote{The contribution from $\Delta_s^{(N)}$ is smaller though not negligible.}

\begin{figure}
\centering
\includegraphics[width=0.495\textwidth]{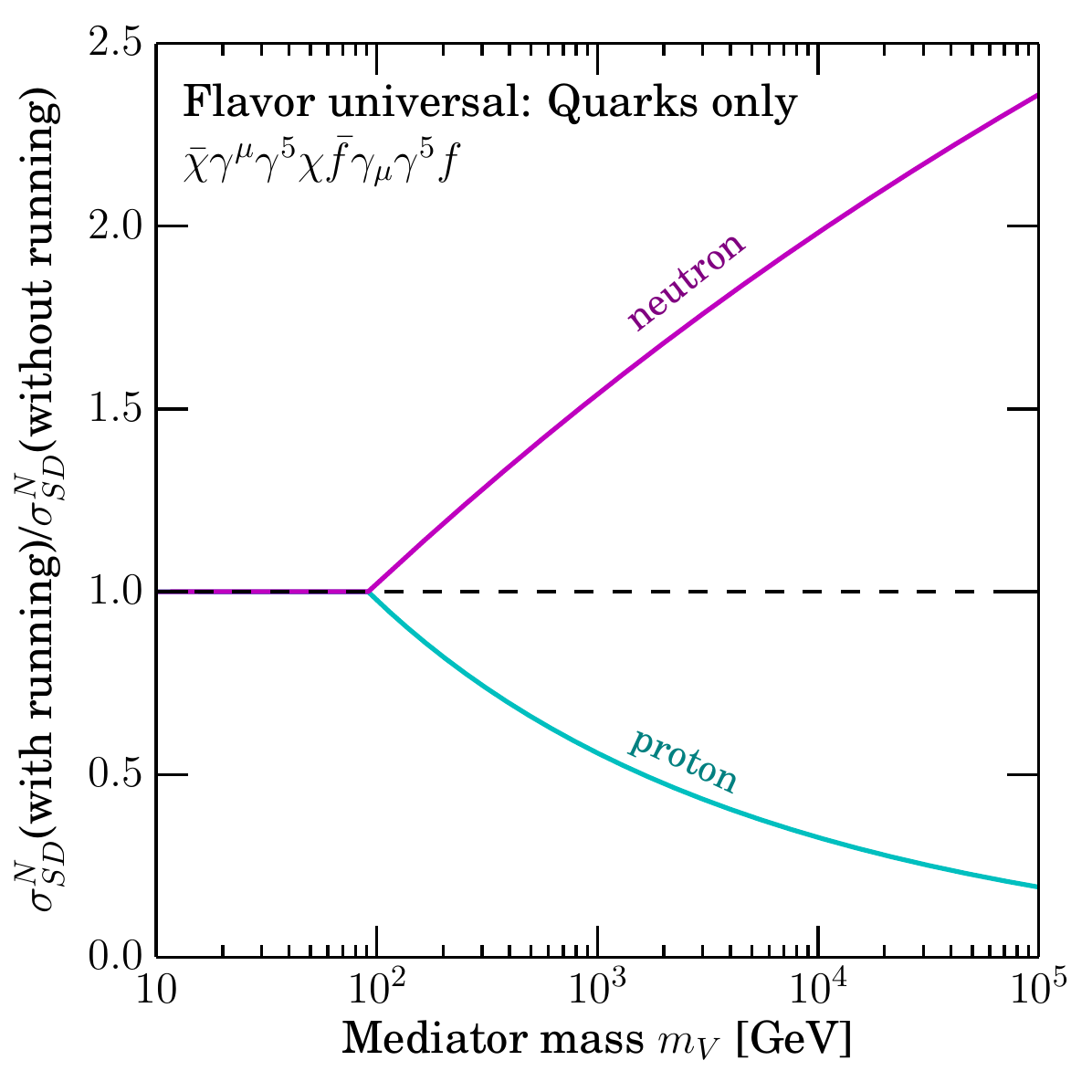} 
\caption{Spin-dependent (SD) DM-nucleon cross section as a function of the mediator mass $m_V$, normalised to the value obtained when the running is not considered. Results for the DM-neutron cross section are shown in magenta, while the DM-proton cross section is shown in cyan. We assume that at high energy DM couples universally to all quarks through the axial-vector interaction $\overline{\chi} \gamma^\mu \gamma^5 \chi \overline{f} \gamma_\mu \gamma^5 f$.}
\label{fig:sigmaSD}
\end{figure}

This is illustrated in Fig.~\ref{fig:sigmaSD}, where we show the proton and neutron SD cross sections as a function of the mediator mass. For each value of the mediator mass, we normalise the cross section relative to its na\"ive value without taking running into account. The origin of the differences in Fig.~\ref{fig:LHClimits} is now clear; the running of the couplings enhances the DM-neutron coupling, leading to weaker limits from the \LHC\ (and vice-versa for DM-proton couplings). 
It should also be clear from this plot that any choice of the ratio between DM-neutron and DM-proton couplings at the nuclear energy scale must necessarily be fine tuned in this framework. The seemingly natural choice of equal SD couplings to protons and neutrons will correspond to a different choice of couplings at the energy scale of the mediator. In other words, running of the couplings will naturally induce isospin-violation for SD couplings~\cite{Crivellin:2014qxa,D'Eramo:2014aba}.


We have focused here on the case of axial-vector couplings, finding that the mapping from \LHC\ limits to the DD plane should be corrected by a factor of roughly 2 when including the effects of running. However, in some cases, the standard lore is that \LHC\ limits on the mediator mass cannot be mapped onto DD limits, as the corresponding DD rate is small.
Interactions of the form $\overline{\chi}\gamma^\mu \gamma^5 \chi \overline{q} \gamma_\mu q$ and $\overline{\chi}\gamma^\mu  \chi \overline{q} \gamma_\mu \gamma^5 q$ lead to velocity- and momentum-suppressed rates in DD experiments and are therefore not expected to be competitive with DD limits. However, as we have shown in Sec.~\ref{sec:BenchmarkI}, mixing effects can induce substantial rates which are comparable to or even greater than the standard SD rate (see Fig.~\ref{fig:universalQ_fermionDM}). In these cases, the inclusion of RG effects is therefore essential for a correct comparison between the \LHC\ and DD results.


\section{Conclusions}
\label{sec:conclusions}

In this work, we have performed a systematic study of the direct detection of SM-singlet DM which interacts with SM fermions through the exchange of a heavy vector mediator. We have considered both scalar and fermion DM with general, gauge-invariant couplings to SM fermions. We have also taken into account all relevant non-relativistic DM-nucleon interactions in order to obtain current and projected limits on the mass scale of the mediator. Most importantly, however, we have accounted for RG effects connecting the couplings defined at high energy and those relevant for low-energy DM-nucleon scattering. 

If the mediator couples universally to all quarks (Sec.~\ref{sec:BenchmarkI}), there are two key effects. The first is to change the size of the couplings at low energy with respect to the values defined at the energy scale of the mediator mass. When the mediator couples to quark vector currents, this effect is driven by electromagnetic loops and is therefore small. In contrast, if the mediator couples to quark axial-vector currents, the effect is dominated by the top Yukawa and is therefore larger. This effect is most pronounced when DM also couples through the axial-vector current, leading to spin-dependent DM-nucleon scattering. 

The second effect is operator mixing, inducing new interactions at low energy that are not present at high energy. Though the size of the mixing may be relatively small, the DD rate can be increased by several orders of magnitude if the new interaction is not suppressed by powers of the velocity or momentum-transfer. For couplings to quarks, the result is that irrespective of the DM spin and interaction structure, DD experiments can always exclude mediators lighter than $m_V \sim 200 \, \, \mathrm{GeV}$, and in some cases this exclusion extends up to $m_V \sim 10 \, \, \mathrm{TeV}$.

In cases where the mediator does not couple to light quarks, DM-nucleon scattering is not possible at tree level. However, SM loops generically induce these couplings, allowing DD experiments to set limits on such models, as demonstrated in Sec.~\ref{sec:BenchmarkII}~and~\ref{sec:BenchmarkIII}. Such limits are typically stronger if the mediator couples to the top quark axial-vector current, due to the large top Yukawa coupling giving large mixing effects. The only model we have considered which is not constrained by current DD experiments is the case in which the mediator couples to the axial-vector currents of both the DM and SM leptons. However, even in this case, future experiments such as \LZ\ should be able to explore such models.

Of course, the benchmarks we have considered are not exhaustive. However, they have allowed us to highlight the key features which appear when the running of operators is correctly accounted for. For more general sets of couplings, Sec.~\ref{sec:RGE} gives the recipe for connecting the high energy Lagrangian with the low-energy DD observables. A key ingredient in this recipe, the low-energy light quark couplings as a function of the mediator mass, is approximately given by the analytical expressions in Appendix~\ref{app:analytical}. We also distribute together with this work the code \runDM, which can be downloaded at \href{https://github.com/bradkav/runDM/}{this http URL}, with which a more rigorous RGE can be performed. Once we have derived the low-energy couplings, exclusion limits for any arbitrary choice of couplings can be obtained by adopting the rescaling procedure described in \Sec{sec:rescaling}. We emphasise once again that these RG effects arise only from SM loops and are therefore not optional. The tools we provide should facilitate the inclusion of these effects in all future studies of vector mediated simplified models.

Finally, we have briefly examined the comparison between \LHC\ and DD limits. We have found that for the case of a mediator with axial-vector interactions, \LHC\ bounds translated into the $(m_\chi, \, \sigma_{\chi N})$ plane can be altered by a factor of 2 when the effects of running are included. For interactions which na\"ively give no limits in DD experiments, we have also demonstrated that operator mixing can have a significant impact, meaning that \LHC\ limits on such interactions can in fact be complementary to DD searches. We urge the \LHC\ experimental collaborations to include these effects when comparing with DD experiments in order to correctly map the complementarity between the different search strategies.

\section*{Acknowledgments} 

The authors thank U. Haisch, W. Shepherd,  F. Sala and J. Silk for discussions and S. Profumo for helpful comments on this manuscript. FD thanks J. Silk for his kind hospitality at the Institut d'Astrophysique de Paris where part of this work was completed. PP thanks S. Profumo for his kind hospitality at UC Santa Cruz, where this work was initiated. FD is supported by the U.S. Department of Energy grant number DE-SC0010107 (FD). P.P. acknowledges support of the European Research Council project 267117 hosted by Universit\'e Pierre et Marie Curie-Paris 6, PI J. Silk. BJK is supported by the John Templeton Foundation Grant 48222 and by the European Research Council ({\sc Erc}) under the EU Seventh Framework Programme (FP7/2007-2013)/{\sc Erc} Starting Grant (agreement n.\ 278234 --- `{\sc NewDark}' project).


\appendix


\section{A simple analytical solution for the RGE}
\label{app:analytical}

The complete procedure to properly connect the energy scales is rather involved, as described in the first part of Sec.~\ref{sec:RGE}. The study of the benchmark models in Sec.~\ref{sec:benchmarks} was performed by implementing the full RGE without any approximation. Our results in Figs.~\ref{fig:universalQ_fermionDM} to \ref{fig:universalThird_scalarDM} can be qualitatively understood with the help of analytical equations expressing the low-energy couplings in terms of the ones at high-energy. These formulae are the results of a fixed-order calculation, which capture the size of these RG effects. Providing these equations is the goal of this appendix. The accuracy of these approximate solutions is quantified in \App{app:code} for the benchmark model of a mediator coupled to quarks, where RG effects are maximal due to the large top Yukawa coupling. A more refined study can be performed by using the \runDM code.

We start by quantifying the approximations made to derive the analytical expressions given in this Appendix. The system of RG equations as given in \Ref{D'Eramo:2014aba}, both above and below the EWSB scale, schematically reads
\be
\frac{d c_i(\mu)}{d \ln \mu} = \gamma_{ij}(\mu) c_j(\mu) \ .
\ee
Here, $\mu$ is the renormalization scale and $c_i$ is an array of Wilson coefficients. The entries of the anomalous dimension matrix $\gamma_{ij}(\mu)$ are numerical coefficients and SM couplings. The renormalization scale dependence in the anomalous dimension matrix can only be implicit through the running SM couplings, since we work in a mass independent scheme ($\overline{\rm MS}$). 

Our first approximation is to ignore the scale dependence of the SM couplings, and we take them evaluated at the renormalization scale $\mu = m_Z$.\footnote{This is the approximation made in Appendix~D of \Ref{D'Eramo:2014aba}. Here, we push this approximation one step further as described in the following.} The RG system becomes
\be
\frac{d c_i(\mu)}{d \ln \mu} = \gamma_{ij}^{(0)} c_j(\mu) \ ,
\label{eq:RGsimplified}
\ee
where $\gamma_{ij}^{(0)} =  \gamma_{ij}(m_Z)$. The linear system in \Eq{eq:RGsimplified} can be solved analytically, since
\be
\exp\left[ - \gamma^{(0)} \, \ln \mu  \right]_{ij} c_j(\mu) = {\rm const} \ .
\label{eq:approx1}
\ee
It is straightforward to check the validity of this result by taking the derivative with respect to $d \ln \mu$ and then using \Eq{eq:RGsimplified}. This relation allows a simple connection between the couplings $c_j(\mu)$ at two different scales. However, it involves the exponential of the anomalous dimension, which is a 16 $\times$ 16 matrix~\cite{D'Eramo:2014aba}, and therefore it cannot be translated into simple analytical equations. We make one further approximation: Taylor expand the exponential
\be
\left[1 - \gamma^{(0)} \, \ln \mu  \right]_{ij} c_j(\mu) = {\rm const} \ .
\label{eq:approx2}
\ee
This is justified as long as we do not have large logarithms in our solution. The largest entry in the anomalous dimension matrix is the one proportional to the square of the top Yukawa coupling, therefore our approximation is valid as long as
\be
3 \frac{\lambda_t^2}{8 \pi^2} \ln(m_V / m_Z) \ll 1 \ .
\ee

In spite of the complicated initial RG system, these two approximations yield an extremely simple final result. Here, we give the low-energy couplings appearing in the effective Lagrangian in \Eq{eq:LagEFTmuN} defined at the nuclear scale $\mu_N$, since they are the only ones relevant to the calculations of direct detection rates. We consider the most general choice of SM couplings to the mediators, namely we keep the $15$ couplings appearing in \Eq{eq:JmuSM} completely general. For each coupling, the final result takes the form
\be
 \mathcal{C}_{V,A}^{(u,s,d)} = \left.  \mathcal{C}_{V,A}^{(u,s,d)} \right|_{\rm tree} + 
 \left.  \mathcal{C}_{V,A}^{(u,s,d)} \right|_{\rm e.m.} + \left.  \mathcal{C}_{V,A}^{(u,s,d)} \right|_{\rm Yukawa}  \ . 
\ee
The tree-level contribution, if present, is the coupling we would have without considering RG effects. The two other terms are one-loop interactions induced by the RGE, and within our approximation their effects are driven by electromagnetic and Yukawa interactions. 

We start from vector currents, with direct detection rates only sensitive to the coupling to the up and down quarks. For the up quark we have the induced coupling at low-energy
\be
\begin{split}
\left. \mathcal{C}_{V}^{(u)} \right|_{\rm tree}  = & \, \frac{c_{q}^{(1)} + c_{u}^{(1)}}{2}  \ , \\
\left. \mathcal{C}_{V}^{(u)} \right|_{\rm e.m.}  = & \, 
- 8 \frac{\alpha_{\rm em}}{9 \pi} \left[ \left(\frac{c_{q}^{(3)} + c_{u}^{(3)}}{2} \right) \ln(m_V/m_Z) + \sum_{i =1}^2 \left(\frac{c_{q}^{(i)} + c_{u}^{(i)}}{2}\right) \ln(m_V/\mu_N) \right] + \\ &
4 \frac{\alpha_{\rm em}}{9 \pi} \sum_{i =1}^{3}  \left[\frac{c_{q}^{(i)} + c_{d}^{(i)}}{2} + \frac{c_{l}^{(i)} + c_{e}^{(i)}}{2}   \right]  \ln(m_V/\mu_N) \ , \\ 
\left. \mathcal{C}_{V}^{(u)} \right|_{\rm Yukawa}  = & \, \frac{\alpha_t}{2 \pi} (3 - 8 s_w^2) \left(\frac{- c_{q}^{(3)} + c_{u}^{(3)}}{2} \right) \ln(m_V/m_Z)  + \\ &
- (3 - 8 s_w^2) \left[ \frac{\alpha_b}{2 \pi}  \left(\frac{- c_{q}^{(3)} + c_{d}^{(3)}}{2} \right) 
+ \frac{\alpha_\tau}{6 \pi} \left(\frac{- c_{l}^{(3)} + c_{e}^{(3)}}{2} \right) \right] \ln(m_V/\mu_N)  \ .
\end{split}
\label{eq:CvupAnalytical}
\ee
The tree-level coupling in the first equation is already present at the mediator scale $m_V$. The first row of the one-loop induced coupling driven by electromagnetic interactions has contributions from the up-type quarks. The top quark effects are only present down to the weak scale, since for lower energies the top is not a propagating degree of freedom. In the second row of the electromagnetic term we have contributions to the RGE by the down-type quarks and leptons. The Yukawa terms are dominated by third generation fermions, and again the RG driven by the top quark only contributes down to the weak scale. Likewise, for the down quark we have
\be
\begin{split}
\left. \mathcal{C}_{V}^{(d)} \right|_{\rm tree}  = & \, \frac{c_{q}^{(1)} + c_{d}^{(1)}}{2}  \ , \\
\left. \mathcal{C}_{V}^{(d)} \right|_{\rm e.m.}  = & \,  4 \frac{\alpha_{\rm em}}{9 \pi} 
\left[\left(\frac{c_{q}^{(3)} + c_{u}^{(3)}}{2} \right) \ln(m_V/m_Z) +  \sum_{i =1}^2 \left(\frac{c_{q}^{(i)} + c_{u}^{(i)}}{2}\right) \ln(m_V/\mu_N)\right] + \\ & - 2 \frac{\alpha_{\rm em}}{9 \pi} \sum_{i =1}^{3}  \left[\frac{c_{q}^{(i)} + c_{d}^{(i)}}{2} + \frac{c_{l}^{(i)} + c_{e}^{(i)}}{2}   \right]  \ln(m_V/\mu_N)  \ , \\ 
\left. \mathcal{C}_{V}^{(d)} \right|_{\rm Yukawa}  = & \,  - \frac{\alpha_t}{2 \pi} (3 - 4 s_w^2) \left(\frac{- c_{q}^{(3)} + c_{u}^{(3)}}{2} \right) \ln(m_V/m_Z) + \\ & 
(3 - 4 s_w^2) \left[ \frac{\alpha_b}{2 \pi}  \left(\frac{- c_{q}^{(3)} + c_{d}^{(3)}}{2} \right) 
+ \frac{\alpha_\tau}{6 \pi} \left(\frac{- c_{l}^{(3)} + c_{e}^{(3)}}{2} \right) \right] \ln(m_V/\mu_N) \ .
\end{split}
\label{eq:CvdownAnalytical}
\ee

We now consider the induced axial-vector currents of light quarks. The solutions are even simpler for these cases, since the electromagnetic effects are absent 
\be
\left. \mathcal{C}_{A}^{(u,d,s)} \right|_{\rm e.m.}   = 0 \ .
\label{eq:CAemAnalytical}
\ee
For the up quark we have the following tree-level and Yukawa terms
\be
\begin{split}
\left. \mathcal{C}_{A}^{(u)} \right|_{\rm tree}  = & \, \frac{- c_{q}^{(1)} + c_{u}^{(1)}}{2}  \ , \\
\left. \mathcal{C}_{A}^{(u)} \right|_{\rm Yukawa}  = & \, - 3 \frac{\alpha_t}{2 \pi} \left(\frac{- c_{q}^{(3)} + c_{u}^{(3)}}{2}\right) \ln(m_V/m_Z) + \\ &  
\left[3 \frac{\alpha_b}{2 \pi}  \left(\frac{- c_{q}^{(3)} + c_{d}^{(3)}}{2} \right) +
\frac{\alpha_\tau}{2 \pi}  \left(\frac{- c_{l}^{(3)} + c_{e}^{(3)}}{2} \right) \right] \ln(m_V/\mu_N)  \ .
\end{split}
\label{eq:CaupAnalytical}
\ee
Likewise, for the down and strange quarks we have
\be
\begin{split}
\left. \mathcal{C}_{A}^{(d)} \right|_{\rm tree}  = & \, \frac{- c_{q}^{(1)} + c_{d}^{(1)}}{2}  \ , \\
\left. \mathcal{C}_{A}^{(s)} \right|_{\rm tree}  = & \, \frac{- c_{q}^{(2)} + c_{d}^{(2)}}{2}  \ , \\
\left. \mathcal{C}_{A}^{(d)} \right|_{\rm Yukawa} = & \, \left. \mathcal{C}_{A}^{(s)} \right|_{\rm Yukawa} = 
- \left. \mathcal{C}_{A}^{(u)} \right|_{\rm Yukawa}  \ .
\end{split}
\label{eq:CadownAnalytical}
\ee
The RG induced couplings driven by Yukawa interactions are the same for the down and the strange quarks, and they are the opposite of the one for the up quark.

A few comments are in order. The relations in Eqs.~(\ref{eq:CvupAnalytical})-(\ref{eq:CadownAnalytical}) are approximate, since the running of the SM couplings is neglected. The error made by using these solutions is quantified in \App{app:code}. Furthermore, the above equations are only valid for $m_V > m_Z$. For mediator mass below the weak scale, we cannot consider the coupling to the top quark, since it is not consistent to integrate-out the mediator and keep the top quark in the spectrum. The above equations can still be used for $m_V < m_Z$, but without the contributions arising from loops of the top quark. In this case, the effective couplings defined at the scale $m_V$ are understood as arising from integrating-out the mediator $V$ and the heavy SM degrees of freedom. 

We conclude this appendix by rewriting the analytical solutions in terms of vector and axial-vector currents of SM fermions as given in \Eq{eq:JprimemuSM}. We emphasize again that this choice is not gauge invariant, and the couplings have to respect the conditions listed in \Eq{eq:VAfromGaugeInv}. Nevertheless, it is still useful to rewrite the solutions within this language, since they take a particularly simple form. The low-energy vector current of up quark reads
\be
\begin{split}
\left. \mathcal{C}_{V}^{(u)} \right|_{\rm tree}  = & \, c_{Vu}^{(1)}   \ , \\
\left. \mathcal{C}_{V}^{(u)} \right|_{\rm e.m.}  = & \, 
- 8 \frac{\alpha_{\rm em}}{9 \pi} \left[ c_{Vu}^{(3)} \, \ln(m_V/m_Z) + \left(\sum_{i =1}^{2} c_{Vu}^{(i)} - \frac{1}{2} \sum_{i =1}^{3}  \left(c_{Vd}^{(i)} + c_{Ve}^{(i)}\right)  \right)  \ln(m_V/\mu_N) \right]    \ , \\ 
\left. \mathcal{C}_{V}^{(u)} \right|_{\rm Yukawa}  = & \, \frac{\alpha_t}{2 \pi} (3 - 8 s_w^2) c_{Au}^{(3)} \ln(m_V/m_Z) - (3 - 8 s_w^2) \left[ \frac{\alpha_b}{2 \pi}   c_{Ad}^{(3)} + \frac{\alpha_\tau}{6 \pi}  c_{Ae}^{(3)} \right] \ln(m_V/\mu_N)  \ .
\end{split}
\label{eq:CvupAnalytical2}
\ee
The one for the down quarks reads
\be
\begin{split}
\left. \mathcal{C}_{V}^{(d)} \right|_{\rm tree}  = & \, c_{Vd}^{(1)}  \ , \\
\left. \mathcal{C}_{V}^{(d)} \right|_{\rm e.m.}  = & \,  4 \frac{\alpha_{\rm em}}{9 \pi}\left[ c_{Vu}^{(3)} \, \ln(m_V/m_Z) + \left(\sum_{i =1}^{2} c_{Vu}^{(i)} - \frac{1}{2} \sum_{i =1}^{3}  \left(c_{Vd}^{(i)} + c_{Ve}^{(i)}\right)  \right)  \ln(m_V/\mu_N) \right]  \ , \\ 
\left. \mathcal{C}_{V}^{(d)} \right|_{\rm Yukawa}  = & \,  - \frac{\alpha_t}{2 \pi} (3 - 4 s_w^2) c_{Au}^{(3)}  \ln(m_V/m_Z)  +
 (3 - 4 s_w^2) \left[ \frac{\alpha_b}{2 \pi} c_{Ad}^{(3)} + \frac{\alpha_\tau}{6 \pi} c_{Ae}^{(3)} \right] \ln(m_V/\mu_N)\ .
\end{split}
\label{eq:CvdownAnalytical2}
\ee
The expression for the low-energy axial-vector currents are even simpler
\be
\begin{split}
\left. \mathcal{C}_{A}^{(u)} \right|_{\rm tree}  = & \, c_{Au}^{(1)}  \ , \\
\left. \mathcal{C}_{A}^{(u)} \right|_{\rm Yukawa}  = & \,- 3 \frac{\alpha_t}{2 \pi} c_{Au}^{(3)} \ln(m_V/m_Z) + \left[3 \frac{\alpha_b}{2 \pi}  c_{Ad}^{(3)} + \frac{\alpha_\tau}{2 \pi}  c_{Ae}^{(3)} \right] \ln(m_V/\mu_N)  \ ,
\end{split}
\label{eq:CaupAnalytical2}
\ee
and
\be
\begin{split}
\left. \mathcal{C}_{A}^{(d)} \right|_{\rm tree}  = & \, c_{Ad}^{(1)}   \ , \\
\left. \mathcal{C}_{A}^{(s)} \right|_{\rm tree}  = & \, c_{Ad}^{(2)}   \ , \\
\left. \mathcal{C}_{A}^{(d)} \right|_{\rm Yukawa} = & \, \left. \mathcal{C}_{A}^{(s)} \right|_{\rm Yukawa} = 
- \left. \mathcal{C}_{A}^{(u)} \right|_{\rm Yukawa}  \ .
\end{split}
\label{eq:CadownAnalytical2}
\ee
for up and down quarks, respectively. 

The usefulness of the solutions as in Eqs.~(\ref{eq:CvupAnalytical2})-(\ref{eq:CadownAnalytical2}), other than being manifest expressions for the low-energy couplings in the benchmark discussed in Sec.~\ref{sec:benchmarks}, consists in providing us with a simple {\it vademecum} to quantify RG effects. If the mediator couples to vector currents of SM fermions, RG effects only generates vector currents to light quarks at low-energy. If this is the case, these effects are relevant only if the mediator does not couple to light quarks at tree-level (e.g. leptophilic, heavy quarks), otherwise we only have $\mathcal{O}(1 \%)$ corrections to the couplings, since the effect is driven electromagnetic interactions. On the contrary, if the mediator couples to axial-vector currents of SM fermions, RG effects generate both vector and axial-vector currents with light quarks at low-energy. In this case the effects is driven by Yukawa couplings, therefore it is dominated by loops of heavy SM fermions. 


\section{\runDM: a code for the RGE}
\label{app:code}

We release together with this paper the public code \runDM. This code can RG evolve the Wilson coefficients of the simplified models discussed in this work between two arbitrary energy scales. As a specific application, \runDM can evolve the couplings from the mediator mass scale down to the nuclear scale and provide us with the low-energy couplings relevant to direct detection rates. The code is available at \href{https://github.com/bradkav/runDM/}{this http URL} together with its documentation. In this appendix we describe how the RGE is implemented in \runDM. We conclude by comparing the output of the code with the full numerical solution of the RG system and the analytical solutions provided in \App{app:analytical}. We perform the comparison in the first benchmark model discussed in this work, where RG effects are maximal due to the large top Yukawa coupling. We find that \runDM reproduces the full results with extreme accuracy, and we quantify the error made by using the analytical expressions.

Following the framework developed in \Ref{D'Eramo:2014aba}, the RGE is divided in two regimes, above and below the EWSB scale. At energy scale above the $Z$ boson mass, the effective interactions between DM and SM are described by the Lagrangian
\be
\begin{split}
\mathcal{L}^{(m_Z < \mu < m_V)}_{\rm EFT} =  &\, - \frac{1}{m_V^2} \, J^\mu_{{\rm DM}} \,
 \sum_{i = 1}^3 \left( c_{q}^{(i)} \; \overline{q_L^i} \gamma^\mu q_L^i + c_{u}^{(i)} \;  \overline{u_R^i} \gamma^\mu u_R^i +  c_{d}^{(i)} \;  \overline{d_R^i} \gamma^\mu d_R^i  \right) + \\ & 
  - \frac{1}{m_V^2} \, J^\mu_{{\rm DM}} \,\sum_{i=1}^3 \left( c_{l}^{(i)} \; \overline{l_L^i} \gamma^\mu l_L^i + c_{e}^{(i)} \; \overline{e_R^i} \gamma^\mu e_R^i \right) - \frac{1}{m_V^2}  \, J^\mu_{{\rm DM}} \, c_H H^\dagger i \overleftrightarrow D_\mu H \ .
\end{split}
\label{eq:LagEFTaboveAPP}
\ee
Although we have always assumed $c_H(m_V) = 0$ throughout this work (i.e.~that the mediator is not coupled to the Higgs), \runDM can deal with this more general case as well. We organize the Wilson coefficients in the 16-dimensional array
\be
\scriptsize   c^T(\mu) = \left(\begin{array}{cccccccccccccccc}
c_{q}^{(1)} & c_{u}^{(1)} & c_{d}^{(1)} & c_{l}^{(1)} & c_{e}^{(1)} & 
c_{q}^{(2)} & c_{u}^{(2)} & c_{d}^{(2)} & c_{l}^{(2)} & c_{e}^{(2)} &
c_{q}^{(3)} & c_{u}^{(3)} & c_{d}^{(3)} & c_{l}^{(3)} & c_{e}^{(3)} &
c_{H} \end{array}\right) \ ,
\label{ed:cdef}
\ee
and the connection between the couplings at two different energy scales (both above the Z boson mass) can be found by solving the system of RG equations
\be
\frac{d \, c}{d \ln \mu}  = \gamma_{{\rm SM}_\chi} c \ .
\label{RGsystem1}
\ee
The explicit form of $\gamma_{{\rm SM}_\chi}$ can be found in \Ref{D'Eramo:2014aba}. Likewise, at energies below the 
EWSB scale we have the effective couplings
\be
\begin{split}
\mathcal{L}^{(\mu_N < \mu < m_Z)}_{\rm EFT} = & \, 
- \frac{1}{m_V^2} \, J^\mu_{{\rm DM}} \, \sum_{i = 1}^2  \mathcal{C}_{Vu}^{(i)} \; \overline{u^i} \gamma^\mu u^i 
- \frac{1}{m_V^2} \, J^\mu_{{\rm DM}} \, \sum_{i = 1}^3 \left[ \mathcal{C}_{Vd}^{(i)} \; \overline{d^i} \gamma^\mu d^i + \mathcal{C}_{Ve}^{(i)} \; \overline{e^i} \gamma^\mu e^i \right] + \\ & 
- \frac{1}{m_V^2} \, J^\mu_{{\rm DM}} \sum_{i = 1}^2 \mathcal{C}_{Au}^{(i)} \; \overline{u^i} \gamma^\mu \gamma^5 u^i - \frac{1}{m_V^2} \, J^\mu_{{\rm DM}} \sum_{i = 1}^3 \left[ \mathcal{C}_{Ad}^{(i)} \; \overline{d^i} \gamma^\mu \gamma^5 d^i + \mathcal{C}_{Ae}^{(i)} \; \overline{e^i} \gamma^\mu \gamma^5 e^i  \right] \ .
\label{eq:LagEFT5}
\end{split}
\ee
In the above equation there is no contribution from the top quark couplings, since at such low energy scales the top is integrated out. The Wilson coefficients in the above Lagrangian can be arranged in the array
\be
\scriptsize  \mathcal{C}^T(\mu) = \left(\begin{array}{cccccccccccccccc}
\mathcal{C}_{Vu}^{(1)} & \mathcal{C}_{Vd}^{(1)} & \mathcal{C}_{Vu}^{(2)} & \mathcal{C}_{Vd}^{(2)} & \mathcal{C}_{Vd}^{(3)} & \mathcal{C}_{Ve}^{(1)} & \mathcal{C}_{Ve}^{(2)}  & \mathcal{C}_{Ve}^{(3)}  & \mathcal{C}_{Au}^{(1)} & \mathcal{C}_{Ad}^{(1)} & \mathcal{C}_{Au}^{(2)} & \mathcal{C}_{Ad}^{(2)} & \mathcal{C}_{Ad}^{(3)} & 
\mathcal{C}_{Ae}^{(1)} & \mathcal{C}_{Ae}^{(2)}  &\mathcal{C}_{Ae}^{(3)} \end{array}\right) \ .
\label{ed:cdef2}
\ee
The connection between the couplings at two arbitrary energy scales below the Z boson mass is achieved by solving the RG system
\be
\frac{d \, \mathcal{C}}{d \ln \mu}  = \gamma_{{\rm EMSM}_\chi} \mathcal{C} \ ,
\label{RGsystem2}
\ee
where the explicit form of $\gamma_{{\rm EMSM}_\chi}$ can also be found in \Ref{D'Eramo:2014aba}. If we want to connect two energy scales above and below the $Z$ boson mass, there is a further intermediate step: a matching between the two EFTs at the EWSB scale. This is obtained by
\be
\mathcal{C}(m_Z) = U_{\rm match} \, c(m_Z)\ , 
\ee
where the explicit form for $U_{\rm match}$ is given in \Ref{D'Eramo:2014aba}.

The RG systems in \Eqs{RGsystem1}{RGsystem2} allow for a connection between two arbitrary energy scales in each EFT. We formally define the evolution matrix $U_{{\rm SM}_\chi}(\mu_a, \mu_b)$ above the EWSB scale, which evolves the array of Wilson coefficients in \Eq{ed:cdef} from the energy scale $\mu_b$ to the energy scale $\mu_a$. More explicitly, the evolution matrix acts as follows
\be
c(\mu_a) = U_{{\rm SM}_\chi}(\mu_a, \mu_b) \, c(\mu_b) \ .
\ee
We define the analogous quantity $U_{{\rm EMSM}_\chi}(\mu_a, \mu_b)$ for the RG evolution below the EWSB scale. The general evolution matrix between two energy scales thus reads
\be
U(\mu_a, \mu_b) = \left\{ \begin{array}{lccccc}
U_{{\rm SM}_\chi}(\mu_a, \mu_b) & & & & & \mu_a, \mu_b > m_Z \\
U_{{\rm EMSM}_\chi}(\mu_a, m_Z) \; U_{\rm match} \; U_{{\rm SM}_\chi}(m_Z, \mu_b) & & & & & \mu_a < m_Z < \mu_b \\
U_{{\rm EMSM}_\chi}(\mu_a, \mu_b) \; U_{\rm match}  & & & & & \mu_a, \mu_b < m_Z
\end{array} \right.  \ .
\label{eq:Uevol}
\ee
In the above equation, we assume the boundary condition at $\mu_b$ to be expressed in terms of the array in \Eq{ed:cdef}, and the final output at $\mu_a$ in terms of the array in \Eq{ed:cdef} (\Eq{ed:cdef2}) for $\mu_a$ greater (smaller) than the $Z$ boson mass. This is the reason why we include the matrix $U_{\rm match}$ in the last row of the above equation. As a specific application, we can run from the mediator mass scale down to the nuclear scale $\mu_N$, and we have the evolution matrix
\be
U_{\rm DD}(m_V) = \left\{ \begin{array}{lccccc}
U_{{\rm EMSM}_\chi}(\mu_N, m_Z) \; U_{\rm match} \; U_{{\rm SM}_\chi}(m_Z, m_V) & & & & & m_V > m_Z \\
U_{{\rm EMSM}_\chi}(\mu_N, m_V) \; U_{\rm match}  & & & & & m_V \leq m_Z
\end{array} \right.  \ .
\label{eq:Uevol}
\ee

All we have left to do is to evaluate the evolution matrices. We follow the prescription given in Appendix~C of \Ref{D'Eramo:2014aba}, where the running of SM couplings was treated as a perturbation. The set-up is formally identical to time-dependent perturbation theory in quantum mechanics, where the running of SM couplings acts as the potential interaction. The system can be solved by defining the equivalent of the ``interaction-picture'' variables, and expressing the solution of the resulting ``Schwinger-Tomonaga'' equation in terms of a Dyson series. We produce evolution matrices accounting for the running SM couplings up to the 1st order in the Dyson series. These are the evolution matrices used by \runDM.

We conclude this Appendix with a comparison among different methods to solve the system of RG equations. The purpose of this test is twofold: validate the evolution matrices used by \runDM and quantify the accuracy of the analytical solutions given in \App{app:analytical}. We perform the comparison in the benchmark model where the mediator is coupled to quarks only, and we take flavor universal couplings. This is the case where RG effects are expected to be maximal, as a consequence of the large top Yukawa coupling. We consider both vector and axial-vector couplings, and we solve the RG system with 3 different methods: the full solution by solving the differential equations of the RG system, the analytical solutions in \App{app:analytical} and the solution to the ``Schwinger-Tomonaga'' equation up to the 1st order in the Dyson series (as used in the \textsc{runDM} code).

\begin{figure}
\begin{center}
\includegraphics[width=0.5\textwidth]{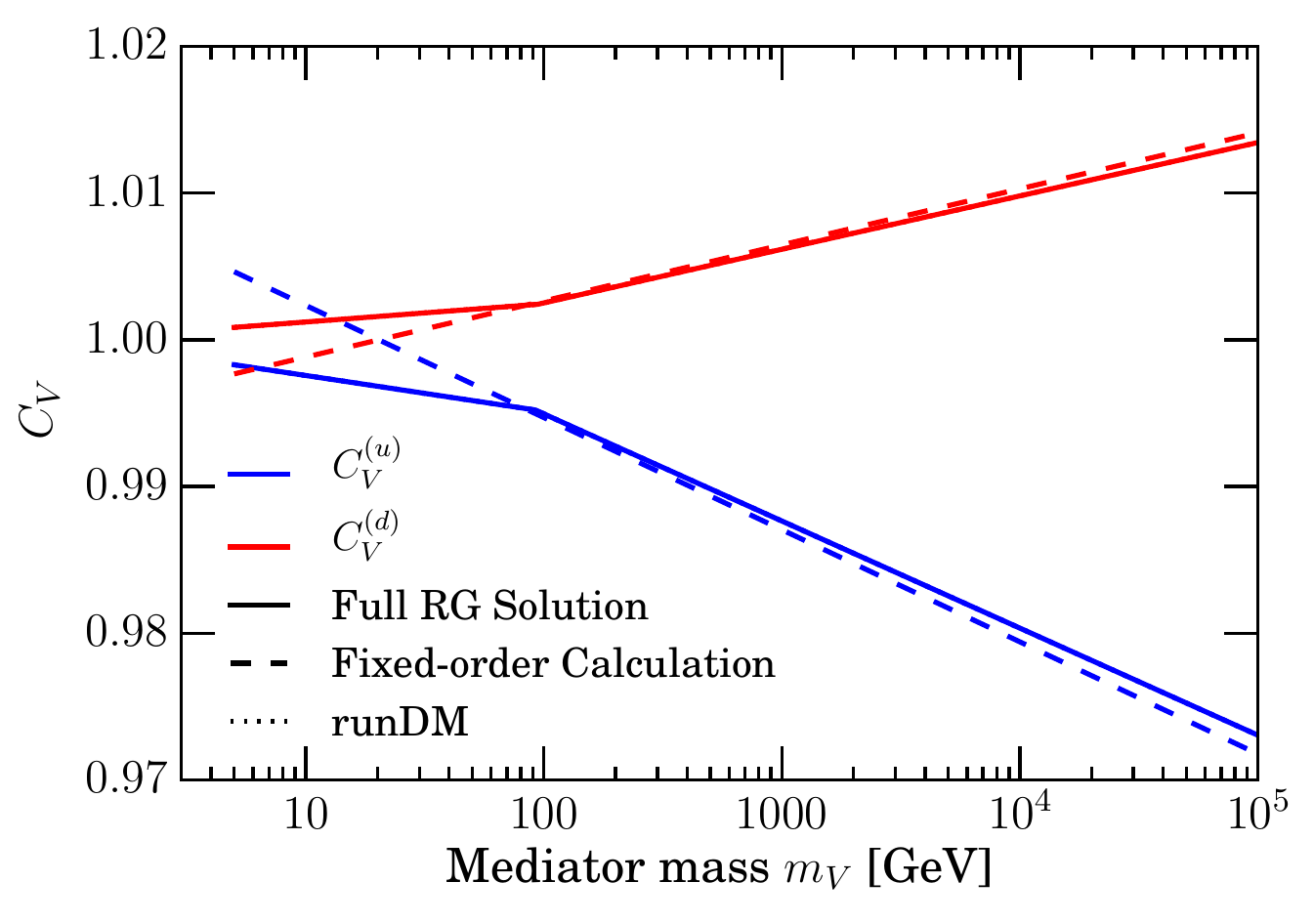} 
\end{center}
\caption{RG solutions for the benchmark of flavor universal quark vector currents.}
\label{fig:comparisonvector}
\end{figure}

We start from couplings to quark vector currents as given in \Eq{eq:bench1a}. In this case there are no induced axial-vector currents at low-energy. We focus on the induced vector currents, which have both a tree-level part as well as a one-loop RG contribution driven by electromagnetic interactions. We show results in \Fig{fig:comparisonvector}. The analytical solution is obtained by fixing the couplings at the $Z$ boson mass, thus it does not work too well for mediator masses away from the $Z$ pole. However, the \textsc{runDM} solution captures these effects, and the solutions in the plot cannot be distinguished from the full RG solution.

\begin{figure}
\begin{center}
\includegraphics[width=0.465\textwidth]{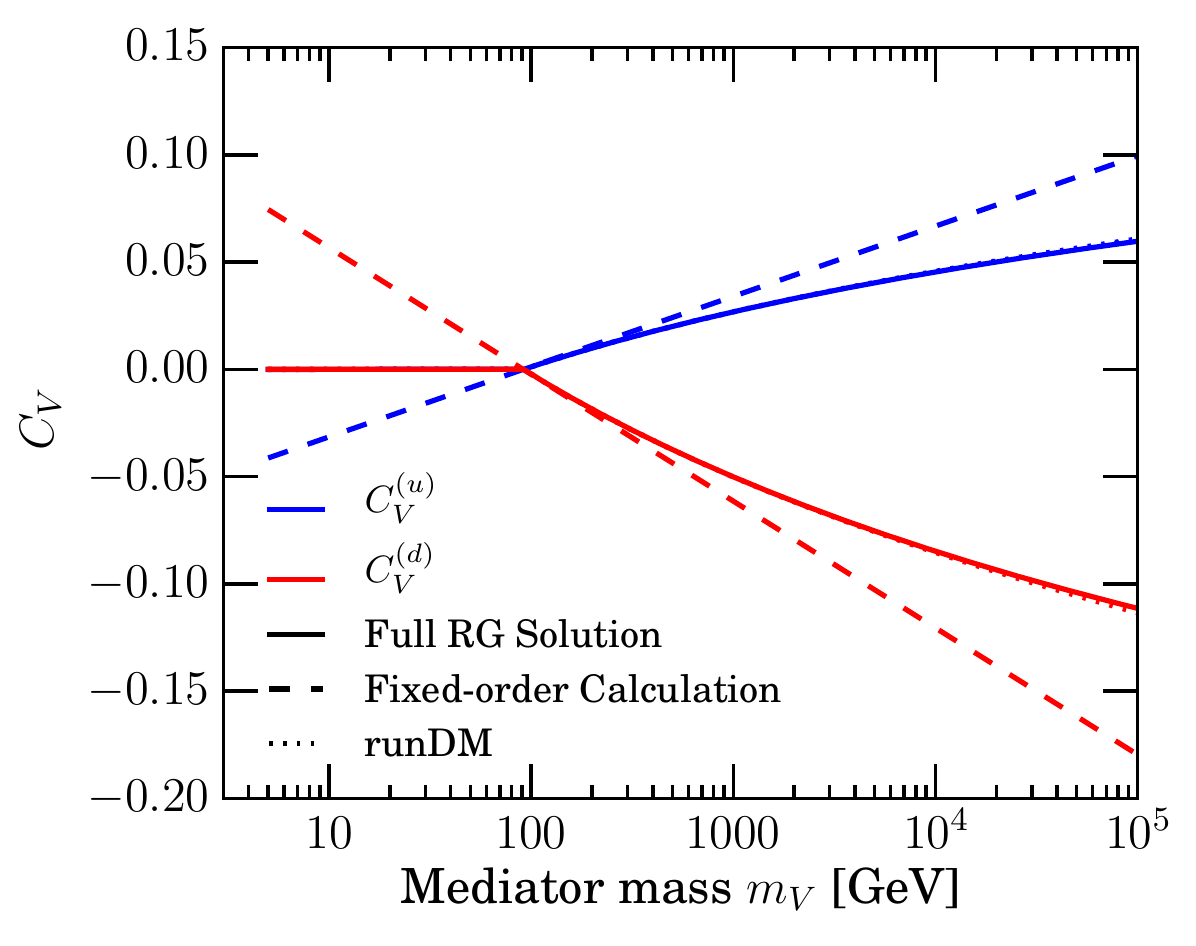} $\qquad$
\includegraphics[width=0.465\textwidth]{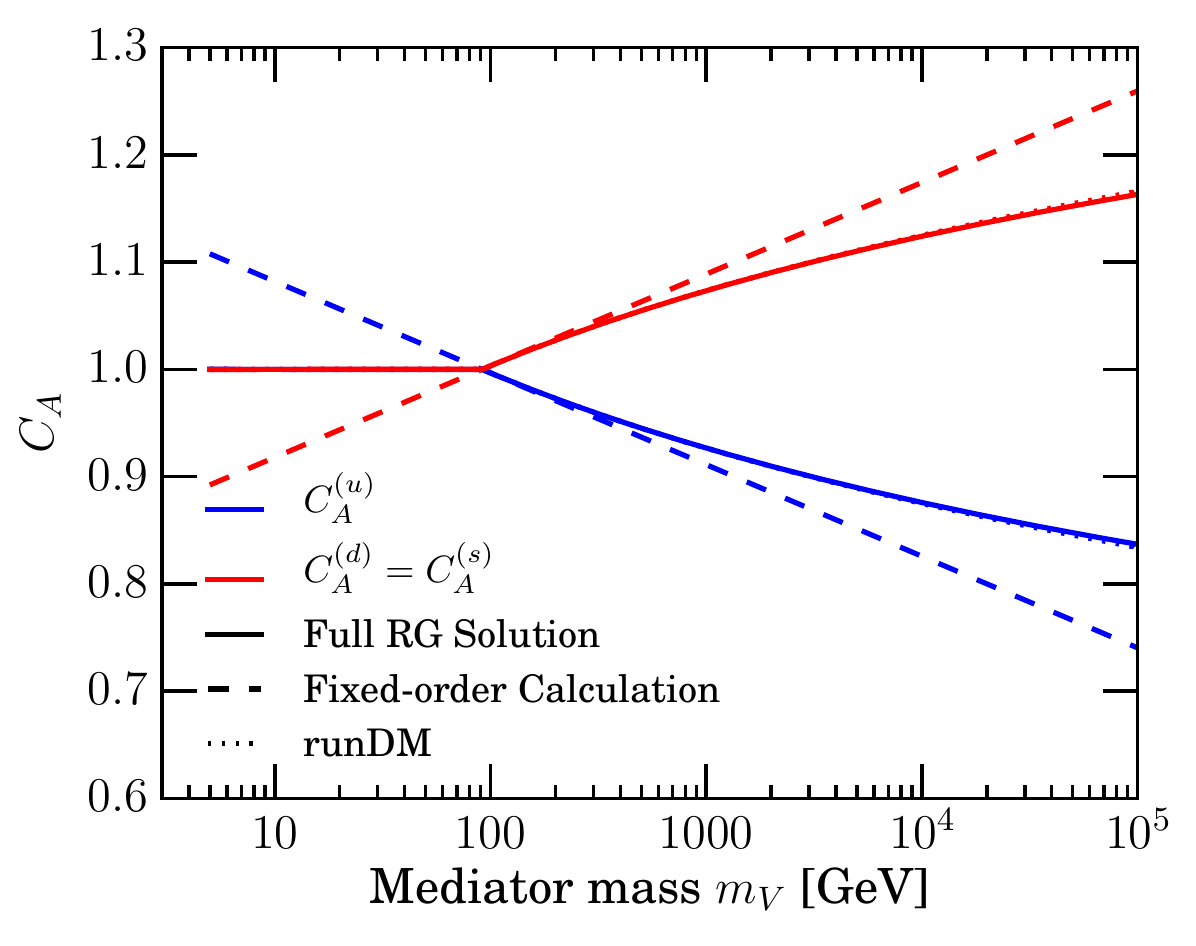}
\end{center}
\caption{RG solutions for the benchmark of flavor universal quark axial-vector currents.}
\label{fig:comparisonaxial}
\end{figure}

We also consider the case of quark axial-vector currents as in \Eq{eq:bench1a}, with results shown in \Fig{fig:comparisonaxial} for the low-energy induced vector (left) and axial-vector (right) currents. Not surprisingly, also in this case the analytical solution works in the vicinity of the weak scale, since we choose to fix the couplings around that scale. However, this time the disagreement is large also for heavy mediator masses, since the effect is drive by the top Yukawa and the running is substantial. Remarkably, the \textsc{runDM} solution still captures these effects.

\section{Details of the matching at the nuclear  scale}
\label{app:NRoperators}
In this appendix we explicitly show, in full generality, the steps to connect the Lagrangian in Eq.~\eqref{eq:LagEFTnucleon} with the main observables in DD searches. In particular, the most important outputs we provide are the generic expressions for the NR coefficients and operators that fully encapsulate all the information coming from the nature of the DM-nucleus interactions and the NR nuclear physics.   

The starting point is the nucleon-level Lagrangian in Eq.~\eqref{eq:LagEFTnucleon} that can be cast as
\be
\mathcal{L}^{(\mu_N)}_{\phi(\chi) \, N} =  \sum_{k} \zeta_{\phi(\chi) \, k}^{(N)} \mathcal{O}_{\phi(\chi) \, k}^{(N)}\, \ ,
\label{eq:LagEFTnucleoncompact}
\ee
where the couplings $\zeta_{\phi(\chi) \, k}^{(N)}$ and dimension six effective operators $\mathcal{O}_{\phi(\chi) \, k}^{(N)}$  are 
\be\label{eq:LaglevelOp&coeffScalar}
\begin{split}
&\zeta_{\phi \, 3}^{(N)}= - c_\phi  \mathcal{C}_{V}^{(N)} /m_V^2 \quad\qquad \mathcal O_{\phi \, 3}^{(N)} =  \phi^\dag \, \overleftrightarrow{\partial}_\mu \phi \, \overline{N} \gamma^\mu N \\
&\zeta_{\phi \, 4}^{(N)}= - c_\phi  \mathcal{C}_{A}^{(N)} /m_V^2 \quad\qquad \mathcal O_{\phi \, 4}^{(N)} =  \phi^\dag \, \overleftrightarrow{\partial}_\mu \phi \, \overline{N} \gamma^\mu \gamma^5 N
\end{split}  \qquad\qquad \text{scalar DM} \ , 
\ee
\be\label{eq:LaglevelOp&coeffMajorana}
\begin{split}
&\zeta_{\chi \, 6}^{(N)}= - c_{\chi A}   \mathcal{C}_{V}^{(N)} /(2m_V^2) \quad\qquad \mathcal O_{\chi \, 6}^{(N)} =  \overline{\chi} \gamma^\mu \gamma^5 \chi \, \overline{N} \gamma^\mu N \\
&\zeta_{\chi \, 8}^{(N)}= - c_{\chi A}   \mathcal{C}_{A}^{(N)} /(2m_V^2) \quad\qquad \mathcal O_{\chi \, 8}^{(N)} =  \overline{\chi} \gamma^\mu \gamma^5 \chi \, \overline{N} \gamma^\mu \gamma^5 N
\end{split}  \qquad\qquad \text{Majorana DM} \ , 
\ee
\be\label{eq:LaglevelOp&coeffDirac}
\begin{split}
&\zeta_{\chi \, 5}^{(N)}= - c_{\chi V}   \mathcal{C}_{V}^{(N)} /m_V^2 \quad\qquad \mathcal O_{\chi \, 5}^{(N)} =  \overline{\chi} \gamma^\mu \chi \, \overline{N} \gamma^\mu N \\
&\zeta_{\chi \, 6}^{(N)}= - c_{\chi A}   \mathcal{C}_{V}^{(N)} /m_V^2 \quad\qquad \mathcal O_{\chi \, 6}^{(N)} =  \overline{\chi} \gamma^\mu \gamma^5 \chi \, \overline{N} \gamma^\mu N \\
&\zeta_{\chi \, 7}^{(N)}= - c_{\chi V}   \mathcal{C}_{A}^{(N)} /m_V^2 \quad\qquad \mathcal O_{\chi \, 7}^{(N)} =  \overline{\chi} \gamma^\mu \chi \, \overline{N} \gamma^\mu \gamma^5 N \\
&\zeta_{\chi \, 8}^{(N)}= - c_{\chi A}   \mathcal{C}_{A}^{(N)} /m_V^2 \quad\qquad \mathcal O_{\chi \, 8}^{(N)} =  \overline{\chi} \gamma^\mu \gamma^5 \chi \, \overline{N} \gamma^\mu \gamma^5 N
\end{split}  \qquad\qquad \text{Dirac DM} \ .
\ee
We follow the same numbering as in Ref.~\cite{DelNobile:2013sia} and we introduce the subscript $\phi(\chi)$ as we explicitly write the DM currents in Eq.~\eqref{eq:JmuDM} which are different for  scalar and fermionic fields.

\smallskip
The DM-nucleon ME expressed in terms of NR operators is obtained by expanding the solution of the Dirac equation in the NR limit. It reads
\be
\Mel_{\phi(\chi) \, N}  \equiv \,  _{\rm out}\langle \phi(\chi) N |\mathcal{L}^{(\mu_N)}_{\phi(\chi) \, N } | \phi(\chi) N \rangle_{\rm in} \simeq  \sum_i \mathfrak{c}_{\phi(\chi) \, i}^{(N)} \mathcal O_i^{\rm NR} \ ,
\label{eq:MEL}
\ee
where now the sum runs  over the relevant NR operators.  Recalling that $\langle \chi | J_\mu^{\rm DM} | \chi \rangle$ for a Majorana field is two times that of either a Dirac or complex scalar one, the NR effective field theory coefficients $ \mathfrak{c}_{\phi(\chi) \, i}^{(N)}$ and operators $\mathcal O_i^{\rm NR}$ in Eq.~\eqref{eq:MEL} are 
\be\label{eq:MEOp&coeffScalar}
\begin{split}
&\langle  \zeta_{\phi \, 3}^{(N)} \mathcal O_{\phi \, 3}^{(N)}  \rangle \simeq  \mathfrak{c}_{\phi \, 1}^{(N)} \mathcal O_1^{\rm NR}  \, , \qquad 
\left\{\begin{array}{l}
\mathfrak{c}_{\phi \, 1}^{(N)} = - 4 m_\phi \, m_N \, c_\phi   \mathcal{C}_{V}^{(N)} /m_V^2 \\
\mathcal O_1^{\rm NR} =  \unop
\end{array}\right. \\
&\langle  \zeta_{\phi \, 4}^{(N)} \mathcal O_{\phi \, 4}^{(N)}  \rangle \simeq  \mathfrak{c}_{\phi \, 7}^{(N)} \mathcal O_7^{\rm NR} \, ,  \qquad 
\left\{\begin{array}{l}
\mathfrak{c}_{\phi \, 7}^{(N)} = 8 m_\phi \, m_N \, c_\phi   \mathcal{C}_{A}^{(N)} /m_V^2 \\
\mathcal O_7^{\rm NR} =   \vec{s}_N \cdot \vec{v}^\perp
\end{array}\right.
\end{split}  \qquad \text{scalar DM} \ , 
\ee
\be\label{eq:MEOp&coeffMajorana}
\begin{split}
&\langle  \zeta_{\chi \, 6}^{(N)} \mathcal O_{\chi \, 6}^{(N)}  \rangle \simeq  \mathfrak{c}_{\chi \, 8}^{(N)} \mathcal O_8^{\rm NR} +  \mathfrak{c}_{\chi \, 9}^{(N)} \mathcal O_9^{\rm NR}  \, , \qquad 
\left\{\begin{array}{l}
\mathfrak{c}_{\chi \, 8}^{(N)} = - 8 m_\chi \, m_N \, c_{\chi A}   \mathcal{C}_{V}^{(N)} /m_V^2 \\
\mathfrak{c}_{\chi \, 9}^{(N)} = - 8 m_\chi  \, c_{\chi A}   \mathcal{C}_{V}^{(N)} /m_V^2 \\
\mathcal O_8^{\rm NR} =  \vec{s}_\chi \cdot \vec{v}^\perp \\
\mathcal O_9^{\rm NR} = \vec{s}_\chi \cdot (\vec{s}_N \times \vec{q}_{\rm R}) 
\end{array}\right. \\
&\langle  \zeta_{\chi \, 8}^{(N)} \mathcal O_{\chi \, 8}^{(N)}  \rangle \simeq  \mathfrak{c}_{\chi \, 4}^{(N)} \mathcal O_4^{\rm NR} \, ,  \qquad 
\left\{\begin{array}{l}
\mathfrak{c}_{\chi \, 4}^{(N)} = 16 m_\chi \, m_N \, c_{\chi A}   \mathcal{C}_{A}^{(N)} /m_V^2 \\
\mathcal O_4^{\rm NR} = \vec{s}_\chi \cdot \vec{s}_N
\end{array}\right.
\end{split}  \qquad \text{Majorana DM} \ , 
\ee
\be\label{eq:MEOp&coeffDirac}
\begin{split}
&\langle  \zeta_{\chi \, 5}^{(N)} \mathcal O_{\chi \, 5}^{(N)}  \rangle \simeq  \mathfrak{c}_{\chi \, 1}^{(N)} \mathcal O_1^{\rm NR} \, ,  \qquad 
\left\{\begin{array}{l}
\mathfrak{c}_{\chi \, 1}^{(N)} = -4 m_\chi \, m_N \, c_{\chi V}   \mathcal{C}_{V}^{(N)} /m_V^2 \\
\mathcal O_1^{\rm NR} = \unop
\end{array}\right. \\
&\langle  \zeta_{\chi \, 6}^{(N)} \mathcal O_{\chi \, 6}^{(N)}  \rangle \simeq  \mathfrak{c}_{\chi \, 8}^{(N)} \mathcal O_8^{\rm NR} +  \mathfrak{c}_{\chi \, 9}^{(N)} \mathcal O_9^{\rm NR}  \, , \qquad 
\left\{\begin{array}{l}
\mathfrak{c}_{\chi \, 8}^{(N)} = - 8 m_\chi \, m_N \, c_{\chi A}   \mathcal{C}_{V}^{(N)} /m_V^2 \\
\mathfrak{c}_{\chi \, 9}^{(N)} = - 8 m_\chi  \, c_{\chi A}   \mathcal{C}_{V}^{(N)} /m_V^2 \\
\mathcal O_8^{\rm NR} =  \vec{s}_\chi \cdot \vec{v}^\perp \\
\mathcal O_9^{\rm NR} = \vec{s}_\chi \cdot (\vec{s}_N \times \vec{q}_{\rm R}) 
\end{array}\right. \\
&\langle  \zeta_{\chi \, 7}^{(N)} \mathcal O_{\chi \, 7}^{(N)}  \rangle \simeq  \mathfrak{c}_{\chi \, 7}^{(N)} \mathcal O_7^{\rm NR} +  \mathfrak{c}_{\chi \, 9}^{(N)} \mathcal O_9^{\rm NR}  \, , \qquad 
\left\{\begin{array}{l}
\mathfrak{c}_{\chi \, 7}^{(N)} = - 8 m_\chi \, m_N \, c_{\chi V}   \mathcal{C}_{A}^{(N)} /m_V^2 \\
\mathfrak{c}_{\chi \, 9}^{(N)} = - 8 m_N  \, c_{\chi V}   \mathcal{C}_{A}^{(N)} /m_V^2 \\
\mathcal O_7^{\rm NR} =  \vec{s}_N \cdot \vec{v}^\perp \\
\mathcal O_9^{\rm NR} = \vec{s}_\chi \cdot (\vec{s}_N \times \vec{q}_{\rm R}) 
\end{array}\right. \\
&\langle  \zeta_{\chi \, 8}^{(N)} \mathcal O_{\chi \, 8}^{(N)}  \rangle \simeq  \mathfrak{c}_{\chi \, 4}^{(N)} \mathcal O_4^{\rm NR} \, ,  \qquad 
\left\{\begin{array}{l}
\mathfrak{c}_{\chi \, 4}^{(N)} = 16 m_\chi \, m_N \, c_{\chi A}   \mathcal{C}_{A}^{(N)} /m_V^2 \\
\mathcal O_4^{\rm NR} = \vec{s}_\chi \cdot \vec{s}_N
\end{array}\right.
\end{split}  \qquad \text{Dirac DM} \ .
\ee
where again we use the same numbering as in Ref.~\cite{DelNobile:2013sia}.  Here,  $\vec{s}_\chi$ and $\vec{s}_N$ are the DM and nucleon spins, $\vec{q}_{\rm R}$ is the nuclear recoil momentum and the transverse velocity is given by $\vec{v}^\perp = \vec{v} + \frac{\vec{q}_{\rm R}}{\mu_{\chi N}}$, where $\vec{v}$ is the DM-nucleon relative velocity.

\smallskip
The spin-averaged amplitude squared for scattering off a target nucleus $\mathcal N$ with mass $m_{\mathcal N}$, can be constructed in terms of a finite set of nuclear response functions $F_{i,j}^{(N,N')}(v^2, q_{\rm R}^2)$ . It reads~\cite{Fitzpatrick:2012ix,DelNobile:2013sia}
\be
\overline{\left|\Mel_{\phi(\chi), \mathcal N}\right|^2} =  \frac{m_{\mathcal N}^2}{m_N^2} \sum_{i,j} \sum_{N, N'} \mathfrak{c}_{\phi(\chi) \, i}^{(N)} \mathfrak{c}_{\phi(\chi) \, j}^{(N')} \, F_{i,j}^{(N,N')}(v^2, q_{\rm R}^2) \ .
\label{eq:MELNucleus}
\ee

A complete set of these functions, for each pair of NR operators ($i,j$),  each pair of nucleons ($N,N'$) and for some target nuclei $\mathcal N$ has been provided in numerical form in the appendix of Ref.~\cite{Fitzpatrick:2012ix}.  
For example, according to the numbering in Eq.~\eqref{eq:MEL}, the nuclear response functions for the standard SI and SD interactions are $F_{1,1}^{(N,N')}$ and $F_{4,4}^{(N,N')}$ respectively.

\smallskip
From the DM-nucleus ME in Eq.~\eqref{eq:MELNucleus} the general expression for the differential scattering cross section is~\cite{Fitzpatrick:2012ix, DelNobile:2013sia, Panci:2014gga}
\be
\frac{{\ud \sigma_{\mathcal N}}}{\ud \ER} \equiv \frac{1}{32\pi}\frac{1}{m_{\rm DM}^2 m_{\mathcal N}} \frac1{v^2} \, \overline{\left|\Mel_{\phi(\chi), \mathcal N}\right|^2} = \frac{1}{32\pi}\frac{m_{\mathcal N}}{m_{\rm DM}^2 m_N^2} \frac1{v^2} \sum_{i,j} \sum_{N, N'} \mathfrak{c}_{\phi(\chi) \, i}^{(N)} \mathfrak{c}_{\phi(\chi) \, j}^{(N')} \, F_{i,j}^{(N,N')}(v^2, q_{\rm R}^2) \ ,
\label{eq:diffXS}
\ee
where $\ER = q_{\rm R}^2/(2m_{\mathcal N})$ is the nuclear recoil energy. 

\smallskip
Having at our disposal the most general relation for the NR differential cross section, the rate of nuclear recoils and in turn the expected number of events in a given detector can be evaluated for all possible types of NR structures coming from Eq.~\eqref{eq:LagEFTnucleon}. Since the detectors in direct searches can be composed of different nuclides with abundances $\xi_{\mathcal N}/m_{\mathcal N}$, the differential rate of nuclear recoils reads
\be
\frac{{\ud R_{\mathcal N}}}{\ud \ER} = \frac{\xi_{\mathcal N}}{m_{\mathcal N}}\frac{\rho_\odot}{m_{\rm DM}}\int_{\vmin(\ER)}^{\vesc} \ud^3 v \, v f_{\rm E}(\vec v) \, \frac{{\ud \sigma_{\mathcal N}}}{\ud \ER}(v, \ER) \ ,
\ee
where $\rho_\odot\simeq0.3$ GeV/cm$^3$ is the local DM energy density and $f_{\rm E}(\vec v)$ is the DM velocity distribution in the Earth's frame. Here, the extremes of integration $\vmin$ and $\vesc$ are the minimal DM velocity providing a nuclear recoil $\ER$  and the DM escape velocity from the Milky Way respectively. In our work, we use the customary Maxwell-Boltzmann velocity distribution with velocity dispersion $v_0 = 220$ km/s. We truncate $f_{\rm E}(\vec v)$ at $\vesc = 544$ km/s.


\section{Projected LZ limits}
\label{app:LZ}

In order to calculate the projected limits from \LUX-\ZEPLIN\ (\LZ), we consider the experimental parameters specified in the Conceptual Design Report \cite{Akerib:2015cja}. Based on this, we assume a total exposure of  $w = 5600$ tonne-days, a uniform efficiency of  $\epsilon = 50\%$ in the energy range from $E_\mathrm{min} = 6 \, \mathrm{keV}_\mathrm{NR}$ to $E_\mathrm{max} = 30 \, \mathrm{keV}_\mathrm{NR}$. We note that the precise details of the efficiency and threshold energy are only relevant at low WIMP masses $m_\chi \lesssim 10 \, \mathrm{GeV}$, so the \LZ\ limits we calculate should be accurate projections over most of the WIMP mass range we consider. 

The total rate of signal events is given by:
\begin{equation}
R = \int_{E_\mathrm{min}}^{E_\mathrm{max}} \epsilon(E_R) \sum_{\mathcal N} \frac{\mathrm{d}R_\mathcal{N}}{\mathrm{d}E_\mathrm{R}} \, \mathrm{d}E_R\,,
\end{equation}
where the sum is over the different species of target nuclei $\mathcal{N}$ in the detector. The differential recoil rate is described in Appendix~\ref{app:NRoperators}. The total number of signal events is simply the rate times the exposure: $N_S = w R$.

We determine the limits based on the total number of events observed (rather than including energy information about individual events). This allows us to easily rescale the limits for different combinations of DM-nucleon operators, as described in Ref.~\cite{DelNobile:2013sia}. The total number of observed events $N_O$ is simply Poisson-distributed, so the upper limit on the number of signal events at the $\gamma\%$ confidence level, $N_S^{\gamma\%}$, is defined by:
\begin{equation}
\sum_{k = 0}^{N_O} P(k | N_{BG} + N_S^{\gamma\%}) = (100 - \gamma)\%\,.
\end{equation}
Here, $P(k | N_{BG} + N_S)$ is the Poisson probability of observing $k$ events, where the expected number of events is the sum of the expected background $N_{BG}$ and signal $N_S$. The expected background for the total exposure is 2.37 events and we present limits assuming that 2 events are observed during the run. The upper limit on the number of signal events $N_S^{\gamma\%}$ can then be converted into an upper limit on the cross section or, equivalently, the mediator mass $m_V$. 

\bibliography{ZprimeDD}

\begin{thebibliography}{10}

\bibitem{Bertone:2004pz}
G.~Bertone, D.~Hooper, and J.~Silk,
\newblock Phys. Rept. {\bf 405}, 279 (2005), [\hhref{hep-ph/0404175}].

\bibitem{Feng:2010gw}
J.~L. Feng,
\newblock Ann. Rev. Astron. Astrophys. {\bf 48}, 495 (2010),
  [arXiv:\hhref{1003.0904}].

\bibitem{Goldberg:1983nd}
H.~Goldberg,
\newblock Phys. Rev. Lett. {\bf 50}, 1419 (1983),
\newblock [Erratum: Phys. Rev. Lett.103,099905(2009)].

\bibitem{Ellis:1983ew}
J.~R. Ellis, J.~S. Hagelin, D.~V. Nanopoulos, K.~A. Olive, and M.~Srednicki,
\newblock Nucl. Phys. {\bf B238}, 453 (1984).

\bibitem{Servant:2002aq}
G.~Servant and T.~M.~P. Tait,
\newblock Nucl. Phys. {\bf B650}, 391 (2003), [\hhref{hep-ph/0206071}].

\bibitem{Cheng:2002ej}
H.-C. Cheng, J.~L. Feng, and K.~T. Matchev,
\newblock Phys. Rev. Lett. {\bf 89}, 211301 (2002), [\hhref{hep-ph/0207125}].

\bibitem{Birkedal:2006fz}
A.~Birkedal, A.~Noble, M.~Perelstein, and A.~Spray,
\newblock Phys. Rev. {\bf D74}, 035002 (2006), [\hhref{hep-ph/0603077}].

\bibitem{Lee:1977ua}
B.~W. Lee and S.~Weinberg,
\newblock Phys. Rev. Lett. {\bf 39}, 165 (1977).

\bibitem{Scherrer:1985zt}
R.~J. Scherrer and M.~S. Turner,
\newblock Phys. Rev. {\bf D33}, 1585 (1986),
\newblock [Erratum: Phys. Rev.D34,3263(1986)].

\bibitem{Srednicki:1988ce}
M.~Srednicki, R.~Watkins, and K.~A. Olive,
\newblock Nucl. Phys. {\bf B310}, 693 (1988).

\bibitem{Kopp:2009et}
J.~Kopp, V.~Niro, T.~Schwetz, and J.~Zupan,
\newblock Phys.Rev. {\bf D80}, 083502 (2009), [arXiv:\hhref{0907.3159}].

\bibitem{Hill:2011be}
R.~J. Hill and M.~P. Solon,
\newblock Phys. Lett. {\bf B707}, 539 (2012), [arXiv:\hhref{1111.0016}].

\bibitem{Frandsen:2012db}
M.~T. Frandsen, U.~Haisch, F.~Kahlhoefer, P.~Mertsch, and K.~Schmidt-Hoberg,
\newblock JCAP {\bf 1210}, 033 (2012), [arXiv:\hhref{1207.3971}].

\bibitem{Haisch:2013uaa}
U.~Haisch and F.~Kahlhoefer,
\newblock JCAP {\bf 1304}, 050 (2013), [arXiv:\hhref{1302.4454}].

\bibitem{Hill:2013hoa}
R.~J. Hill and M.~P. Solon,
\newblock Phys. Rev. Lett. {\bf 112}, 211602 (2014), [arXiv:\hhref{1309.4092}].

\bibitem{Vecchi:2013iza}
L.~Vecchi,
\newblock (2013), [arXiv:\hhref{1312.5695}].

\bibitem{Kopp:2014tsa}
J.~Kopp, L.~Michaels, and J.~Smirnov,
\newblock JCAP {\bf 1404}, 022 (2014), [arXiv:\hhref{1401.6457}].

\bibitem{Crivellin:2014qxa}
A.~Crivellin, F.~D'Eramo, and M.~Procura,
\newblock Phys. Rev. Lett. {\bf 112}, 191304 (2014), [arXiv:\hhref{1402.1173}].

\bibitem{Crivellin:2014gpa}
A.~Crivellin and U.~Haisch,
\newblock Phys.Rev. {\bf D90}, 115011 (2014), [arXiv:\hhref{1408.5046}].

\bibitem{Hill:2014yxa}
R.~J. Hill and M.~P. Solon,
\newblock Phys. Rev. {\bf D91}, 043505 (2015), [arXiv:\hhref{1409.8290}].

\bibitem{D'Eramo:2014aba}
F.~D'Eramo and M.~Procura,
\newblock JHEP {\bf 04}, 054 (2015), [arXiv:\hhref{1411.3342}].

\bibitem{Berlin:2015njh}
A.~Berlin, D.~S. Robertson, M.~P. Solon, and K.~M. Zurek,
\newblock (2015), [arXiv:\hhref{1511.05964}].

\bibitem{D'Eramo:2016mgv}
F.~D'Eramo, J.~de~Vries, and P.~Panci,
\newblock (2016), [arXiv:\hhref{1601.01571}].

\bibitem{Goodman:1984dc}
M.~W. Goodman and E.~Witten,
\newblock Phys. Rev. {\bf D31}, 3059 (1985).

\bibitem{Drukier:1986tm}
A.~K. Drukier, K.~Freese, and D.~N. Spergel,
\newblock Phys. Rev. {\bf D33}, 3495 (1986).

\bibitem{Fitzpatrick:2012ix}
A.~L. Fitzpatrick, W.~Haxton, E.~Katz, N.~Lubbers, and Y.~Xu,
\newblock JCAP {\bf 1302}, 004 (2013), [arXiv:\hhref{1203.3542}].

\bibitem{DeSimone:2016fbz}
A.~De~Simone and T.~Jacques,
\newblock (2016), 1603.08002.

\bibitem{Dudas:2009uq}
E.~Dudas, Y.~Mambrini, S.~Pokorski, and A.~Romagnoni,
\newblock JHEP {\bf 08}, 014 (2009), [arXiv:\hhref{0904.1745}].

\bibitem{Goodman:2011jq}
J.~Goodman and W.~Shepherd,
\newblock (2011), [arXiv:\hhref{1111.2359}].

\bibitem{An:2012va}
H.~An, X.~Ji, and L.-T. Wang,
\newblock JHEP {\bf 07}, 182 (2012), [arXiv:\hhref{1202.2894}].

\bibitem{Frandsen:2012rk}
M.~T. Frandsen, F.~Kahlhoefer, A.~Preston, S.~Sarkar, and K.~Schmidt-Hoberg,
\newblock JHEP {\bf 07}, 123 (2012), [arXiv:\hhref{1204.3839}].

\bibitem{Dreiner:2013vla}
H.~Dreiner, D.~Schmeier, and J.~Tattersall,
\newblock Europhys. Lett. {\bf 102}, 51001 (2013), [arXiv:\hhref{1303.3348}].

\bibitem{Buchmueller:2013dya}
O.~Buchmueller, M.~J. Dolan, and C.~McCabe,
\newblock JHEP {\bf 01}, 025 (2014), [arXiv:\hhref{1308.6799}].

\bibitem{Alves:2013tqa}
A.~Alves, S.~Profumo, and F.~S. Queiroz,
\newblock JHEP {\bf 04}, 063 (2014), [arXiv:\hhref{1312.5281}].

\bibitem{Arcadi:2013qia}
G.~Arcadi, Y.~Mambrini, M.~H.~G. Tytgat, and B.~Zaldivar,
\newblock JHEP {\bf 03}, 134 (2014), [arXiv:\hhref{1401.0221}].

\bibitem{Lebedev:2014bba}
O.~Lebedev and Y.~Mambrini,
\newblock Phys. Lett. {\bf B734}, 350 (2014), [arXiv:\hhref{1403.4837}].

\bibitem{Abdullah:2014lla}
M.~Abdullah {\em et~al.},
\newblock Phys. Rev. {\bf D90}, 035004 (2014), [arXiv:\hhref{1404.6528}].

\bibitem{Bell:2014tta}
N.~F. Bell, Y.~Cai, R.~K. Leane, and A.~D. Medina,
\newblock Phys. Rev. {\bf D90}, 035027 (2014), [arXiv:\hhref{1407.3001}].

\bibitem{Buchmueller:2014yoa}
O.~Buchmueller, M.~J. Dolan, S.~A. Malik, and C.~McCabe,
\newblock JHEP {\bf 01}, 037 (2015), [arXiv:\hhref{1407.8257}].

\bibitem{Harris:2014hga}
P.~Harris, V.~V. Khoze, M.~Spannowsky, and C.~Williams,
\newblock Phys. Rev. {\bf D91}, 055009 (2015), [arXiv:\hhref{1411.0535}].

\bibitem{Alves:2015pea}
A.~Alves, A.~Berlin, S.~Profumo, and F.~S. Queiroz,
\newblock (2015), [arXiv:\hhref{1501.03490}].

\bibitem{Jacques:2015zha}
T.~Jacques and K.~Nordström,
\newblock JHEP {\bf 06}, 142 (2015), [arXiv:\hhref{1502.05721}].

\bibitem{Chala:2015ama}
M.~Chala, F.~Kahlhoefer, M.~McCullough, G.~Nardini, and K.~Schmidt-Hoberg,
\newblock JHEP {\bf 07}, 089 (2015), [arXiv:\hhref{1503.05916}].

\bibitem{Alves:2015mua}
A.~Alves, A.~Berlin, S.~Profumo, and F.~S. Queiroz,
\newblock (2015), [arXiv:\hhref{1506.06767}].

\bibitem{Alves:2015dya}
A.~Alves and K.~Sinha,
\newblock Phys. Rev. {\bf D92}, 115013 (2015), 1507.08294.

\bibitem{Gupta:2015lfa}
A.~Gupta, R.~Primulando, and P.~Saraswat,
\newblock JHEP {\bf 09}, 079 (2015), [arXiv:\hhref{1504.01385}].

\bibitem{Autran:2015mfa}
M.~Autran, K.~Bauer, T.~Lin, and D.~Whiteson,
\newblock Phys. Rev. {\bf D92}, 035007 (2015), [arXiv:\hhref{1504.01386}].

\bibitem{Bai:2015nfa}
Y.~Bai, J.~Bourbeau, and T.~Lin,
\newblock JHEP {\bf 06}, 205 (2015), [arXiv:\hhref{1504.01395}].

\bibitem{DelNobile:2013sia}
M.~Cirelli, E.~Del~Nobile, and P.~Panci,
\newblock JCAP {\bf 1310}, 019 (2013), [arXiv:\hhref{1307.5955}].

\bibitem{Bell:2015sza}
N.~F. Bell, Y.~Cai, J.~B. Dent, R.~K. Leane, and T.~J. Weiler,
\newblock Phys. Rev. {\bf D92}, 053008 (2015), [arXiv:\hhref{1503.07874}].

\bibitem{Bell:2015rdw}
N.~F. Bell, Y.~Cai, and R.~K. Leane,
\newblock JCAP {\bf 1601}, 051 (2016), [arXiv:\hhref{1512.00476}].

\bibitem{Haisch:2016usn}
U.~Haisch, F.~Kahlhoefer, and T.~M.~P. Tait,
\newblock (2016), [arXiv:\hhref{1603.01267}].

\bibitem{Buttazzo:2013uya}
D.~Buttazzo {\em et~al.},
\newblock JHEP {\bf 12}, 089 (2013), [arXiv:\hhref{1307.3536}].

\bibitem{Xing:2011aa}
Z.-z. Xing, H.~Zhang, and S.~Zhou,
\newblock Phys. Rev. {\bf D86}, 013013 (2012), [arXiv:\hhref{1112.3112}].

\bibitem{Pivovarov:2000cr}
A.~A. Pivovarov,
\newblock Phys. Atom. Nucl. {\bf 65}, 1319 (2002), [\hhref{hep-ph/0011135}],
\newblock [Yad. Fiz.65,1352(2002)].

\bibitem{Cheng:2012qr}
H.-Y. Cheng and C.-W. Chiang,
\newblock JHEP {\bf 07}, 009 (2012), [arXiv:\hhref{1202.1292}].

\bibitem{Fitzpatrick:2012ib}
A.~L. Fitzpatrick, W.~Haxton, E.~Katz, N.~Lubbers, and Y.~Xu,
\newblock (2012), [arXiv:\hhref{1211.2818}].

\bibitem{Akerib:2013tjd}
LUX, D.~S. Akerib {\em et~al.},
\newblock Phys. Rev. Lett. {\bf 112}, 091303 (2014), [arXiv:\hhref{1310.8214}].

\bibitem{Akerib:2015cja}
LZ, D.~S. Akerib {\em et~al.},
\newblock (2015), [arXiv:\hhref{1509.02910}].

\bibitem{D'Ambrosio:2002ex}
G.~D'Ambrosio, G.~F. Giudice, G.~Isidori, and A.~Strumia,
\newblock Nucl. Phys. {\bf B645}, 155 (2002), [\hhref{hep-ph/0207036}].

\bibitem{Boveia:2016mrp}
G.~Busoni {\em et~al.},
\newblock (2016), [arXiv:\hhref{1603.04156}].

\bibitem{Aaboud:2016tnv}
ATLAS, M.~Aaboud {\em et~al.},
\newblock (2016), [arXiv:\hhref{1604.07773}].

\bibitem{Amole:2015lsj}
PICO, C.~Amole {\em et~al.},
\newblock Phys. Rev. Lett. {\bf 114}, 231302 (2015),
  [arXiv:\hhref{1503.00008}].

\bibitem{Aaboud:2016uro}
ATLAS, M.~Aaboud {\em et~al.},
\newblock (2016), 1604.01306.

\bibitem{WebPlot}
A.~Rohatgi,
\newblock Webplotdigitizer v3.10,
\newblock \url{http://arohatgi.info/WebPlotDigitizer},
\newblock Accessed: 30 March 2016.

\bibitem{Panci:2014gga}
P.~Panci,
\newblock Adv. High Energy Phys. {\bf 2014}, 681312 (2014),
  [arXiv:\hhref{1402.1507}].

\end{thebibliography}
\bibliographystyle{h-physrev}

\end{document}